\RequirePackage{fix-cm}
\documentclass[12pt,a4paper]{book}
\usepackage{sourcesanspro} 
\DeclareRobustCommand{\sourcesanspro}{\fontfamily{SourceSansPro-LF}\selectfont}
\renewcommand{\sfdefault}{lmss} 

\usepackage{packages}
\newcommand{\expnumber}[2]{%
  #1\mathrm{e}\,%
  \IfBeginWith{#2}{-}
    {\text{-}\StrGobbleLeft{#2}{1}}
    {#2}%
}

\DeclareMathOperator*{\argmax}{argmax}
\newcommand{\unitmatrix}{\mathds{1}}

\newgeometry{
a4paper,
top=20mm,
bottom=20mm,
lmargin=15mm,
rmargin=15mm
}
\savegeometry{MAIN_DOCUMENT}

\titleformat{\chapter}
  {\bfseries\huge}
  {\LARGE Chapter \thechapter.}
  {1em}
  {\LARGE\raggedright}
  [\vspace{8pt}\titlerule]

\titlespacing*{\chapter}
  {0pt}
  {-20pt}  
  {15pt}   
  
\newcommand{\minichapter}[1]{%
  \par\vspace{0pt}%
  {\bfseries\Large\raggedright
   #1\par}%
  \vspace{8pt}\hrule\vspace{10pt}%
}

\newif\ifmainmatterstarted
\mainmatterstartedfalse
\pretocmd{\chapter}{\mainmatterstartedtrue}{}{}

\makeatletter
  \def\vhrulefill#1{\leavevmode\leaders\hrule\@height#1\hfill \kern\z@}
\makeatother

\excludecomment{hidepages} 
\includecomment{hidepages}

\makeatletter
\patchcmd{\markboth}{\protected@xdef\@themark}{\protected@xdef\@themark{\protect\trimspaces}}{}{}
\makeatother

\makeatletter
\renewcommand{\chaptermark}[1]{%
  \ifmainmatterstarted
    \ifnum\value{chapter}>0
      \markboth{{\sourcesanspro \textsc{Chapter \thechapter.\ #1}}}{}%
    \else
      \markboth{{#1}}{}%
    \fi
  \else
    \markboth{}{}%
  \fi
}

\newif\ifinappendix
\inappendixfalse

\definecolor{bluetitle}{RGB}{29,141,176}
\definecolor{blueaff}{RGB}{0,0,128}
\definecolor{blueline}{RGB}{82,189,236}
\definecolor{KUL_dark}{RGB}{18,60,117}
\begin{document}

\thispagestyle{empty}

\newcommand{\form}[1]{\scalebox{1.087}{\boldmath{#1}}}
\sffamily
\begin{textblock}{191}(-24,-11)
\colorbox{bluetitle}{\hspace{139mm}\ \parbox[c][18truemm]{52mm}{\textcolor{white}{FACULTY OF SCIENCE}}}
\end{textblock}
\begin{textblock}{70}(-18,-19)
\textblockcolour{}
\includegraphics*[height=19.8truemm]{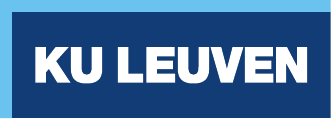}
\end{textblock}
\begin{textblock}{160}(-6,63)
\textblockcolour{}
\vspace{-\parskip}
\flushleft
\fontsize{30}{33}\selectfont \textcolor{bluetitle}{Reduced Order Modelling for Nuclear Linear Response and the Incompressibility of \ce{Pb-208}}\\[1.5mm]
\fontsize{20}{22}\selectfont {}
\end{textblock}
\begin{textblock}{82}(50,103)
\textblockcolour{}
\vspace{-\parskip}
\flushleft
\end{textblock}
\begin{textblock}{160}(8,153)
\textblockcolour{}
\vspace{-\parskip}
\flushright
\fontsize{14}{16}\selectfont \textbf{Emma VANCAYSEELE}
\end{textblock}
\begin{textblock}{70}(-6,191)
\textblockcolour{}
\vspace{-\parskip}
\flushleft
Supervisor:\\ \; Dr.~Wouter~Ryssens \\[-2pt]
\textcolor{KUL_dark}{\; Université~libre~de~Bruxelles}\\[5pt]
Co-supervisor:\\\; Prof.~Dr.~Riccardo~Raabe\\[-2pt]
\textcolor{KUL_dark}{\; KU Leuven}\\[5pt]
Mentor:\\\; Dr.~Pepijn~Demol\\[-2pt]
\textcolor{KUL_dark}{\; Université~libre~de~Bruxelles}\\
\end{textblock}
\begin{textblock}{160}(8,191)
\textblockcolour{}
\vspace{-\parskip}
\flushright
Thesis presented in\\[4.5pt]
fulfillment of the requirements\\[4.5pt]
for the degree of Master of Science\\[4.5pt]
in Physics\\
\end{textblock}
\begin{textblock}{160}(8,232)
\textblockcolour{}
\vspace{-\parskip}
\flushright
Academic year 2025-2026
\end{textblock}
\begin{textblock}{191}(-24,248)
{\color{blueline}\rule{550pt}{5.5pt}}
\end{textblock}
\vfill
\newpage


\renewcommand{\familydefault}{\sfdefault}

\setcounter{page}{0}
\pagenumbering{roman}

\vspace*{\fill}
\begin{center}
© Copyright by KU Leuven\par
\vspace{0.5em}
Without written permission of the promotors and the authors it is forbidden to reproduce or adapt in any form or by any means any part of this publication. Requests for obtaining the right to reproduce or utilize parts of this publication should be addressed to KU Leuven, Faculteit Wetenschappen, Celestijnenlaan 200H -- bus 2100, 3001 Leuven (Heverlee), Telephone +32 16 32 14 01.\par
\vspace{0.5em}
A written permission of the promotor is also required to use the methods, products, schematics and programs described in this work for industrial or commercial use, and for submitting this publication in scientific contests.
\end{center}

\newpage

\minichapter{Contribution statement}
The research topic was conceptualized by Wouter Ryssens, who also provided supervision with Pepijn Demol. The MOCCa software, used for mean-field and (Q)FAM calculations and serving as the starting point of the emulator development, was provided by W.R., P.D., and Luis González-Miret Zaragoza. The reference data using the Gogny interaction for testing the emulator was provided by L.G.-M.Z., and using the Skyrme interactions by P.D. on Lucia (a large-scale Tier-1 supercomputer). 

I contributed to all remaining aspects of the work, in particular the development and implementation of the emulator, the analysis of the results, benchmarking, and writing of the manuscript.

\minichapter{Declaration on the use of GenAI}
I used the generative AI assistance tools ChatGPT, GitHub Copilot and Google AI Studio. 

Generative AI was used as a language assistant for
reviewing or improving texts I wrote myself. Generative AI was used for generating programming code. Specifically, it was used to: (i) assist in generating plotting scripts, (ii) refactor original, author-written python code to significantly improve performance with Numba, (iii) generate parser scripts for handling different data formats and (iv) assist in documenting code. The text/code/images in this thesis are my own (unless otherwise specified) and generative AI has only been used in accordance with the KU Leuven guidelines and appropriate references have been added. I have reviewed and edited the content as needed and I take full responsibility for the content of the thesis.

\begin{hidepages}
\newpage
\minichapter{Abstract}
Linear response theory provides essential information regarding the excitations of many-body systems, such as atomic nuclei. It yields ground state transition probabilities, or strength functions, from which reaction rates and cross-sections can be derived. These quantities are for example a critical input for astrophysical simulations and modelling of $\beta$-decay. Currently, the most general theoretical framework for modelling global nuclear properties is Energy Density Functional (EDF) theory. 
 
Modern approaches for linear response employ the quasiparticle random-phase approximation (QRPA) on top of a mean-field vacuum. This can be done by using conventional matrix-QRPA formulations, but can be sped up substantially by using the finite amplitude method (FAM). Nevertheless, obtaining highly-resolved response functions over the complete nuclear chart remains computationally demanding, which limits large-scale applications.

In this work, we propose a Reduced Order Modelling (ROM) approach to emulate FAM-QRPA calculations. A single FAM-QRPA calculation yields the response to a given excitation operator at a given excitation energy. The emulator is constructed from a small set of FAM-QRPA solutions at specific excitation energies, which we refer to as snapshots. Once built, the emulator is capable of accurately interpolating over the complete excitation energy range for a negligible numerical cost.

We investigate several ways to construct a ROM emulator and benchmark them against full FAM-QRPA strength functions. We find that a modified 2D-greedy strategy, which combines snapshots at different points in the complex-frequency plane, provides a good balance between information gain and computation speed. Although our benchmark underestimates the efficiency, the proposed approach already achieves a $\times$20 speed-up compared to FAM-QRPA evaluation of the complete strength spectrum. Furthermore, the methodology is validated for different types of energy density functionals, parametrisations, computational bases, nuclei and for monopole, dipole and quadrupole electric operators.

A second objective of this work is to investigate the correlation between the infinite nuclear matter property $K_\infty$ and the ISGMR centroid position of \ce{Pb-208}, specifically for EDF forms and parametrisations developed in Brussels: the BSk(G)-family. The $K_\infty$ is commonly used as a constraint in the fitting procedure of the EDF parameters, despite its empirical value not being known up to large precision. Previous studies, e.g.~\textcite{Colo2004}, reported a strong correlation between $K_\infty$ and the ISGMR centroid energy in \ce{Pb-208} for standard Skyrme parameters. Following their analysis, we examine whether the same correlation persists within the BSk(G) family. Our results indicate that the centroid energy no longer exhibits a clear correlation with $K_\infty$. Consequently, EDF parametrisations with substantially different values of $K_\infty$ may reproduce similar centroid energies. 

We conclude by proposing to employ the centroid position of \ce{Pb-208} instead of $K_\infty$ the fitting procedure. Although, finding this centroid position would significantly increase the computational cost of the fitting procedure, even with future improvements in the FAM-QRPA method. 

\newpage
\minichapter{Summary for the general audience}
All matter is made out of atoms. These atoms consist of a tiny nucleus surrounded by a cloud of electrons. While the electron cloud determines the chemical properties, the nucleus determines the total mass and stability of an atom. To understand the nucleus, scientists study the \textit{response}, i.e. how the nucleus reacts when it is disturbed by an external probe, such as being hit by a high-energy photon. 

We can learn about nuclear response via theory and computer simulations, which is particularly useful when experimental data is unavailable. The nuclear response can be mathematically described via linear response theory, which yield so-called \textit{strength functions}. Such functions are a map of the nucleus excitation probabilities for different energy values; i.e.~how \textit{strong} the nucleus reacts to a given type of excitation at a specific energy. These are essential for calculating reaction rates, which are primary inputs for astrophysical simulations and for understanding $\beta$-decay. 

However, obtaining high-resolution strength functions for all nuclei requires significant computing resources. So far, efforts have been made to make these computations more feasible. For example, the Finite Amplitude Method finds the strength for a specific excitation energy reasonably efficiently. However, doing this for the full energy spectrum and all nuclei remains difficult. 

This thesis aims to improve the calculations even further by developing an emulator. This is a highly efficient tool that approximates the output of expensive simulations. Specifically, this uses only a few expensive calculations at specific energy values, after which it can find the strength across the whole energy spectrum with negligible effort. 

A second objective of the thesis is related to the nuclear incompressibility. This tells us how much nuclear matter can (or cannot) be compressed. This is highly relevant for astrophysics, as this determines for example the size and density of neutron stars. A crucial quantity is the incompressibility of infinite matter, which is only defined for an infinite system. Since it cannot be measured directly, its value must be inferred from theoretical models that connect it to finite nuclear matter, which can be studied experimentally. This relation is not well understood and leads to large uncertainties on the value, or even, conflicting results depending on the context. Nevertheless, the value of the incompressibility has also been used for modelling nuclei. Consequently, also a correlation between a strength function centroid position of \ce{Pb-208} and the infinite matter value was found in literature, for a standard model type.

In this thesis, we aim to verify whether or not this correlation still holds for non-standard models, and question the way how infinite matter is used currently.

\newpage
\minichapter{List of abbreviations}
\begin{flushleft}
     \renewcommand{\arraystretch}{1.1}
  \begin{tabularx}{\textwidth}{@{}p{40mm}X@{}}
  \toprule
    \text{\bfseries Abbreviation}   & Description \\
    \midrule
    BSk & Brussels-Montreal Skyrme (HO basis) \\
    BSkG & Brussels-Montreal Skyrme on a Grid \\
    D1M & Parametrisation for Gogny interaction\\
    EDF & Energy Density Functional \\
    EoS & Equation of State \\
    FAM & Finite Amplitude Method \\
    G&Galerkin\\
    HF(B)& Hartree-Fock(-Bogoliubov)\\
    INM & Infinite Nuclear Matter \\
    ISGMR & IsoScalar Giant Monopole Resonance\\
    ISGQR & IsoScalar Giant Quadrupole Resonance \\
    IVGDR & IsoVector Giant Dipole Resonance \\
    MOCCa &Modular Cranking Code\\
    NLO & Next-to-Leading Order \\
    N2LO & Next-to-Next-to-Leading Order \\
    PG&Petrov-Galerkin\\
    POD&Proper Orthogonal Decomposition\\
    QRPA & Quasiparticle Random Phase Approximation \\
    RMS&Root mean square\\
    ROM & Reduced Order Model \\
    RPA & Random Phase Approximation \\
    SLy4 & Skyrme-Lyon 4 parametrisation \\
    SVD& Singular Value Decomposition\\
    TDHF(B) & Time-Dependent Hartree-Fock(-Bogoliubov)\\
    \bottomrule
  \end{tabularx}
\end{flushleft}

\newpage
\minichapter{List of important symbols}
\begin{flushleft}
 \renewcommand{\arraystretch}{1.1}
  \begin{tabularx}{\textwidth}{@{} >{$}p{25mm}<{$} X p{30mm} @{}}
  \toprule
    \text{\bfseries Symbol} & Description & Units \\
    \midrule
    \hat{H} & Nuclear Hamiltonian & MeV \\
    \hat{H}_0 & Mean field Hamiltonian & MeV \\
    \hat{V}_\mathrm{res} &Residual interaction & MeV \\
    \mathcal{E} & Energy density & MeV fm$^{-3}$ \\
    F & Excitation operator (monopole, dipole, quadrupole) & (fm$^2$, fm$^1$, fm$^2$)\\
    X_{\mu\nu} & Forward FAM amplitude&[$F$]$^2$MeV$^{-1}$\\
    Y_{\mu\nu} & Backward FAM amplitude& [$F$]$^2$MeV$^{-1}$\\
    S_\mathrm{FAM} & FAM strength function& [$F$]$^2$MeV$^{-1}$\\
    S_\mathrm{ISGMR} & Nuclear isoscalar monopole strength & fm$^4$\,MeV$^{-1}$\\
    S_\mathrm{IVGDR} & Nuclear isovector dipole strength & fm$^2$\,MeV$^{-1}$ \\
    S_\mathrm{ISGQR} & Nuclear isoscalar quadrupole strength &  fm$^4$\,MeV$^{-1}$\\
    \gamma & Smearing parameter & MeV \\
    \omega_R & Real part of the complex frequency ($\omega = \omega_R+i\gamma$) & MeV \\
    \midrule
     \mathcal{X} & Snapshot vector& [$F$]$^2$MeV$^{-1}$\\
     \mathcal{F} &External force vector& [$F$] \\
     \mathcal{V},\, \mathcal{U} & Snapshot matrix,\, SVD& [$F$]$^2$MeV$^{-1}$\\
     \alpha(\omega),\, \beta(\omega) &Coefficients in the reduced space,\, SVD&1\\
     \mathbf{W},\,\mathbf{W}_i & Greedy search space along real axis & MeV\\
     \mathbf{W}_r & Greedy search space along complex axis& MeV\\
     \eta &Error estimator& 1\\
     r(\omega) & Residual of the FAM equation & [$F$] \\
     \epsilon_\mathrm{RMS} & Root mean square error & [$F$]$^2$ MeV$^{-1}$\\
     \tau & threshold value for pseudo 2D-greedy& 1 \\
     \midrule
    \rho & Nuclear matter density & fm$^{-3}$ \\
    \rho_*& Infinite nuclear matter density & fm$^{-3}$\\
    \rho_n & Neutron nuclear matter density & fm$^{-3}$\\
    \rho_p & Proton nuclear matter density & fm$^{-3}$ \\
    \rho_0 & Total nuclear matter density & fm$^{-3}$\\
    \rho_1 & Nuclear matter density asymmetry & fm$^{-3}$ \\
    \tau & Nuclear kinetic density &  fm$^{-5}$\\
    Q & Higher-order kinetic density & fm$^{-7}$\\
    \mathcal{E}_K& Nuclear kinetic energy density &  MeV\,fm$^{-3}$\\
    K_A & Nuclear finite matter incompressibility & MeV \\
    K_\infty & Nuclear infinite matter incompressibility & MeV\\
    E_\mathrm{ISGMR} & Centroid position of the monopole resonance & MeV\\
    \bottomrule
\end{tabularx}
\end{flushleft}

\end{hidepages}

\makeatletter
\clearpage
\minichapter{Contents}
\@starttoc{toc}  
\makeatother

\newpage

\setcounter{page}{0}
\pagenumbering{arabic}

\loadgeometry{MAIN_DOCUMENT}
\fontsize{12}{14}
\selectfont
\pagestyle{fancy} 

\chapter{Introduction}\label{chap:1}
The field of nuclear physics provides the framework for understanding atomic nuclei, which account for more than 99.9\% of the visible mass in our universe. Understanding nuclei is important for astrophysics, as their properties determine the structure and evolution of stars, the synthesis of elements and the dynamics of neutron stars. Beyond its theoretical importance, a precise understanding of atomic nuclei is also essential for the advancement of technologies ranging from nuclear medicine to energy production through fusion and fission processes.

\section{Nuclear modelling}
Our understanding of the nucleus is defined by observables such as nuclear masses, charge radii, excitation spectra, lifetimes and transition probabilities. Ideally, these properties would be measured experimentally, tabulated and used as high-precision inputs for larger-scale simulations. 

\begin{figure}
    \centering
    \includegraphics[width=0.85\linewidth]{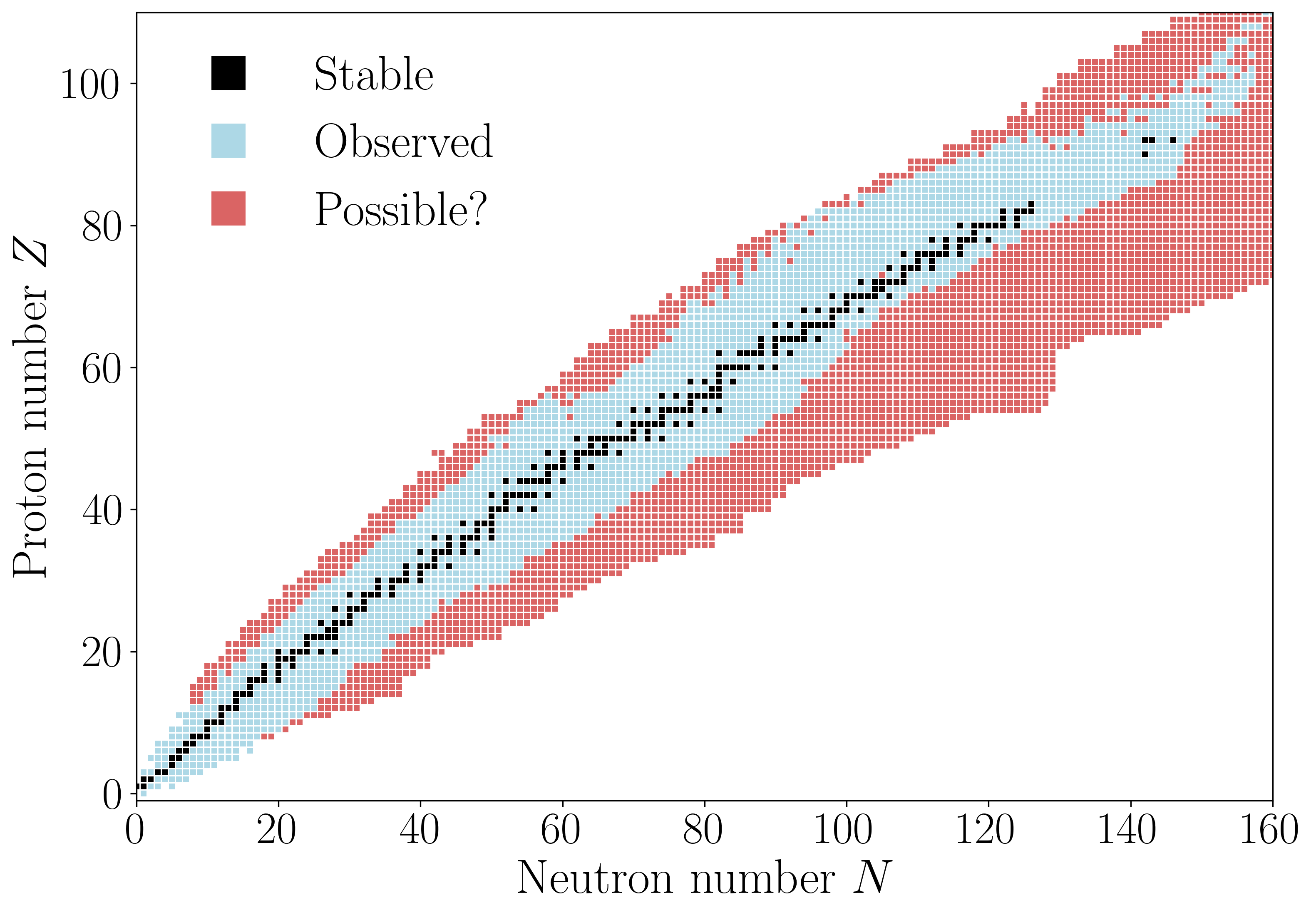}
    \caption{Nuclear chart indicating stable and observed nuclei and theoretically possible nuclei that have not yet been observed experimentally. Courtesy of W.R.~\cite{Wryssens_private}.}
    \label{fig:nuclear_chart}
\end{figure}

As illustrated in the nuclear chart in Fig.~\ref{fig:nuclear_chart}, we have observed only a fraction of the isotopes that can exist theoretically. While extensive data exist for several stable and long-lived nuclei, measuring the properties of highly unstable, short-lived isotopes remains extremely difficult. Consequently, experimental data are either limited or completely unavailable for vast regions of the nuclear chart. For example, to model the rapid-neutron capture process, the reaction and decay rates of several thousand extremely neutron-rich isotopes are required~\cite{Grams2026}. 

This lack of data creates a bottleneck for astrophysical simulations, which must instead rely on theoretical predictions. Therefore, robust and reliable theoretical models are necessary. Only if such theoretical models can be trusted, can we extrapolate and obtain new data that was otherwise unachievable. Furthermore, to achieve a global understanding of nuclear phenomena, we require frameworks that are computationally feasible and applicable across the entire nuclear chart. 

At this moment, the most general theoretical framework for modelling global nuclear properties is Energy Density Functional (EDF) theory~\cite{Bender2003}. It is similar, though not equivalent, to density functional theory (DFT) used in condensed matter and atomic physics. 

\section{Energy density functionals and collective motion}
A nucleus is a complex quantum many-body system. Solving the many-body Schrödinger equation exactly for heavy nuclei is computationally prohibitive. While static mean-field theory describes the ground state of a nuclear system, we are interested in its dynamic response to external perturbations.

This response is characterised by the strength function, $S(E)$, which describes the transition probabilities as a function of the excitation energy. The strength function represents the response of a nucleus to a specific external operator, for example an electromagnetic multipole operator or the Fermi and Gamow-Teller operators in $\beta$-decay.  At low energies, peaks in the strength function correspond to discrete bound states, whereas at higher energies, they represent resonances. These resonances are considered collective when they arise from coherent motion of many nucleons. A giant resonance is a specific type of collective motion, characterised by a broad peak in the strength function.  For instance, the giant monopole resonance represents a "breathing mode" where the nucleus as a whole periodically increases and decreases in size, while the giant quadrupole resonance involves oscillations in the shape of the nucleus. 

Modelling these strength functions is essential for predicting photo-absorption cross-sections, where the dominant contributions are from E1 and M1 transitions and the associated giant resonances.  These cross sections are, in turn, critical for calculating reaction rates in nucleosynthesis~\cite{Wiedeking}.

To model these resonances, we go beyond the static mean-field and apply the Quasiparticle Random Phase Approximation (QRPA)~\cite{RingSchuck1980}. While this is a powerful framework that builds excitations on top of the mean-field vacuum, it is computationally expensive as it requires the manipulation of extremely large matrices. 

To make these calculations manageable, the Finite Amplitude Method (FAM) is used~\cite{Nakatsukasa2007, Avogadro2011}. FAM allows solving the QRPA equations without explicitly constructing such matrices, significantly reducing the computational cost. However, even with FAM, calculating the strength functions for every excitation operator and every isotope across the nuclear chart remains a difficult task. So far, only a limited number of global evaluations of the photon strength functions across the nuclear chart are available~\cite{daoutidis2012, gonzalez-miretzaragoza2025, goriely2004, goriely2016, goriely2025, martini2016, xu2021, Wiedeking, goriely2019reference}, and even less so for $\beta$-decay~\cite{marketin2016, martini2014, ney2020, ravlic2026}.

Linear response theory can be applied in many fields of physics. While we focus on strength functions relevant for nuclear reactions and decay processes, a similar formalism appears in quantum chemistry, condensed matter physics or in materials science when studying electronic excitations and ground-state correlation energies~\cite{chen, Ziegler, Ren_2012}\cite[Sec. 5.3]{DFT}.

\section{EDF fitting and nuclear incompressibility}
Building an EDF consists of two parts. First, a specific ansatz in terms of densities has to be proposed, which determines what terms are included in the functional. Second, these terms are multiplied by coupling parameters, and these have to be fitted to nuclear and astrophysical data. One of the astrophysical inputs is the infinite nuclear matter incompressibility $K_\infty$, which describes how the energy of nuclear matter changes under compression. 

The infinite matter incompressibility is an important parameter in the  equation of state for neutron stars, and is used for modelling astrophysical effects such as core-collapse supernova explosions~\cite{Garg2018} or the signals emitted from neutron stars and gravitational waveforms following the merger of two dense stars~\cite{Margueron2026}.

Historically, $K_\infty$ was constrained by the centroid position of the Isoscalar Giant Monopole Resonance (ISGMR), particularly in \ce{Pb-208}. While $K_\infty$ is a property of infinite matter, the resonance observed experimentally is influenced by surface, Coulomb, and asymmetry effects. Extracting $K_\infty$ from finite nuclei is not straightforward. However, the ISGMR can be directly related to the finite nucleus incompressibility $K_A$, which in turn can be related to $K_\infty$ via theoretical models. 

In the literature, many attempts have been made to establish a relationship between the finite-nucleus $K_A$ and the infinite-matter $K_\infty$, yet these efforts often yield model-dependent and, consequently, incompatible results~\cite{Colo2004, Garg2018, stone, Khan2012}. By creating different models using a fixed $K_\infty$ value, a linear correlation between the ISGMR centroid energies and $K_\infty$ was found for standard and relativistic Skyrme EDF forms~\cite{Colo2004, Sagawa2019, Zamora2024}. Consequently, the bounds on the $K_\infty$ parameter naturally arise.

Although its value is not well understood, including $K_\infty$ as a constraint to the EDF fitting procedure is crucial to reduce the degrees of freedom. Hence, conventionally, $K_\infty$ is taken with large uncertainties around 240$\pm$20\,MeV~\cite{Colo2004, Garg2018, stone, Khan2012}. 

\section{Aims and structure of the thesis}
Over the last few years, increased attention has been drawn towards emulation to bypass computational bottlenecks. An emulator is a tool designed to mimic the output that would be obtained by the original methodology for a fraction of the computational cost. An emulator needs to be fast and within reasonable accuracy in order to be effective. 

In this work, we propose the use of a Reduced Order Modelling (ROM) technique to create an emulator for the FAM-QRPA framework. By projecting the problem onto a significantly smaller, optimized subspace, we aim to bypass additional FAM calculations that would be needed for obtaining the full strength function. 

As a second goal, we aim to verify the correlation that was found in~\cite{Colo2004} between the \ce{Pb-208} ISGMR centroid and $K_\infty$ for the the BSk(G) parametrisation. These are parametrisations of extended-Skyrme functionals, and crucially, are different from the standard Skyrmes considered in~\cite{Colo2004}. Nevertheless, the parameter fitting procedure on these extended Skyrme functionals included the same $K_\infty$ constraint, although it was never thoroughly verified; if said correlation does not hold, the possibility arises that the EDF parameters are over-constrained, and that an unconstrained parametrisation might reproduce a correct \ce{Pb-208} centroid while yielding a $K_\infty$ value outside of the conventional range. 

The thesis is structured as follows. Following an introductory overview in  Chap.~\ref{chap:1}, the necessary theoretical background is established Chap.~\ref{chap:2}. Chap.~\ref{chap:emulator} constitutes the main body of this work, detailing the development and benchmarking of the FAM-QRPA emulator. The developed code is available on GitHub~\cite{code_github}. In Chap.~\ref{chap:4}, we check if the correlation with the \ce{Pb-208} ISGMR centroid persists for the BSk(G) family. The work concludes in Chap.~\ref{chap:5} with a summary of the results and suggestions for future research. 
\chapter{Theoretical background}\label{chap:2}

Accurately modelling nuclei is challenging. First, the exact form of the nuclear Hamiltonian, $\hat{H}$, is not known from first principles. Second, even for a chosen model of the interaction, the resulting many-body Schrödinger equation is often impossible to solve. 

From a theoretical perspective, the nuclear problem consists of finding the eigenstates $\ket{\Psi_i}$ and energies $E_i$ that satisfy:
\begin{equation}
\hat{H}\ket{\Psi_i} = E_i\ket{\Psi_i} \;.\label{eq:nuclear_hamiltionian_basicbasic}
\end{equation}
The nuclear Hamiltonian consists of the standard kinetic and potential terms; the latter includes short-ranged nucleon-nucleon interactions. In principle, a microscopic description can be achieved by including the interaction between every nucleon and every other nucleon explicitly. Such approaches (often referred to as \textit{ab initio} methods) are feasible for describing properties of light nuclei, e.g. \ce{^3H} and \ce{^3He}~\cite{abinitio:chen, abinitio:friar}, where the number of interactions is limited. As more nucleons are added, the many-body equation becomes analytically unsolvable, and numerical solutions quickly become computationally prohibitive. Ab initio nuclear physics is an active field of research mainly due to its predictive power, stemming from the use of fundamental interactions rather than empirical fitting. While these methods have been extended to certain heavy nuclei up to $A\approx100$~\cite{abinitio:navratil, soma2018} and last year for doubly magic \ce{Pb-208} and \ce{Pb-266}~\cite{abinitio:Pb}, the computational cost remains a significant barrier for many systems.

Alternatively, while effective theories allow for a reduced computational cost, they compromise predictive power by making a phenomenological ansatz for the Hamiltonian that must somehow include many-body correlations. Energy density functional theory directly postulates the total energy as a function of the nuclear density, allowing for more freedom to incorporate many-body correlations~\cite{soma2018}. Additionally, the simplicity of the EDF form eases inclusion of symmetry-breaking and restoration~\cite{soma2018}. However, the main drawback of EDF theory is that functionals are not constructed in a systematic way, such that there is no logical way to improve a functional.

In Sec.~\ref{sec:mf_hfb}, we assume a nuclear Hamiltonian is known and focus on solving the Schrödinger equation by applying the mean-field approximation. Although this theory is well-established in the literature, we repeat it here as it is fundamental for the remainder of the thesis. Next, we derive the Random Phase Approximation (RPA) equations starting from linear response, describing nuclear transitions in Sec.~\ref{sec:LR_QRPA}. The Finite Amplitude Method (FAM) is used to solve the RPA equations and is discussed in Sec.~\ref{sec:FAM}. Next, EDF theory is introduced in Sec.~\ref{sec:EDF} using the Skyrme functional and the relevant parametrisations for Chap.~\ref{chap:emulator} and~\ref{chap:4} are summarised.

\section{Mean-Field and Hartree-Fock-Bogoliubov Theory}\label{sec:mf_hfb}
\subsection{Mean-Field}
We formulate the  mean-field theory within the second quantisation formalism. We first introduce a vacuum state $\ket{0}$, which represents a state without particles. The vacuum state is normalised by construction: $\bra{0}\ket{0}=1$. We now consider a set of single-particle wave functions $\{\phi_\mu\}$ which are solutions to the eigenvalue problem of the nuclear single-particle Hamiltonian. These are characterised by (good) nuclear quantum numbers, e.g. $(n, \ell, j, m, \tau)$ for spherical systems, labelled by the index $\mu$~\cite{Schunck2019}. Here, $n$ is the principal quantum number, $\ell$ is the orbital angular momentum, $j$ the total angular momentum ($j=\ell \pm \tfrac{1}{2}$), $m$ the projection of the total angular momentum, $\tau$ the isospin and the parity $\pi=(-1)^\ell$~\footnote{Not to be confused with atomic quantum numbers, which are typically $(n,\ell,m_{\ell},m_s)$.}.

We can write a one-body state $\ket{\mu}$ in coordinate space as follows~\cite[Chap. 1.1]{Schunck2019}~\cite{saclay_notes}:
\[
\ket{\mu} = \int d\vec{r} \sum_{\sigma}\sum_{\tau} \ket{\vec{r}\sigma \tau} \phi_\mu(\vec{r}\sigma\tau)\;,
\]
where $\phi_\mu = \bra{\vec{r}\sigma\tau}\ket{\mu}$ is the wave function of the state $\ket{\mu}$, introduced earlier. The sum is over the isospin ($\tau = \pm 1$) and over the spin (up/down).

To each single-particle state $\mu$ we associate a creation and annihilation operator $a_\mu^\dagger, a_\mu$ respectively, such that:
\begin{align*}
a_\mu^\dagger \ket{0} &= \ket{\mu}\qquad\mathrm{and}\qquad
a_\mu \ket{\mu} = \ket{0}\;.
\end{align*}
These operators obey the anti-commutation relations:
\[
\{a_\mu^\dagger, a_\nu^\dagger \} = 0\,,\qquad \{a_\mu ,a_\nu \} = 0\,,\qquad \{a_\mu,a_\nu^\dagger \} = \delta_{\mu\nu} \;,
\]
which ensure that Pauli's exclusion principle is satisfied:
\[
 \qquad a_\mu^\dagger \ket{\mu} = 0 \;\qquad \mathrm{and} \qquad a_\mu \ket{\nu} = \delta_{\mu\nu}\ket{0} = 0 \;\;\mathrm{ if }\;\;\mu\neq\nu \;.
\]
A one-body operator $\hat{F}$ acts on each particle of a many-body state independently. Following~\cite{RingSchuck1980}, in the single-particle bases defined earlier, we assign (App.~\ref{appendix:F_one-body}):
\[
\hat{F} = \sum_{\mu\nu}F_{\mu\nu}a^\dagger_\mu a_\nu\;, \qquad F_{\mu\nu} = \bra{\mu}\hat{F}\ket{\nu}\;.
\]

The nuclear Hamiltonian (for $N$ particles) takes a specific form:
\begin{align*}
    \hat{H} = \sum_{i=1}^N \frac{\left.\vec{p}_i\right.^2}{2m} + \frac{1}{2}\sum^N_{i\neq j} V_2(i,j) + \frac{1}{6}\sum^N_{i\neq j\neq k}V_3(i,j,k) +\ldots 
\end{align*}
where the interaction potential $V_n$ represents all $n$-body interactions. 

In the mean-field approximation, we take a one-body operator $\hat{U}=\sum_i \hat{U}_i$ that can maximally represent the physics of the system~\cite{Schunck2019}. This allows us to partition the nuclear Hamiltonian as follows:
\begin{align}
    \hat{H}_0 &\equiv \sum_{i=1}^N \left(\frac{\left.\vec{p}_i\right.^2}{2m} + \hat{U}_i\right) \;,\nonumber\\
    \hat{H} &=  \hat{H}_0 + \left( \frac{1}{2}\sum^N_{i\neq j} V_2(i,j) + \frac{1}{6}\sum^N_{i\neq j\neq k}V_3(i,j,k) +\ldots - \sum_{i=1}^N \hat{U}_i\right) \nonumber \\ &\equiv  \hat{H}_0 + \hat{V}_\mathrm{res}\label{eq:mean_field_main} \;,
\end{align}
where $\hat{H}_0$ is the mean-field Hamiltonian and $\hat{V}_\mathrm{res}$ denotes the residual interaction. If $\hat{U}$ is chosen appropriately, the residual interaction is small $|\hat{V}_\mathrm{res}| \ll  |\hat{H}_0|$. Since $\hat{H}_0$ is a sum of independent one-body operators, it is typically written as $\hat{H}_0=\sum_i h_i$, with $h_i$ the single-particle Hamiltonian for the $i$-th particle.

\subsection{Solutions for Mean-Field}
For a system of $N$ identical fermions, the many-body wave function must be antisymmetric under the exchange of any two particles. The simplest wave function satisfying this property is a Slater determinant, which is a product state of $N$ single-particle orbitals $\{\phi_i\}$~\cite{saclay_notes}:
\begin{equation}
    \ket{\psi} = \frac{1}{\sqrt{N}}\det | \phi_1 \dots \phi_N| = \prod_{i=1}^N \hat{a}^\dagger_i \ket{0}\;.
\end{equation}
Solutions to $\hat{H}=\hat{H}_0$ within the HF framework are Slater determinants (or generalised Slaters in HFB). Note that the exact solution to the full Hamiltonian $\hat{H}$ is \textit{not} a Slater determinant due to the presence of $\hat{V}_\mathrm{res}$, inducing many-body correlations.

Crucially, $\hat{H}_0=\sum_i h_i$ is a function of the one-body density $h[\rho]$. The density matrix is defined by the expectation value of the one-body density operator ($\hat{\rho}_{\mu\nu}=a^\dagger_\nu a_\mu$) acting on the many-body state:
\[
\rho_{\mu\nu} = \bra{\psi}a^\dagger_\nu a_{\mu}\ket{\psi} \;.
\]

\subsection{Static Hartree-Fock}
The Hartree-Fock method is a variational approach used to find the `best' possible Slater determinant by minimizing the expectation value of the energy. In other words, by varying the single-particle orbitals to minimize $\bra{\psi}\hat{H}\ket{\psi}$, 
we effectively find an optimal choice for the mean-field potential $\hat{U}$ such that $\hat{V}_\mathrm{res}$ is minimized. 
 
Following~\cite[Chap.~5]{RingSchuck1980} and~\cite{saclay_notes}, we define the Hartree-Fock (HF) energy:
\[
E^{(HF)}_0 \equiv \min_{\ket{\psi}} \bra{\psi}H\ket{\psi}\;.
\]
The minimization ensures that the HF energy is as close as possible to the real ground state energy $E_0$, while $E_0 \le E_0^{(HF)}$ always holds. One can find that
\[
\bra{\psi}H\ket{\psi}=\sum_{\mu\nu}t_{\mu\nu}\rho_{\mu\nu}+\frac{1}{2}\sum_{\mu\nu\delta\sigma} \bar{v}_{\mu\nu\delta\sigma}\rho_{\mu\delta}\rho_{\nu\sigma} \equiv E[\rho]\;.
\]
It can be shown that, by introducing a Hartree-Fock potential $\Gamma$, the energy functional can be written as~\cite{RingSchuck1980}:
\[
E[\rho] = \Tr(t\rho)+\frac{1}{2}\Tr(\rho\Gamma\rho)\;.
\]
The ground state density matrix is found by minimizing the energy. The stationary condition $\delta E = 0$ subject to the constraint $\rho^2=\rho$, yields~\cite[Eq.~(5.35)]{RingSchuck1980}:
\begin{equation}
    [h[\rho],\rho] = 0\label{eq:HF-stationary} \;,
\end{equation}
where $h$ is the single-particle Hamiltonian $h=\frac{\delta E}{\delta \rho}$.

Since the solution $\ket{\psi}$ is assumed to be a Slater determinant, $\rho$ is constructed directly from the single-particle eigenvectors of $h[\rho]$. This circular dependency results in a self-consistent problem, which must be solved using iterative methods.   

The standard Hartree-Fock method is insufficient for describing open-shell nuclei, as it fails to account for pairing correlations. To treat such short-range interactions, it is generalised to the Hartree-Fock-Bogoliubov theory.

\clearpage
\subsection{Static Hartree-Fock-Bogoliubov}
The Hartree-Fock method can be interpreted as the zero-pairing limit of the more general Hartree-Fock-Bogoliubov (HFB) method. HFB mixes particle-hole states by applying an unitary transformation to the creation and annihilation operators, resulting in quasiparticle operators. 

Following the notation from~\cite{saclay_notes, Schunck2019}, the Bogoliubov transformation is defined as:
\[
\begin{pmatrix} \beta \\ \beta^\dagger\end{pmatrix} = \mathcal{W}^\dagger\begin{pmatrix}    a \\ a^\dagger\end{pmatrix}\;,
\qquad 
\begin{pmatrix} a \\a^\dagger\end{pmatrix} = \mathcal{W}\begin{pmatrix}    \beta \\ \beta^\dagger\end{pmatrix}\;,
\qquad \mathrm{where}
\qquad
\mathcal{W}= \begin{pmatrix}
    U & V^*\\V&U^*
\end{pmatrix}\;.
\]
The complex matrices $U$ and $V$ satisfy orthonormality conditions to ensure that $\mathcal{W}$ is unitary and that the quasiparticle operators obey the standard fermionic commutation relations. The resulting HFB many-body state $\ket{\phi_\mathrm{HFB}}$ is defined as the vacuum for these quasiparticles~\cite[Sec. 3.1.2]{Schunck2019}:
\[
\beta_\mu \ket{\phi_\mathrm{HFB}} = 0\;, \qquad \forall \mu \;.
\]
The one-body density matrix $\rho$ and pairing tensor $\kappa$ are defined as:
\[
 \kappa_{kl}=\bra{\phi_\mathrm{HFB}}a_l a_k\ket{\phi_\mathrm{HFB}} \;,  \qquad \rho_{kl}=\bra{\phi_\mathrm{HFB}}a_l^\dagger a_k\ket{\phi_\mathrm{HFB}}\;.
\]
The system is no longer described solely by the density matrix $\rho$, but by a generalized density matrix $\mathcal{R}$ incorporating this pairing tensor $\kappa$:
\begin{equation*}
\mathcal{R} = \begin{pmatrix}
\rho & \kappa \\
-\kappa^* & 1-\rho^*
\end{pmatrix}\;.
\end{equation*}

Concerning the HFB equation, the HFB Hamiltonian can be expressed in the particle basis $(a,a^\dagger)$ as a $2\times 2$ block matrix:
\[
H = \sum_\mu E_\mu \beta^\dagger_\mu \beta_\mu = \frac{1}{2}\begin{pmatrix}
    a^\dagger&a
\end{pmatrix}\begin{pmatrix} h-\lambda & \Delta \\ -\Delta^*&-h^*+\lambda   
\end{pmatrix}\begin{pmatrix}
    a\\a^\dagger
\end{pmatrix} \;,
\]
where $h$ is the single-particle Hamiltonian, $\Delta$ are the pairing gaps, and $\lambda$ is the chemical potential required to constrain the average particle number~\cite[Chap.~3.1.3]{Schunck2019},
and:
\begin{equation}
h_{kl}[\rho,\kappa,\kappa^*] = \frac{\partial \mathcal{E}}{\partial \rho_{lk}}\;,\qquad \Delta_{kl}[\rho,\kappa,\kappa^*]= \frac{\partial \mathcal{E}}{\partial \kappa^*_{kl}} \;.\label{eq:h_and_delta}
\end{equation}

\subsection{Time-dependent HFB}
When we introduce time dependence, we allow for the HFB Hamiltonian to change as a function of time. Consequently, the quasiparticles become time-dependent as well:
\[
\begin{pmatrix} \beta(t) \\ \beta^\dagger(t)\end{pmatrix} = \mathcal{W}^\dagger(t)\begin{pmatrix}    a \\ a^\dagger\end{pmatrix}\;,
\qquad 
\begin{pmatrix} a \\a^\dagger\end{pmatrix} = \mathcal{W}(t)\begin{pmatrix}    \beta(t) \\ \beta^\dagger(t)\end{pmatrix}\;,
\qquad \mathrm{where}
\qquad
\mathcal{W}= \begin{pmatrix}
    U(t) & V(t)^*\\V(t)&U(t)^*
\end{pmatrix}\;.
\]
and from Eq.~\eqref{eq:h_and_delta} also the pairing field $\Delta(t)$, $\kappa(t)$ and density $\rho(t)$ become time-dependent. The time-dependent-Hartree-Fock-Bogoliubov (TDHFB) equation is given by:
\begin{equation}
    i \frac{d}{dt}\mathcal{R} = [\mathcal{H}(t),\mathcal{R}] \label{eq:TDHFB} \;,
\end{equation}
which is equal to the TDHF equation setting $\Delta(t) = 0$.

\newpage
\subsection{Other useful definitions}
\subsubsection{Operators and selection rules}
\begin{wrapfigure}[26]{r}{0.45\linewidth}
\vspace{-3em}
    \centering
    \includegraphics[width=\linewidth]{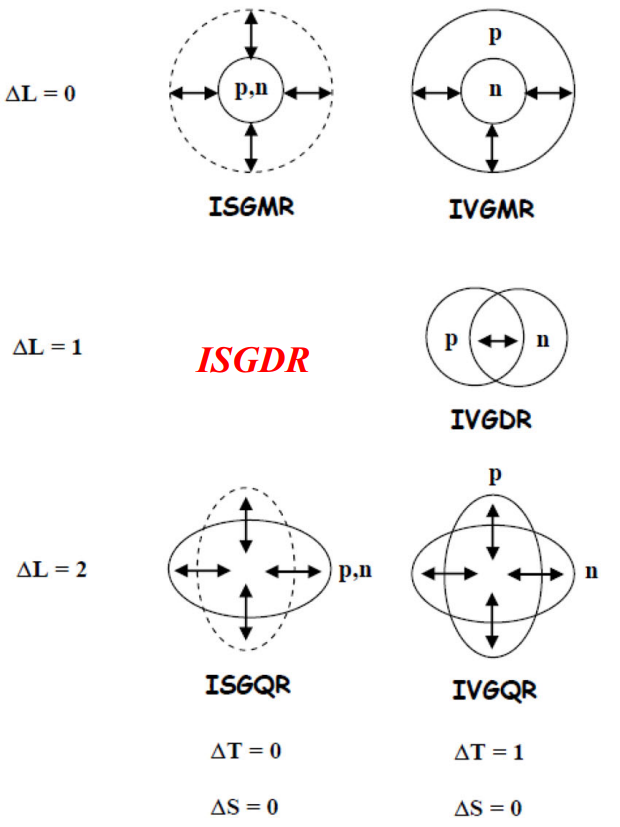}
    \caption{Figure from~\cite{Harakeh2025_GR_fig}. Illustration of the isoscalar and isovector monopole, dipole and quadrupole excitations, which are related to the $E_0, E_1$ and $E_2$ transitions. Note that the ISGDR corresponds to spurious center-of-mass motion (translation)~\cite{garg1999_ISGDR}.}
    \label{fig:E_transitions}
\end{wrapfigure}

Nuclear transitions are classified in terms of parity $\pi$ and change in angular momentum $L \equiv \Delta L = L_\mathrm{final} - L_\mathrm{initial}$. The latter is often called the multipolarity of a transition, where a monopole transition is $L = 0$, dipole $L = 1$, quadrupole $L = 2$, and so on. During de-excitation, the nucleus transitions from an excited state to a lower energy state, conserving energy through the emission of electromagnetic radiation, e.g. a gamma ray with angular momentum $L$. This transition must satisfy the conservation of angular momentum:
\[
|J_\mathrm{initial}-J_\mathrm{final}| \leq L \leq J_\mathrm{initial} + J_\mathrm{final}\;.
\]
The transitions are further categorized as either electric ($EL$) or magnetic ($ML$). The parity change $\Delta \pi$ determines which transitions are allowed, more precisely,  $\Delta \pi = (-1)^L$ for $EL$ and  $\Delta \pi = (-1)^{L+1}$ for $ML$ transitions. Henceforth, we focus on electric transitions. 

Beyond multipolarity, we differentiate excitations by the relative motion of protons and neutrons. If both move in phase, the transitions is \textit{isoscalar} (IS). If they move out of phase, it is \textit{isovector} (IV).

As illustrated in Fig.~\ref{fig:E_transitions}, we can identify each multipolarity with a specific type of motion. The collective monopole transition corresponds to a `breathing mode' of the nucleus. The dipole resonance represents an oscillation where the density increases on one side of the nucleus and decreases on the other side while keeping the centre of mass fixed~\cite{Harakeh2025_GR_fig}.
The quadrupole transition shows a cross-like oscillation. 

The electric isoscalar and isovector operators are defined as follows~\cite{Kortelainen2015}:
\begin{align*}
f^\mathrm{IS}_{LK}&=e_\mathrm{IS} \sum_i^A f_{LK}(\mathbf{r}_i)\;,\\
f^\mathrm{IV}_{LK}&=e_\mathrm{IV,\tau_i} \sum_i^A \tau_i f_{LK}(\mathbf{r}_i)\;,
\end{align*}
where $\tau_i = \pm 1$ for neutrons and protons. The isoscalar and isovector charges are denoted as $e_\mathrm{IS}$ and $e_\mathrm{IV,\tau_i}$, and $f_{LK}(\mathbf{r}) = r^L Y_{LK}(\mathbf{r})$ depend on the spherical harmonics. One can assume $K\geq 0$ by setting
\[
f_{L,K} = (f_{L,K}+f_{L,-K})/\sqrt{2-\delta_{K,0}}\;.
\]

\subsubsection{Transition probability and sum rules}\label{sec:sumrules}
Nuclear transitions are related to a transition operator $F$. For a transition from the ground state $\ket{0}$ to an excited state $\ket{\mu}$ with energy $\Omega_\mu$, the strength of the transition is given by the square of the matrix element $|\!\bra{\mu} F \ket{0}\!|^2$. The distribution of this strength across the energy spectrum is described by the strength function:
\begin{equation}
    S(\omega, F) = \sum_\mu |\!\bra{\mu} F\ket{0}\!|^2\delta(\Omega_\mu-\omega)\;.\label{eq:strength-delta}
\end{equation}
The strength function has units $[F]^2$\,MeV$^{-1}$. While finding $S(\omega)$ is difficult due to the determination of the eigenstates of $H$, this can be done approximately using the random phase approximation discussed in Sec.~\ref{sec:LR_QRPA}.

Next, we consider the moments of the strength function, defined as~\cite[Sec.~3.3]{Blaizot1980}:
\[
m_k = \int_0^\infty \omega^k S(\omega)  d\omega \;,
\]
where $m_1$ is called the \textit{energy-weighted sum-rule}. For discrete energy values $\omega = \omega_0, \dots, \omega_n $, $m_k$ can be approximated with
\[
m_k \approx \sum_{i=1}^{n} \omega_i^{k} S(\omega_i) |\omega_i-\omega_{i-1}|\qquad \text{or with}\qquad m_k \approx \sum_{i=1}^{n-1} \omega_i^k S(\omega_i) \tfrac{1}{2}|\omega_{i+1}-\omega_{i-1}| \;.
\]
This is, however, only accurate for sufficiently dense samples of $\omega$. One could also apply the Simpson rule for integration:
\[
\int_b^a f(x) \mathrm{d}x \approx \frac{1}{3}h\sum_{i=1}^{n+2}(f(x_{2i-2})+4f(x_{2i-1})+f(x_{2i})) \;.
\]
Alternatively, $m_k$ can be found by contour integration of the response function in the complex-energy plane~\cite{Hinohara2015}.

We further define the average excitation energy:
\[
E_\mathrm{av}= \frac{m_1}{m_0}\;,
\]
and for collective resonant spectra, when only a single dominant mode is present, the centroid of the resonance:
\[
E_\mathrm{c} = \sqrt{\frac{m_1}{m_{-1}}}\;.
\]

\section{Linear Response Theory and (Q)RPA}\label{sec:LR_QRPA}
\subsection{Free response}
\begin{figure}
    \centering
    \includegraphics[width=.45\linewidth]{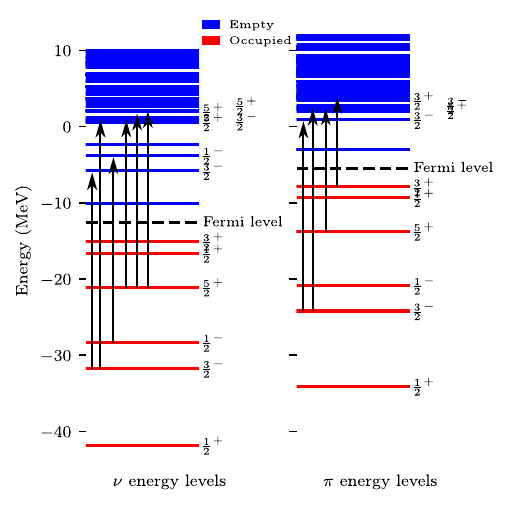}
    \includegraphics[width=0.45\linewidth]{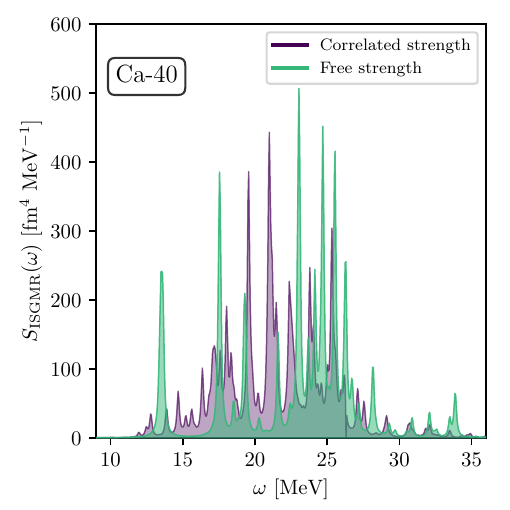}
    \caption{(left) Visualisation of the single-particle energy levels of a mean-field solution of \ce{Ca-40} with possible monopole excitations and the corresponding strength function (right).}
    \label{fig:free_response_Ca}
\end{figure}

In the previous section, we discussed the mean-field approach to approximate the ground state of a nuclear system. To study nuclear dynamics, we have to consider how a mean-field state reacts to external perturbations. In the independent-particle model an excitation consists of a single nucleon being promoted from an occupied state $i$ to an unoccupied state $m$. This is known as a particle-hole (ph) excitation and is illustrated in Fig.~\ref{fig:free_response_Ca} (left). 

We consider first the case without pairing. The `free' response refers to a system where nucleons are in a static potential $h[\rho^{(0)}]$ (we will clarify this notation later). In this case, the excitation energy is simply the difference between single-particle energies $\epsilon$ of a particle state $m$ and a hole state $i$: $\omega_{mi}=\epsilon_m - \epsilon_i$. If we calculate the response of the nucleus to an external operator $F$, the resulting free strength function is a sum of delta-peaks located at the energy differences, as shown in Fig.~\ref{fig:free_response_Ca} (right). 

However, this ignores the fact that as a nucleon excites, the nuclear density $\rho$ changes. Because the mean-field potential $h[\rho]$ is a functional of that density, the potential itself also changes. As shown in Fig.~\ref{fig:free_response_Ca}, this feedback shifts the energy levels themselves leading to a self-consistent problem to be solved.

\subsection{Linear response}
Linear response theory can be applied to describe nuclear transitions. This is a perturbative approach to evaluate the response of a nuclear system in the ground state. Following~\cite[Sec. 8.5]{RingSchuck1980}\cite[Sec. 5.1.3]{Schunck2019}\cite{Nakatsukasa2007}, we derive the RPA equations from the TDHF equation~\eqref{eq:TDHF} in the small-amplitude limit. We begin by introducing a weak, time-dependent external force:
\[
F(t) = \sum_{\omega'} \eta \left(F(\omega')e^{-i\omega' t} + F(\omega')e^{i\omega' t}\right) \;,
\]
where $\eta$ is a small dimensionless parameter. For simplicity, we take $F(t)$ to oscillate with a specific frequency $\omega$ and no longer repeat the summation. In the stationary case discussed previously, the Hartree-Fock density satisfies:
\[
[h[\rho^{(0)}], \rho^{(0)}] = 0\;,
\]
where $\rho^{(0)}=  \bra{\psi}a^\dagger_\nu a_\mu\ket{\psi}$ is the stationary density matrix. Under the influence of the external field, the Slater determinant becomes time-dependent, and the density matrix evolves according to the TDHF equation:
\begin{equation}
[h[\rho(t)] + F(t), \rho(t)] = i \frac{d}{dt}\rho(t)\;. \label{eq:TDHF}
\end{equation}
Assuming the deviation from the ground state is small, we use a linear approximation: 
\begin{equation}
\rho(t) = \rho^{(0)} + \delta \rho(t) = \rho^{(0)}+\eta(\delta \rho(\omega) e^{-i\omega t}+\delta \rho^\dagger(\omega) e^{i\omega t}) + \mathcal{O}(\eta^2)\label{eq:rho_time_dep}
\end{equation}
Note that, while the total fluctuation $\delta \rho(t)$ is (and must be) Hermitian, its Fourier component $\delta \rho(\omega)$ is generally not. Therefore, the Eq.~\eqref{eq:rho_time_dep} does not simplify. Substituting this into~\eqref{eq:TDHF} and retaining only terms linear in $\eta$, we obtain
\begin{align}
i \frac{d}{dt}(\rho^{(0)} + \delta \rho(t)) &= [h[\rho^{(0)} + \delta \rho(t)]+F(t), \rho^{(0)} + \delta \rho(t)]\nonumber\\
\omega \delta \rho(\omega) &= [h[\rho^{(0)}], \delta \rho(\omega)]+[\delta h(\omega), \rho^{(0)}]+[F(\omega), \rho^{(0)}]\label{eq:TDHF_b}\;.
\end{align}
The second equality is obtained by performing a Fourier transform and noting that $\tfrac{d}{dt}\rho^{(0)}=0$ and $h[\rho^0+\delta\rho(\omega)]= h[\rho^0]+\delta h(\omega)$~\cite[Eq. 5.14]{Schunck2019}. Setting $\delta h(\omega) = 0$ would recover the free response.

Now consider a state $\alpha$ with energy $\omega$, we can interpret the transition density as follows:
\[
\delta \rho_{\mu\nu} (\omega) = \bra{\alpha} a^\dagger_\nu a_\mu\ket{\psi}\;.
\]
We define the RPA amplitudes as~\cite{Schunck2019}:
\begin{align*}
    \tilde{X}_{mi}(\omega) &= \delta \rho_{mi}(\omega) \;,\\
    \tilde{Y}_{mi}(\omega) &= \delta \rho_{im}(\omega) \;.
\end{align*}
Here, $X_{mi}$ corresponds to particle-hole excitations, while $Y_{mi}$ corresponds to hole-particle excitations, and are often referred to as forward and backward amplitudes respectively~\cite{Nakatsukasa2007}.

In the HF basis, the stationary density is diagonal $\rho^{(0)}_{ij}=\delta_{ij}$, $\rho^{(0)}_{im}=\rho^{(0)}_{mn}=0$, and the single-particle Hamiltonian reads $h^{(0)}_{ab}=\epsilon_{a}\delta_{ab}$ and $\frac{\partial h_{ab}}{\partial \rho_{cd}} = \bar{v}_{adbc}$~\cite[Eq. 5.14]{Schunck2019}.

We continue with Eq.~\eqref{eq:TDHF_b}, but separate the $im$ and the $mi$ matrix elements:
\begin{align*}
    \bra{m}\omega \delta \rho(\omega) \ket{i} &= \bra{m}[h[\rho^{(0)}], \delta \rho(\omega)]+[\delta h(\omega), \rho^{(0)}]+[F(\omega), \rho^{(0)}]\ket{i}\;.
\end{align*}
Simplifying each term and using that $\rho_{mn}^{(0)}=0$, we obtain
\begin{align*}
       \bra{m}\omega \delta \rho(\omega) \ket{i} &= \omega \delta \rho_{mi}(\omega) = \omega X_{mi}(\omega) \;,\\
        \bra{m}h[\rho^{(0)}]\delta \rho(\omega)-\delta \rho(\omega)h[\rho^{(0)}]\ket{i} &= (\epsilon_m - \epsilon_i)\delta\rho_{mi}(\omega) = (\epsilon_m - \epsilon_i)X_{mi}(\omega) \;,\\
        \bra{m}\delta h(\omega)\rho^{(0)}-\rho^{(0)}\delta h(\omega)\ket{i} &= \sum_{jn}\delta h_{mj}(\omega)\rho^{(0)}_{ji}-\rho^{(0)}_{mn}\delta h_{ni}(\omega) = \delta h_{mi}(\omega) \;,\\
        \bra{m}F(\omega)\rho^{(0)}-F(\omega)\rho^{(0)}\ket{i} &= \sum_{jn}F_{mj}(\omega)\rho^{(0)}_{ji}-\rho^{(0)}_{mn}F_{ni}(\omega) = F_{mi}(\omega) \;.
\end{align*}
Similar calculations can be done for $im$, and together they yield the linear response equations:
\begin{align}
&\begin{dcases}
    (\epsilon_i - \epsilon_m + \omega) X_{im}(\omega) -\delta h_{im}(\omega) &= F_{im}(\omega)\\
     (-\epsilon_i + \epsilon_m + \omega) Y_{im}(\omega) +\delta h_{mi}(\omega) &= -F_{mi}(\omega) \;,
\end{dcases}\nonumber\\[0.5em]
\Leftrightarrow\qquad&\begin{dcases}
    (\epsilon_m - \epsilon_i - \omega) X_{im}(\omega) +\delta h_{im}(\omega) &= -F_{im}\\
     (\epsilon_m - \epsilon_i + \omega) Y_{im}(\omega) +\delta h_{mi}(\omega) &= -F_{mi}\;.
\end{dcases}\label{eq:linear_response_FAM}
\end{align}

By expressing the induced field $\delta h$ in terms of the density matrix variation $\delta \rho $~\cite{Nakatsukasa2007, RingSchuck1980}:
\[
\delta h_{ab} = \sum_{mi} \frac{\partial h_{ab}}{\partial \rho_{mi}} \delta \rho_{mi} + \frac{\partial h_{ab}}{\partial \rho_{im}} \delta \rho_{im} =  \sum_{mi}\bar{v}_{aibm}\delta \rho_{mi}+\bar{v}_{ambi}\delta \rho_{im}\;,
\]
the system can be rewritten in the standard RPA matrix form:
\begin{equation}
    \left\{ \begin{pmatrix}
        A&B\\B^*&A^*
    \end{pmatrix}- \begin{pmatrix}
        \omega&0\\0&-\omega
    \end{pmatrix} \right\} \begin{pmatrix}
        X_{mi}(\omega) \\ Y_{mi}(\omega)
    \end{pmatrix} = -\begin{pmatrix}
        F_{mi}\\ F_{im}
    \end{pmatrix}\label{eq:Linear_response_equation}\;.
\end{equation}
The matrix elements are defined as follows:
\begin{align*}
    A_{mi,nj}&= (\epsilon_m - \epsilon_i)\delta_{mn}\delta_{ij}+\bar{v}_{mjin};,\\
    B_{mi,nj}&= \bar{v}_{mnij}\;.
\end{align*}
Finally, we note that the RPA equations can equivalently be derived using the equations-of-motion method, as shown in~\cite[Sec. 5.1.1]{Schunck2019},~\cite{RingSchuck1980, Bacca2026}. This approach involves defining an excitation operator $Q_\nu^\dagger$ as a function of the forward and backward amplitudes:
\[
Q_\nu^\dagger = \sum_{mi}X_{mi}^\nu a_m^\dagger a_i - \sum_{mi}Y_{mi}^\nu
a^\dagger_i a_m \;.
\]
Under the quasi-boson approximation, assuming that the ground state behaves as a vacuum for these excitation operators, this leads to the same RPA equations.

\subsection{QRPA}
For open-shell nuclei, pairing correlations become significant, and Hartree-Fock must be generalized to the Hartree-Fock-Bogoliubov ansatz. In this case, the stationary density matrix $\rho^{(0)}$ is replaced by the stationary generalized density matrix $\mathcal{R}^{(0)}$, which contains both $\rho$ and the pairing tensor $\kappa$.

Using the Bogoliubov transformation, the single-particle operator $F$ can be written as~\cite{Schunck2019}, (see App.~\ref{appendix:F_quasi}):
\[
F = \frac{1}{2}\begin{pmatrix}
     \beta^\dagger&\beta
\end{pmatrix} \begin{pmatrix}
    F^{11} & F^{20} \\ -F^{02} & -(F^{11})^T
\end{pmatrix}\begin{pmatrix}
    \beta\\\beta^\dagger
\end{pmatrix}
+ \mathrm{const}\;.
\]
Similar to the previous section, the QRPA equations can be derived from the small-amplitude limit of the time-dependent HFB equation. The variation of the generalized density matrix $\mathcal{R}$ needs to be linearised:
\[
\mathcal{R}(t) = \mathcal{R}^{(0)}+\delta \mathcal{R}(t) = \mathcal{R}^{(0)} + \eta\left(\delta\mathcal{R}(\omega) e^{-i\omega t}+\delta\mathcal{R}^\dagger(\omega)e^{i\omega t}\right) +\mathcal{O}(\eta^2)\;.
\]
The linear response equations~\eqref{eq:linear_response_FAM} retain their algebraic form but are expanded to include the pairing channel~\cite[Eq. 5.36]{Schunck2019}. In this formalism, we obtain 2 quasiparticle (2-qp) instead of particle-hole (ph) excitations:
\begin{align}
\begin{dcases}
(E_\mu + E_\nu - \omega) X_{\mu\nu}(\omega) + \delta H^{20}_{\mu\nu}(\omega) &= -F^{20}_{\mu\nu} \\
(E_\mu + E_\nu + \omega) Y_{\mu\nu}(\omega) + \delta H^{02}_{\mu\nu}(\omega) &= -F^{02}_{\mu\nu}\;,
\end{dcases}\label{eq:linear_response_QFAM}
\end{align}
where $E_k$ are quasiparticle energies and $\delta H$ is the variation of the HFB Hamiltonian, which takes the form:
\[
\delta H(\omega) = \begin{pmatrix}
    \delta h & \delta \Delta \\ -\delta \bar{\Delta}^* & -\delta h^T
\end{pmatrix}\;,
\]
and now includes the induced change in both mean-field potential $\delta h(t)=\eta(\delta h(\omega)e^{-\omega t} + \delta h^\dagger (\omega) e^{i\omega t})$ and pairing potential $\delta \Delta(t) = \eta(\delta \Delta(\omega) e^{-i\omega t}+\delta \bar{\Delta}(\omega)e^{i\omega t})$~\cite{Avogadro2011}. One can obtain the same matrix equation as Eq.~\eqref{eq:Linear_response_equation}, but with different definitions for $A$ and $B$ which can be found in~\cite[Eq.~5.38]{Schunck2019}).

\newpage
\section{Finite amplitude method (FAM)}\label{sec:FAM}
The finite amplitude method (FAM), as presented in~\cite{Nakatsukasa2007}, and extended for HFB in~\cite{Avogadro2011}, provides an efficient alternative to the traditional matrix formulation. In a traditional (Q)RPA approach, one must explicitly construct and diagonalise the $A$ and $B$ matrices from Eq.~\eqref{eq:Linear_response_equation}, involving a large number of matrix elements of the residual interaction~\cite{Nakatsukasa2007}. As the single-particle model space increases, the diagonalization scales with $\mathcal{O}(N^6)$, making it computationally expensive for heavy or deformed nuclei\footnote{Without using symmetry or other properties that can simplify the problem.}. Furthermore, the methodology requires large memory storage for all the states. 

FAM bypasses these issues by solving the linear response equations~\eqref{eq:linear_response_FAM} iteratively for a specific external frequency $\omega$. To handle the singularities that occur at the physical transition energies $\Omega_n$ a complex frequency is introduced: $\omega \to \omega_R + i\gamma$, where $\omega_R$ is the excitation energy and $\gamma$ is a smoothing parameter. This `smearing' parameter can also be introduced from a phenomenological perspective: the experimental response function will generally be more spread out compared to the RPA method which neglects correlations beyond the 1p-1h correlations. Including a smearing parameter can account for these effects. This leads to the $S_\mathrm{FAM}$ expression (which is also derived in~\cite{Nakatsukasa2007}):
\begin{equation}
     S_\mathrm{FAM}(\omega_R + i\gamma, F) = - \sum_{n} \frac{|\! \bra{n}F\ket{0} \!|^2}{\Omega_n- (\omega_R+i\gamma)} +  \frac{|\! \bra{n}F^\dagger \ket{0} \!|^2}{\Omega_n+ (\omega_R +i\gamma)}\;.\label{eq:S_FAM_poles}
\end{equation}
One can relate $S_\mathrm{FAM}$ to $S$ by applying the Sokhotski–Plemelj theorem~\cite{Nakatsukasa2007} and taking only the imaginary part, i.e. $\Im\left(\lim_{\gamma \to 0}\tfrac{1}{x\pm i\gamma} \right) = \mp \pi \delta(x)$, 
which yields:
\begin{equation*}
     \Im\left(\lim_{\gamma \to 0} S_\mathrm{FAM}\right) = - \sum_{n} |\! \bra{n}F\ket{0} \!|^2 \pi \delta(\Omega_n-\omega_R) + \sum_{n} |\! \bra{n}F^\dagger \ket{0} \!|^2 \pi \delta(\Omega_n +\omega_R)\;.
\end{equation*}
Taking $\omega_R > 0$ and multiplying by $-\tfrac{1}{\pi}$ yields:
\begin{equation}
     S(\omega_R, F) = -\frac{1}{\pi}\lim_{\gamma \to 0} \left(\Im (S_\mathrm{FAM}(\omega, F))\right) \;,\label{eq:S_im_FAM}
\end{equation}
which, by using Eq.~\eqref{eq:S_FAM_poles}, is equal to
\begin{equation}
    S(\omega_R) = \lim_{\gamma \to 0}\frac{\gamma}{\pi} \left(\sum_{n} \frac{|\! \bra{n}F\ket{0} \!|^2}{(\Omega_n- \omega_R)^2+\gamma^2} -  \frac{|\! \bra{n}F^\dagger \ket{0} \!|^2}{(\Omega_n+ \omega_R)^2 +\gamma^2}\right)\;.\label{eq:S_poles}
\end{equation}
As discussed in~\cite{Nakatsukasa2007, Avogadro2011}, the FAM strength can also be expressed using the FAM amplitudes (equal to the RPA amplitudes for $\eta \to 0$) obtained by solving the (Q)RPA equations at complex frequency $\omega$:
\begin{equation}
S_{\mathrm{FAM}}(\omega) = \mathrm{Tr}(F^\dagger \delta \rho)= -\sum_{\mu <\nu} \left( F_{\mu\nu}^{20*} X_{\mu\nu}(\omega) + F_{\mu\nu}^{02*} Y_{\mu\nu}(\omega) \right) \;.\label{eq:S_FAM_XY}
\end{equation}
Hence, one can calculate the strength from Eq.~\eqref{eq:S_im_FAM} using:
\begin{equation}
    S(\omega_R) = \lim_{\gamma \to 0} \left(-\frac{1}{\pi}\Im\left(\sum_{\mu < \nu} F^{20*}_{\mu\nu}X_{\mu\nu}(\omega)+F^{02*}_{\mu\nu}Y_{\mu\nu}(\omega)\right)\right)\;.\label{eq:S_XY}
\end{equation}

The strength function also satisfies specific symmetries. These can be derived from Eq.~\eqref{eq:S_poles} straightforwardly:
\begin{align*}
S(-\omega) = S(-\omega_R - i\gamma) &= - S(-\omega_R+i\gamma) = -S(-\omega^*)\;,\\
S(\omega^*) = S(\omega_R - i\gamma) &= - S(\omega_R+i\gamma) =  -S(\omega)\;,
\end{align*}
which translates to symmetries on the RPA amplitudes by using Eq.~\eqref{eq:S_XY} and that $F$ is hermitian:
\begin{equation}
X(-\omega) = Y(\omega)\;,\qquad
Y(-\omega) = X(\omega)\;,\qquad
X(-\omega^*) = Y^*(\omega)\;,\qquad Y(-\omega^*) =X^*(\omega)\;. \label{eq:FAM_symmetries}
\end{equation}

\subsubsection{Discussion}
We conclude this section by discussing some properties of the FAM amplitudes. 

\begin{figure}
    \centering
    \includegraphics[width=0.9\linewidth]{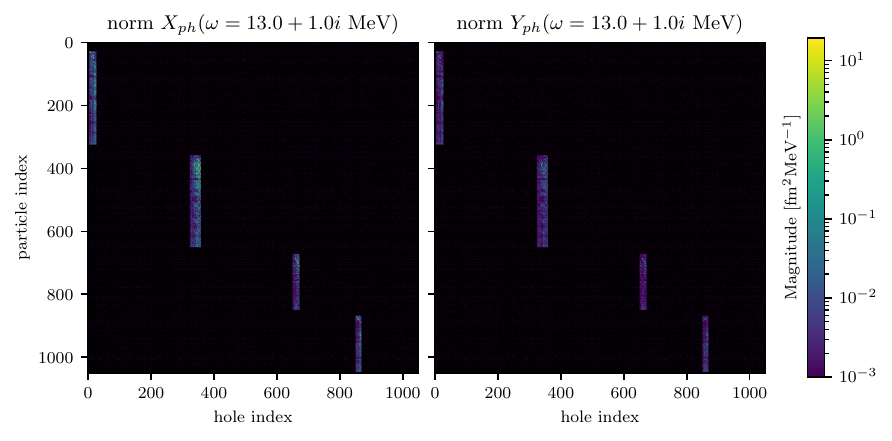}
    \caption{Amplitudes of the RPA matrix elements for \ce{Pb-208} and $\gamma = 1.0$\,MeV.}
    \label{fig:XY_visualisation}
\end{figure}

First, we observe that the RPA matrices are typically sparse due to the selection rules that determine the physically allowed nuclear transitions. As shown in Fig.~\ref{fig:XY_visualisation} for the \ce{Pb-208} monopole transition, sorting the particle-hole indices reveals a clear block structure. 

\begin{figure}
    \centering
    \includegraphics[width=0.33\linewidth]{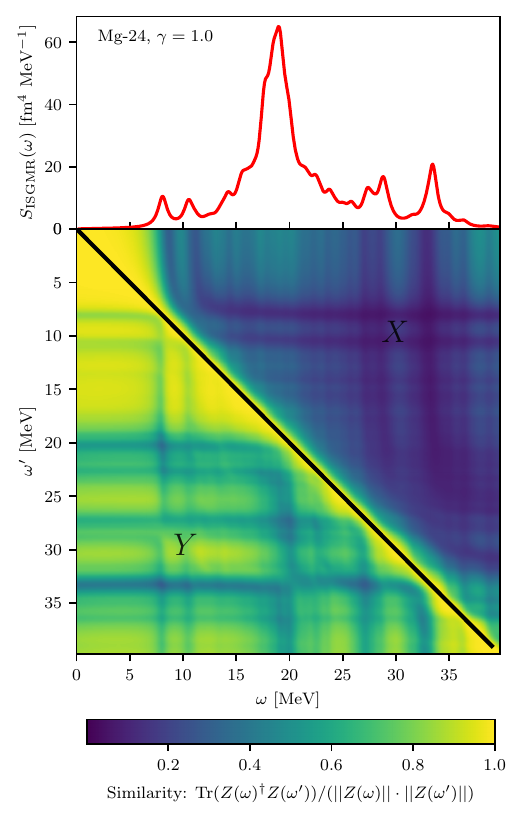}~\includegraphics[width=0.33\linewidth]{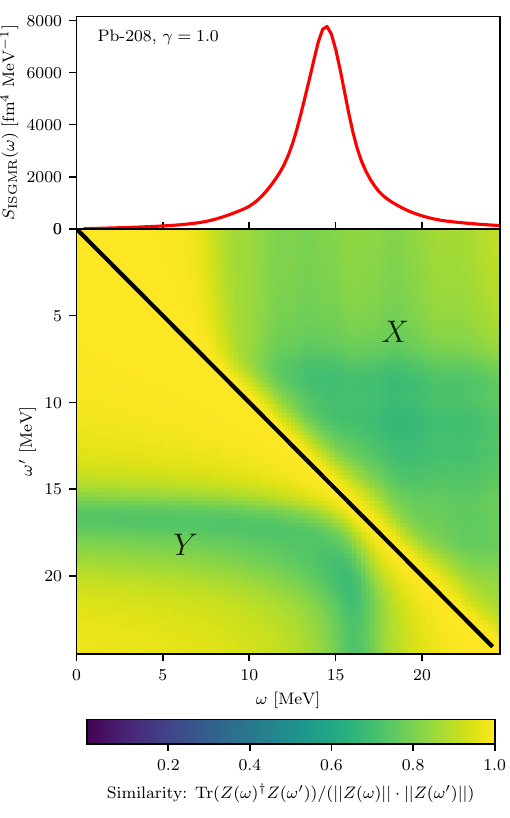}~    \includegraphics[width=0.33\linewidth]{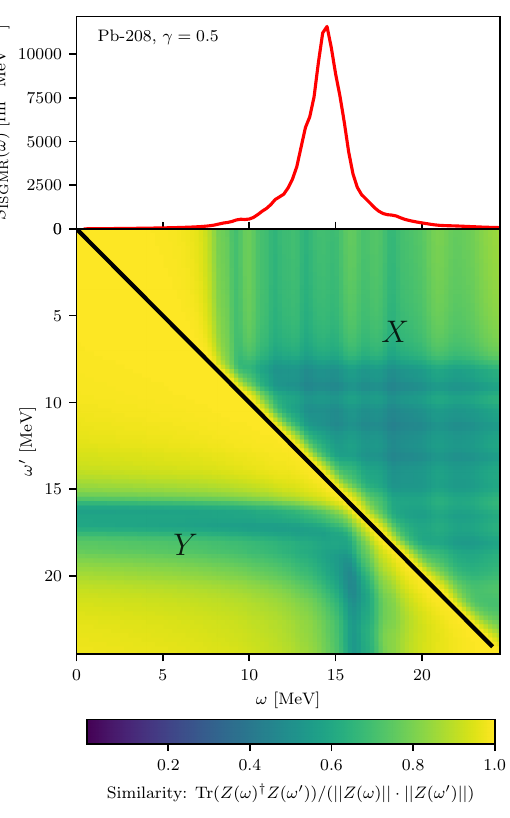}
    \caption{Similarity of the RPA amplitudes ($X, Y$) at frequency $\omega$ with the RPA amplitudes at any other frequency $\omega'$, for \ce{Mg-24} and \ce{Pb-208} and different smearing values. The corresponding strength function is shown to help interpreting the spectrum.}\label{fig:strength_XY_overlap}
\end{figure}

We also investigate how FAM amplitudes vary with $\omega$ using the cosine similarity $\mathrm{Tr}(A^\dagger B) / ||A||\,||B||$. This is typically applied to vectors, and thus ignores the matrix structure. Nevertheless, by vectorizing the FAM matrices, we can measure their overlap across different energies, as shown in Fig.~\ref{fig:strength_XY_overlap}. Far from resonances, where few transitions occur and amplitudes are near zero, the similarity remains close to 1. However, similarity decreases near the spectral peaks. The effect is more pronounced for smaller smearing values. This suggests that FAM amplitudes at a given energy might be accurately represented using neighbouring values. Interestingly, the $X$ amplitudes generally show lower similarity than the $Y$ amplitudes, which exhibit sharp drops only over narrow ranges.

\newpage
\section{Nuclear Energy Density Functional (EDF) Theory}\label{sec:EDF}
In the previous sections, the HFB and QRPA frameworks were presented as general methods for determining static and dynamic properties of a many-body system. However, these methods require a specific definition of the nuclear interaction. While one could start from bare inter-nucleon potentials, the complexity of the many-body problem makes this computationally prohibitive for heavy nuclei. 

Within the framework of nuclear Energy Density Functional (EDF) theory, the focus shifted towards effective interactions (e.g. Skyrme~\cite{skyrme1958, Vautherin1972}, Gogny or relativistic forms)~\cite{Bender2003}. These interactions are formulated as pseudo-potentials containing a finite number of parameters adjusted to reproduce selected nuclear observables.

More recently, the strict connection to a pseudo-potential was abandoned. In this framework, the total energy is directly constructed as a functional of local densities and currents, including all terms compatible with the symmetries of the strong interaction~\cite{soma2018}. This formulation provides additional flexibility, since the functional form is no longer constrained by a specific interaction ansatz.

In this section, we aim to introduce the required background for Chap.~\ref{chap:4}, thereby focussing on the `standard' Skyrme, as presented in~\cite{Vautherin1972}. Some extensions to this model are introduced and the relevant EDF forms using the model parameters will be defined, limited for the usage in the context of INM.

\clearpage
\subsection{Standard Skyrme}
The Skyrme effective interaction is a phenomenological zero-range force that aims to reproduce observables. It was first introduced by Skyrme~\cite{skyrme1958}, and later refined for HF calculations in~\cite{Vautherin1972}. This interaction is typically written in the standard form~\cite{Chamel2009, Bender2003, chabanat1997}:
\begin{align}
V_\mathrm{Sky}(\mathbf{r}_1, \mathbf{r}_2) &= t_0(1+x_0 P_\sigma )\delta(\mathbf{r_1}-\mathbf{r_2}) \label{eq:skyrme_potential}\\
&\qquad + \tfrac{1}{2}t_1(1+x_1P_\sigma)\left[p^2\delta(\mathbf{r_1}-\mathbf{r_2})+\delta(\mathbf{r_1}-\mathbf{r_2})p^2\right] \nonumber\\
   &\qquad + t_2 (1+x_2P_\sigma)\mathbf{p} \cdot \delta(\mathbf{r_1}-\mathbf{r_2}) \mathbf{p}\nonumber\\
   &\qquad + \tfrac{1}{6}t_3(1+x_3P_\sigma)[\rho(\mathbf{r})]^\alpha\delta(\mathbf{r_1}-\mathbf{r_2})\nonumber\\
   &\qquad + iW_0(\boldsymbol{\sigma}_1+\boldsymbol{\sigma}_2) \cdot [\mathbf{p} \delta(\mathbf{r_1}-\mathbf{r_2}) \mathbf{p}]\;,  \nonumber
\end{align}
where
\[ \mathbf{r}= \tfrac{1}{2}(\mathbf{r}_1+\mathbf{r}_2),\qquad   \mathbf{p} = \tfrac{1}{2i}(\nabla_1 - \nabla_2),\qquad P_\sigma = \tfrac{1}{2}(1+\boldsymbol{\sigma}_1\cdot\boldsymbol{\sigma}_2) \;.\]
This form has 10 parameters: $t_0$ and $x_0$ representing the strength and exchange for the zero-range term, $t_{1,2}$ and $x_{1,2}$ for the non-local/momentum dependent term, $t_3$ and $x_3$ for the density-dependent part, $\alpha$ the strength of the density dependence and $W_0$, the strength of the spin-orbit coupling. These parameters are fitted to experimental and microscopic nuclear data. Some (more recent) model adaptations employ an extended spin-orbit term, yielding one extra  $W'_0$ parameter~\cite{Scamps2021}.

\subsubsection{Skyrme EDF forms}

In modern nuclear physics one typically uses the Skyrme energy density functional instead of the interaction. The total energy of the system $E$ is the integral of the energy density $\mathcal{E}$:
\[
E=\int \mathcal{E}(\mathbf{r})d^3r \;,
\]
This energy and also the corresponding density can be decomposed as~\cite{Ryssens2019, Scamps2021, Bender2003}:
\[
\mathcal{E} = \mathcal{E}_\mathrm{K}+\mathcal{E}_\mathrm{Sky}+\mathcal{E}_\mathrm{Coul}+\mathcal{E}_\mathrm{pair}+\mathcal{E}_\mathrm{corr}
\]
where we have the kinetic energy, Skyrme interaction term, Coulomb contributions, pairing term, and correction terms (more info in~\cite{Scamps2021}). The density functional is typically separated in time-odd and time-even currents:
\[
\mathcal{E}_\mathrm{Sky} = \sum_{t=0,1} (\mathcal{E}^\mathrm{even}_t + \mathcal{E}^\mathrm{odd}_t)
\]
and the time-even part is equal to~\cite{Bender2003, Scamps2021, Lesinski2007}:
\begin{align}
\mathcal{E}^\mathrm{even}_t = C^{\rho\rho}_t[\rho_0] \rho^2_t + C_t^{\rho\tau}\rho_t \tau_t + C_t^{\rho\Delta\rho}\rho_t\Delta\rho_t +C^{\rho\nabla J}_t \rho_t \nabla\cdot J_t - C_t^{JJ}\sum_{\mu,\,\nu=xyz} J_{t,\,\mu\nu}J_{t\,\mu\nu}\;.\label{eq:e_dens_even_t}
\end{align}
Note that time-even densities consist of matter density $\rho$, kinetic density $\tau$ and spin-orbit density $J$. For the time-odd part, we refer to \cite[Eq.~(49)]{Bender2003} or~\cite[Eq.~(1.23)]{Schunck2019}. Time-odd densities consists of spin density, current density and spin-kinetic density, which all change sign under time reversal. In some models (e.g. in \cite{Ryssens2019}) the  bilinear form of the spin-current tensor density ($J_{\mu\nu}$) is omitted. We can consider the ground state of nuclear matter to be time-reversal invariant, thus we will only need the time even part in Chap.~\ref{chap:4}. 

In most applications, the coupling constants $\{C\}$ are assumed to be constant, except for $C^{\rho\rho}_t[\rho_0]$ which depends on the density (the $t=0$ part) via the following relation:
\[
C_t^{\rho\rho}[\rho_0] = C_t^{\rho\rho}+C_t^{\rho\rho^\alpha}\rho_0^\alpha \;.
\]

\subsubsection{Coupling constants}
The coupling constants are related to the Skyrme parameters appearing in the potential (Eq.~\eqref{eq:skyrme_potential}). This follows from taking $\bra{\mathrm{HF}}V_\mathrm{Sky}\ket{\mathrm{HF}}$. As some papers prefer the convention using the model parameters (e.g. ~\cite{chabanat1997, Chamel2009}) while others prefer using coupling constants $\{C\}$ (e.g. ~\cite{Scamps2021, Lesinski2007, Ryssens2019}), we would like to derive the relation to go from one convention to the other. Additionally, in Chap.~\ref{chap:4}, it will be more convenient to express the EDF in terms of the model parameters.

Using the following conventions for the kinetic and matter density
\begin{align}
    \rho = \rho_0 &= \rho_n + \rho_p\;, \qquad \rho_1=\rho_n-\rho_p\;,  \nonumber\\
    \tau = \tau_0 &= \tau_n + \tau_p\;, \qquad \tau_1=\tau_n-\tau_p\;,  \nonumber 
\end{align}
we derive that 
\begin{align}
2(\rho_p^2+\rho_n^2) &= \rho_0^2+\rho_1^2\label{eq:rhop^2+rhon^2}\;,\\
\rho_0\tau_0 &= \rho_n\tau_n+\rho_p\tau_p+\rho_n\tau_p+\rho_p\tau_n\;,\nonumber\\
\rho_1\tau_1 &= \rho_n\tau_n+\rho_p\tau_p-\rho_n\tau_p-\rho_p\tau_n\;, \nonumber\\
2(\rho_n\tau_n + \rho_p \tau_p) &= \rho_0\tau_0+\rho_1\tau_1\;.\label{eq:rhontaun+rhoptaup}
\end{align}

We neglect gradient terms and time-odd terms (related to spin or rotation) as we want to investigate the infinite nuclear matter (INM) properties later. INM is homogeneous, isotropic and infinite, so spatial or current-dependent properties average out. Note that we do not restrict ourselves to symmetry in $p$ and $n$. Starting from Eq.~\eqref{eq:e_dens_even_t}:
\begin{align}
    \mathcal{E}^\mathrm{INM}_\mathrm{Sky} &= \sum_{t=0,1} C^{\rho\rho}_t \rho^2_t + C^{\rho\rho^\alpha}_t\rho_0^\alpha \rho_t^2+C^{\rho\tau}_t \rho_t\tau_t + \mathrm{Other\,terms\,we\,will\,neglect}\nonumber\\
    &= C^{\rho\rho}_0 \rho^2_0 + C^{\rho\rho^\alpha}_0\rho_0^{\alpha +2}+C^{\rho\tau}_0 \rho_0\tau_0 + C^{\rho\rho}_1 \rho^2_1 + C^{\rho\rho^\alpha}_1\rho_0^{\alpha }\rho_1^2+C^{\rho\tau}_0 \rho_0\tau_0 \;.\label{eq:temp_link}
\end{align}
Now following~\cite[Eq.~(2.6)]{chabanat1997}, where the EDF is written as:
\begin{equation}
\mathcal{E}=\frac{\hbar^2}{2m}\tau + \mathcal{E}_0 + \mathcal{E}_3 + \mathcal{E}_\mathrm{eff}\;,\label{eq:E_tot_03eff}
\end{equation}
using a zero-range, density-dependent and effective-mass term. We can rewrite these EDF terms as a function of $\rho_0$ and $\rho_1$ by applying Eqs.~\eqref{eq:rhop^2+rhon^2} and~\eqref{eq:rhontaun+rhoptaup}:
\begin{align*}
\mathcal{E}_0 &= \tfrac{1}{4}t_0((2+x_0)\rho^2 - (2x_0+1)(\rho_p^2+\rho_n^2))\\
    &= \tfrac{1}{8}t_0(4\rho_0^2 + 2x_0\rho_0^2 - 2x_0 \rho_0^2-\rho_0^2-2x_0\rho_1^2-\rho_1^2)\\
    &= \tfrac{1}{8}t_0(3\rho_0^2-(2x_0+1)\rho_1^2) \;,\\[0.5em]
\mathcal{E}_3&= \tfrac{1}{24}t_3\rho_0^\alpha ((2+x_3)\rho_0 - (2x_3+1)\tfrac{1}{2}(\rho_0^2+\rho_1^2))\\
    &= \tfrac{1}{48}t_3\rho^\alpha_0(3\rho_0^2-(2x_3+1)\rho_1^2) \;,\\[0.5em]
\mathcal{E}_\mathrm{eff} &= \tfrac{1}{8}t_1(2+x_1)\rho_0\tau_0 -\tfrac{1}{8}t_1(2x_1+1)\tfrac{1}{2}(\rho_0\tau_0+\rho_1\tau_1)+\tfrac{1}{8}t_2(2+x_2)\rho\tau + \tfrac{1}{8}t_2(2x_1+1)(\rho_p\tau_p+\tau_n\rho_n)\\
    &= \tfrac{1}{16}t_1(3\rho_0\tau_0 - (1+2x_1)\rho_1\tau_1)+ \tfrac{1}{16}t_2((5+4x_2)\rho_0\tau_0 + (1+2x_1)\rho_1\tau_1\;.
\end{align*}
Comparing this to Eq.~\eqref{eq:temp_link}, we find the following relations with the $\{C\}$:
\begin{align*}
    &C^{\rho\rho}_0= \tfrac{3}{8} t_0 \;,& & C^{\rho\rho}_1=-\tfrac{1}{4}t_0(\tfrac{1}{2}+x_0) \;,\\
    &C_{0}^{\rho\rho^\alpha}=\tfrac{3}{48}t_3\;,&  &C_{1}^{\rho\rho^\alpha}=-\tfrac{1}{24}t_3(\tfrac{1}{2}+x_3) \;,\\
    &C^{\rho\tau}_0=\tfrac{3}{16}t_1 +\tfrac{1}{4}t_2(\tfrac{5}{4}+x_2) \;,&  &C_{1}^{\rho\tau}=-\tfrac{1}{8}t_1(\tfrac{1}{2}+x_1)+\tfrac{1}{8}t_2(\tfrac{1}{2}+x_2)   \;.
\end{align*}
This is identical to App.~A from~\cite{Scamps2021}\footnote{Except for the typo in $C_{1}^{\rho\tau}$ in that reference.}, and~\cite{Lesinski2007, Grams2025}.

We repeat the final EDF form in terms of model parameters, limited to those relevant for Chap.~\ref{chap:4}:
\begin{align}\nonumber
\mathcal{E}^\mathrm{INM}_\mathrm{Sky} &= \tfrac{1}{8}t_0(3\rho_0^2-(2x_0+1)\rho_1^2) +\tfrac{1}{48}t_3\rho^\alpha_0(3\rho_0^2-(2x_3+1)\rho_1^2)\\&\qquad+\tfrac{1}{16}t_1(3\rho_0\tau_0 - (1+2x_1)\rho_1\tau_1)+ \tfrac{1}{16}t_2((5+4x_2)\rho_0\tau_0 + (1+2x_1)\rho_1\tau_1 \;,
    \label{eq:energy_density_skyrme_parameters}
\end{align}

\subsection{Extended Skyrme 1}\label{sec:t4t5_skyrme}
The standard Skyrme functional typically leads to high-density ferromagnetic instability of neutron stars~\cite{Chamel2009}. To avoid this, additional terms can be added to the standard Skyrme form in Eq.~\eqref{eq:skyrme_potential}. One example we will consider is the $t_4$-$t_5$-extension which generalizes the $t_1$ and $t_2$ to density-dependent terms. The new potential takes the following form~\cite{Chamel2009}:
\begin{align*}
    V'_\mathrm{Sky}(\mathbf{r}_1,\mathbf{r}_2) &= V_\mathrm{Sky}(\mathbf{r}_1,\mathbf{r}_2)  + \tfrac{1}{2}t_4(1+x_4 P_\sigma) \left[p^2 \rho(\mathbf{r})^\beta \delta(\mathbf{r}_1-\mathbf{r}_2) + \delta(\mathbf{r}_1-\mathbf{r}_2)  \rho(\mathbf{r})^\beta p^2\right] \\&\qquad + t_5(1+x_5P_\sigma) \mathbf{p}\cdot\rho(\mathbf{r})^\gamma \delta(\mathbf{r}_1-\mathbf{r}_2) \mathbf{P}
\end{align*}
yielding 6 additional parameters ($t_4, x_4, t_5, x_5, \beta$ and $\gamma$). The EDF changes as follows~\cite[Eq.~(6)]{Grams2023}:
\begin{align*}
    \mathcal{E}'\,^{\mathrm{even}}_t &=  \mathcal{E}^\mathrm{even}_t + C_t^{\nabla \rho \nabla \rho }[\rho_0]\nabla\rho_t \cdot \nabla \rho_t  +C_t^{\rho\nabla \rho \nabla \rho}[\rho_0]\rho_t \nabla \rho_0 \cdot \nabla \rho_t\;.
\end{align*}
The newly introduced coupling constants are density dependent, and, the coefficient $C^{\rho\tau}_t$ from $\mathcal{E}$ in Eq.~\eqref{eq:e_dens_even_t} is now also taken density dependent, which we denote as $\tilde{C}_t^{\rho\tau}$. More precisely~\cite[Eq.(A.30)]{Chamel2009}:
\begin{align*}
\tilde{C}^{\rho\tau}_0 &= C^{\rho\tau}_0+\tfrac{3}{16}t_4\rho^\beta +\tfrac{1}{16}t_5(5+4x_5)\rho^\gamma \;,\\
\tilde{C}^{\rho\tau}_1 &=C^{\rho\tau}_1-\tfrac{1}{16}t_4\rho^\beta (1+2x_4) + \tfrac{1}{16}t_5 \rho^\gamma (1+2x_5)\;.
\end{align*}

Because only the NLO terms are needed in Chap.~\ref{chap:4}, the terms with $\Delta \rho_t \cdot \Delta \rho_t$ are omitted. We obtain:
\begin{align}
 \mathcal{E}'\,^\mathrm{INM}_{\mathrm{Sky}}&=  \mathcal{E}^\mathrm{INM}_{\mathrm{Sky}} + \tfrac{1}{16}t_4\rho^\beta (3\rho_0\tau_0+(1+2x_4)\rho_1\tau_1) + \tfrac{1}{16}t_5\rho^\gamma ((5+4x_5)\rho_0\tau_0+(1+2x_5)\rho_1\tau_1)\;.
    \label{eq:energy_density_skyrme_parameters_t4t5}
\end{align}

\subsection{Extended Skyrme 2}\label{subsec:N2LO}
In this subsection, we consider an alternative extension to the Skyrme functional. The main motivation for this approach is that the standard Skyrme functionals yield insufficiently realistic single-particle spectra. Consequently, they often  fail to reproduce ground state angular momenta of odd-mass and odd-odd systems, odd-even staggering and deformed shell gaps for superheavy nuclei (see~\cite{Grams2026} and references therein). Progress was made by developing new analytical EDF forms, such as the N$\ell$LO ((next-to-)$^ \ell$leading order) Skyrme EDFs. These forms are obtained by including all possible zero-range terms containing up to $2\ell$ derivatives of the matter density. In this context, the standard Skyrme functional corresponds to the NLO version. 

A Skyrme-type EDF can be written by grouping the terms with the same order of derivatives~\cite{Ryssens2019, Grams2025}:
\[
\mathcal{E}_\mathrm{Sky}=\sum_{l=\mathrm{e,o}}\mathcal{E}^{(0)}_l+\mathcal{E}^{(2)}_l+\mathcal{E}^{(4)}_l+\ldots
\]
We only keep terms that are relevant for the INM problem:
\begin{align*}
\mathcal{E}^{\mathrm{INM},(0)}_\mathrm{e} &=\sum_{t=0,1}[C^{\rho\rho}_t\rho^2_t + C^{\rho\rho^\alpha}_t \rho^{\alpha}_0\rho^2_t]\;,\\
\mathcal{E}^{\mathrm{INM},(2)}_\mathrm{e} &=\sum_{t=0,1} C^{\rho\tau}_t\rho_t\tau_t + \mathcal{O}(\rho_t \Delta \rho_t, \rho_t \nabla \cdot J_t, J_t^2)\;,\\
\mathcal{E}^{\mathrm{INM},(4)}_\mathrm{e} &=\sum_{t=0,1} C^{\tau\tau}_t \tau^2_t+C^{\rho Q}_t \rho_t Q_t + 2C_t^{\tau_{\mu\nu}\tau_{\mu\nu}}\sum_{\mu,\nu}\tau^2_{t,\mu\nu} +\mathcal{O}((\Delta \rho_t)^2, \tau_{t,\mu\nu} \nabla_\nu\nabla_\mu\rho_t, \ldots)\;,
\end{align*}
where $Q_t$ is the higher-order kinetic density:
\[
Q_t(\mathbf{r}) = \Delta \Delta' \rho_t(\mathbf{r},\mathbf{r}')|_{\mathbf{r}=\mathbf{r}'}
\;.\]
The N2LO Skyrme EDF incorporates terms up to order $\mathcal{E}^{(4)}$. One of the benefits of this model compared to the $t_4$-$t_5$-extension is that it is a more general or natural extension of the EDF, by simply including higher-order terms while yielding only four additional parameters~\cite{Grams2026}.

\subsection{Parameter fitting}\label{sec:edf-parameters}
Each Skyrme EDF form is defined by a specific parametrisation, i.e. numerical values assigned to the Skyrme parameters. These values are determined by fitting model observables against a selected set of experimental nuclear properties.  

The fitting procedure of the SLy family, such as SLy4~\cite{chabanat1997, Chabanat1998}, uses a minimal set of constraints. These include charge radii and binding energies of (a limited set of) magic and doubly magic nuclei. To ensure reliability in astrophysical applications, constraints are added to reproduce infinite nuclear matter properties, including saturation density, energy per nucleon, incompressibility, symmetry energy and effective mass. 

In contrast, the Brussels-Skyrme (BSk) and BSkG (Brussels-Skyrme-on-a-Grid) models are designed as global models. Their objective is to describe properties of \textit{all} known nuclei with high precision. For this reason, their fitting procedure considers essentially all possible experimental nuclear data. This increases the accuracy on nuclear property prediction and applicability. While the earliest BSk models (BSk1-17) use the standard Skyrme form, more recent models (BSk18-26) incorporate the $t_4$-$t_5$ extension. These additional terms allow to control the stiffness of the Equation of State at high densities, making them effective for neutron star modelling~\cite{HFB-26}. Because this extension contains unconventional terms that introduce considerable complexity to the formalism, the standard Skyrme models are still being improved, e.g.~\cite{HFB-27}.

The BSkG series use a 3D coordinate-space representation, allowing for a better description of deformed nuclei, which are often poorly described by spherical models. The first model in this series, BSkG1~\cite{Scamps2021}, uses a standard Skyrme parametrisation, thus 10 parameters\footnote{Although the true number of parameters can be higher due to additional correction terms.}.  BSkG2~\cite{Ryssens2022, Ryssens2023} allows for time-reversal symmetry breaking. Then, BSkG3~\cite{Grams2023} applies a $t_4$-$t_5$ extension to the standard Skyrme, enabling the prediction of the  EoS of dense matter much better than its predecessors. The next BSkG4~\cite{Grams2025} uses an improved interpolation ansatz for the pairing gap to better reproduce their behaviour at arbitrary asymmetries. BSkG5~\cite{Grams2026} uses another form of functional, namely of the N2LO type, which results in fewer parameters to fit. This model is able to predict nuclear ground-state properties and is consistent with the EoS for pure nucleon matter.

\newpage

\begin{minipage}[t]{0.5\linewidth}
\section{Numerical context}
The ground-state properties and mean-field potentials for this work are obtained using the Modular Cranking Code,  MOCCa~\cite{RyssensMOCCa}. It is a program designed for 3D coordinate-space mean-field and (extended to) linear response calculations. The interaction between nucleons is modelled using Skyrme EDFs. By representing wave functions on a 3D-grid rather than an harmonic oscillator basis which is symmetry restricted, the code allows for a more flexible description of a nuclear system, including deformations. 

\vspace{2em}
\subsection{Solutions with MOCCa}
\subsubsection{Convergence of mean-field}
The convergence of the mean-field solutions in coordinate space has been analysed in detail before~\cite{ryssens2015}. This study concluded that energies and densities are already
stable at small box sizes between (16\;fm)$^3$ and (32\;fm)$^3$. A mesh spacing of 0.8\;fm yields accuracies below 100\,keV for the mean-field energy levels~\cite{ryssens2015}. For this reason, these parameters have been used in the remainder of the thesis. 
\\
\\
Note that, since we are interested in the nuclear strength, we should check the convergence in terms of the response function. This also relies on the convergence of the unbound single-particle states, which is irrelevant for the mean-field solution.
\vspace{1.5em}
\subsubsection{Convergence of strength}
We vary the box size, mesh spacing and number of wave functions in the mean-field solution, and look for convergence in the obtained strength spectra. Fig.~\ref{fig:convergence_Pb} illustrates this for~\ce{Pb-208}. From these figures, we conclude that the result is converged in terms of mesh spacing. The number of wave functions needs to be sufficiently large to represent the single-particle states, and we conclude that taking $(650, 390)$ neutron and proton wave functions suffices. However, considering the box size, we find something interesting: although convergence with respect to larger box sizes is expected from a physical point of view, the spectrum does not seem to converge.

\end{minipage}
\hfill
\begin{minipage}[t]{0.47\linewidth}
\vspace{-1em}
    \centering
\includegraphics[width=\linewidth]{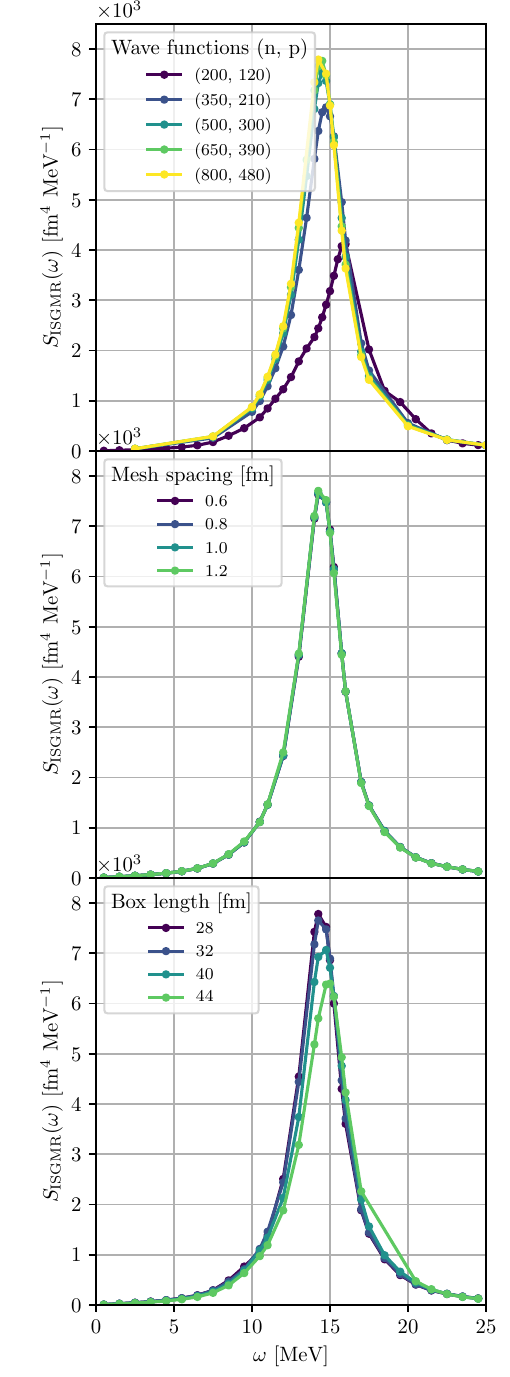}
\vspace{-1em}
\captionof{figure}{Monopole strength function with {$\gamma=1$\,MeV} for \ce{Pb-208} for different values of the box size, mesh spacing and number of wave functions. If not specified by the legend, the parameters are mesh spacing = 1.2\,fm, box size = (32\,fm)$^3$ and wave functions = (650, 390).}
\label{fig:convergence_Pb}
\end{minipage}  

\begin{figure}
\begin{subfigure}[t]{0.45\linewidth}
\centering
\includegraphics[width=\linewidth]{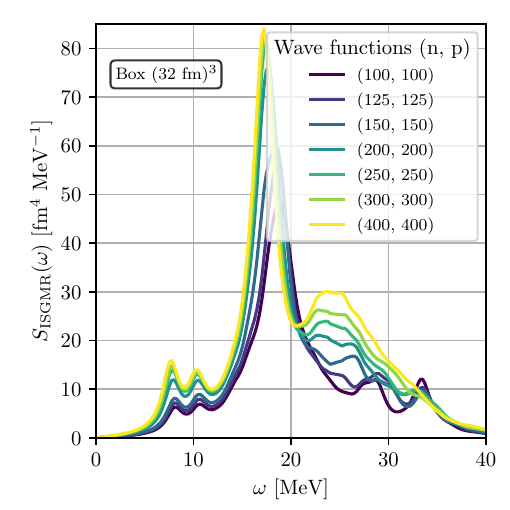}
\caption{Different number of wave functions, box size of (32 fm)$^3$ and mesh spacing of 0.8 fm.}    \label{fig:Mg_convergence_wf_size_16}
\end{subfigure}
\hfill
\begin{subfigure}[t]{0.45\linewidth}
    \centering
    \includegraphics[width=\linewidth]{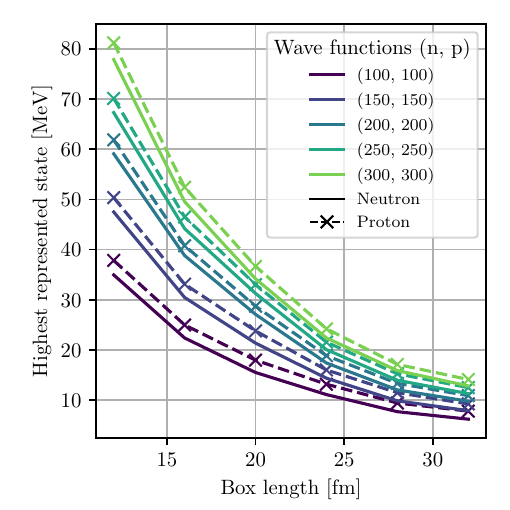}
    \caption{Highest represented state in the mean field solution as a function of box size for \ce{Mg-24}.}
    \label{fig:Mg_convergence_box_size}
\end{subfigure}
\begin{subfigure}[t]{\linewidth}
    \centering
    \includegraphics[width=0.33\linewidth]{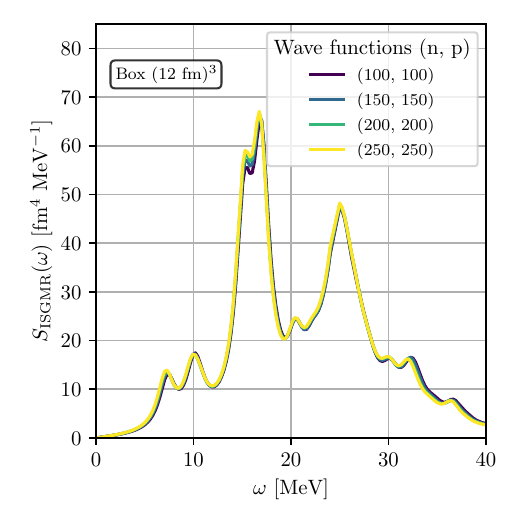}~ 
    \includegraphics[width=0.33\linewidth]{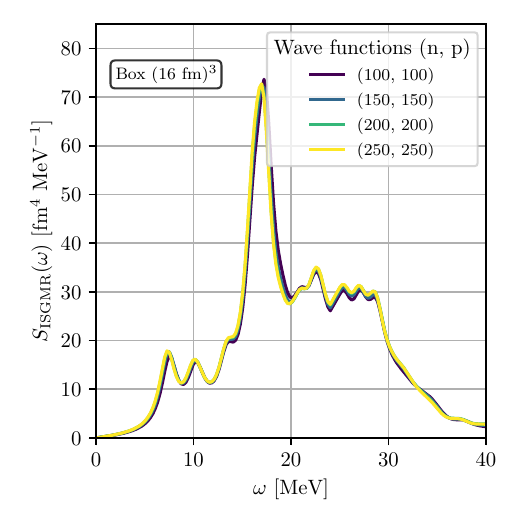}~
    \includegraphics[width=0.33\linewidth]{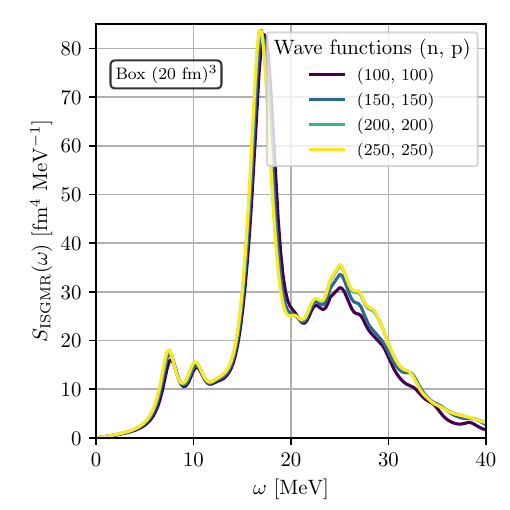}
    \caption{Different box sizes and number of wave functions.}\label{fig:Mg_convergence_wf_sizes}
\end{subfigure}
\caption{Figures related to the convergence analysis of the  ISGMR strength function of \ce{Mg-24} with $\gamma = $1\;MeV using a standard Skyrme (BSkG2).}
\end{figure}

\vspace{1.5em}
In Fig.~\ref{fig:Mg_convergence_wf_size_16} the strength function of $\ce{Mg-24}$ is shown using a fixed box size of (32 fm)$^3$. One can observe (e.g. around $\omega=$ 25\,MeV) that the strength function does not converge in function of number of wave functions. Based on our suspicions related to the box size convergence from Fig.~\ref{fig:convergence_Pb}, we visualised the energy of the highest represented state as a function of the box size in Fig.~\ref{fig:Mg_convergence_box_size}. We concluded that for 32\,fm, it only reaches up to 10\,MeV. The Fermi-level is located around $-$15\,MeV. Adding more wave functions with energies close to the 10\,MeV, allows for more transitions of 25\,MeV, explaining the (incorrect) increase in strength at this energy in Fig.~\ref{fig:Mg_convergence_wf_sizes}.

Further analysis showed that the box size strongly affects the highest represented state for a fixed number of wave functions considered. Intuitively, this corresponds with low, but positive energy wave functions `fitting' inside a large box. We concluded that the issue comes from the incorrect representation of the unbound states in the continuum resulting in incorrect application of the boundary conditions. Fixing this problem (e.g. similar to~\cite{Shlomo_bdry, Inakura_bdry}) requires a fundamental change in the implementation and is outside the scope of this thesis.

\clearpage
\begin{subappendices}
\section{}
\subsection{One-body operator}\label{appendix:F_one-body}
We show a property of a single-particle operator $\hat{F}$. We consider an arbitrary single-particle state $\ket{\mu}$ and insert the identity operator ($\mathbb{I} = \sum_k \ket{k}\bra{k}$) twice:
\begin{align*}
    \hat{F}\ket{\mu} &= \sum_{\mu'} \sum_{\nu'}\ket{\mu'}\bra{\mu'}\hat{F}\ket{\nu'}\bra{\nu'}\ket{\mu}\\
    &= \sum_{\mu'} \sum_{\nu'}a^\dagger_{\mu'}\ket{0}\bra{\mu'}\hat{F}\ket{\nu'}\bra{0}a_{\nu'}\ket{\mu}\\
    &= \sum_{\mu'} \sum_{\nu'}\bra{\mu'}\hat{F}\ket{\nu'}a^\dagger_{\mu'}\ket{0}\bra{0}a_{\nu'}\ket{\mu} \\ 
    &= \sum_{\mu'} \sum_{\nu'}\bra{\mu'}\hat{F}\ket{\nu'}a^\dagger_{\mu'}\delta_{\nu'\mu}\ket{0}\bra{0}\ket{0}= \sum_{\mu'\nu'} f_{\mu'\nu'}a^\dagger_{\mu'}a_{\nu'}\ket{\mu} \\
\end{align*}
where in the last line, we have used that $a_{\nu'}\ket{\mu} = \delta_{\nu'\mu} \ket{0}$ and $\bra{0}\ket{0}=1$.

\subsection{Transformation of single particle operator}\label{appendix:F_quasi}
    We write a hermitian one-body operator in second quantisation: $F = \sum_{ij}f_{ij}a^\dagger_i a_j$. Now using $f_{ij} = f^*_{ji}$ we can write
    \begin{align*}
        F &=\frac{1}{2}\sum_{ij}f_{ij}a^\dagger_i a_j + \frac{1}{2}\sum_{ij}f^*_{ji}a^\dagger_i a_j \\
        &= \frac{1}{2}\sum_{ij}f_{ij}a^\dagger_i a_j + \frac{1}{2}\sum_{ij}f^*_{ji}(\delta_{ij} - a_j a^\dagger_i) \\
        &=  \frac{1}{2}\sum_{i}f_{ii} + \frac{1}{2}\sum_{ij}\left( f_{ji}a^\dagger_i a_j - f^*_{ij} a_i a^\dagger_j \right)\\
        &=  \frac{1}{2}\Tr(f)+ \frac{1}{2}\begin{pmatrix}a^\dagger & a\end{pmatrix}\begin{pmatrix} f & 0 \\ 0 & -f^T\end{pmatrix}\begin{pmatrix}a \\ a^\dagger\end{pmatrix} 
    \end{align*}
    
    If we now instead work with quasiparticle operators, which are related to the single-particle operators through a Bogoliubov transformation, the previous result changes to:
    \[
    F = \frac{1}{2}\Tr(f)+ \frac{1}{2}\begin{pmatrix}a^\dagger & a\end{pmatrix} \mathcal{W}^\dagger\begin{pmatrix} f & 0 \\ 0 & -f^T\end{pmatrix} \mathcal{W}\begin{pmatrix}a \\ a^\dagger\end{pmatrix} 
    \]
    which is more often written (e.g. in \cite{Avogadro2011}) as:
    \[
     \mathcal{W}^\dagger\begin{pmatrix} f & 0 \\ 0 & -f^T\end{pmatrix} \mathcal{W} = \begin{pmatrix}
         F^{11}&F^{20}\\-F^{02}&-(F^{11})^T
     \end{pmatrix}
    \]
    where
    \begin{align*}
    \begin{dcases}
        F^{11} &= U^\dagger f U - V^\dagger f^T V\\
        F^{20} &= U^\dagger f V^* - V^\dagger f^T U^*\\
        -F^{02} &= V^T f U - U^Tf^TV\\
        (-F^{11})^T &= V^TfV^*-U^Tf^TU^*
    \end{dcases}
    \end{align*}
        
\end{subappendices}
\chapter{Emulators for FAM(Q)RPA}\label{chap:emulator}

\section{Introduction}
An emulator (or surrogate model) is a simplified, highly efficient mathematical tool designed to approximate the output of a computationally expensive simulation. While these tools are well-established in many scientific fields, such as engineering and computer science, emulators have only recently been adopted in nuclear physics~\cite{Melendez_2022}. 

There are two main types of emulators: data-driven and model-driven. The former learns patterns from input to output data directly, and often appear in the form of a neural network~\cite{Belley_2026, Melendez_2022}. The main advantage of this approach is that minimal to no understanding of the underlying physics is required, though it necessitates a lot of data to be effective. In practice, such emulators are often used for uncertainty estimation~\cite{Drischler2023, Alnamlah}. 

Model-driven emulators use or adapt a known physical model, keeping its main structure. While their development and/or implementation can be more involved, the main advantage is that the results obtained by the emulation respect the underlying physics, and are able to extrapolate more effectively~\cite{Melendez_2022}. Consequently, these emulators usually need fewer data to obtain a certain accuracy compared to data-driven emulators. One category of these emulators are reduced order models. Recently, this has been applied to model two-body scattering in momentum space~\cite{Giri2025, Maldonado2025} or for uncertainty quantification for elastic scattering processes~\cite{jinlei}.

We aim to emulate the FAM-QRPA equations, i.e.~for a given EDF, parametrisation and nucleus, we want to obtain the full strength spectrum as fast as possible. To do so, we propose the use of a Reduced Order Modelling (ROM) technique. By projecting the FAM-QRPA equation onto a significantly smaller, optimized subspace, we aim to bypass additional expensive FAM calculations that would be needed for obtaining the full spectrum.

The development of the emulator consists of two parts. On the one hand, a suitable projection method has to be chosen. We consider the Galerkin and Least-Square Petrov-Galerkin methods, both reasonably intuitive and standard, see for example~\cite{Giri2025, jinlei}. Potential numerical instabilities are detected and methods to circumvent these are proposed. 

On the other hand, the basis vectors of the emulator's subspace have to be selected in an automated way. We note that straightforward equidistant sampling is not robust (in a sense that it requires a priori information about the resonance region) nor sufficient for efficiently building the emulator. It is also likely to introduce numerical instabilities. Instead, we consider a greedy-1D approach, which is a rather standard procedure~\cite{Hesthaven2016} and is also applied in the literature, e.g.~\cite{Maldonado2025, herbst, Giri2025}. The greedy algorithm requires a definition of a cost function (hereafter denoted as error estimator). These are typically specific to the problem and crucial for the greedy method to work.

Previously, a direct approach to model the response function was done by using a kernel polynomial method~\cite{bjelcic}. It uses Chebyshev polynomials to expand the nuclear response function, and the result are the expansion coefficients. In contrast to our method, the underlying FAM matrices are not obtained and the methodology is fundamentally different.

The first application of ROM techniques to FAM-QRPA calculations is presented in Ref.~\cite{Jin2025}. They aim to build an emulator that can be applied when varying a selected set of EDF parameters. The goals in this thesis are different: we will employ ROM on a fixed EDF parametrisation, emphasizing on obtaining the full strength spectrum with only a \textit{few} FAM-QRPA calculations. By contrast, in Ref.~\cite{Jin2025}, (the closest equivalence of) one snapshot requires the full strength spectrum obtained by FAM-QRPA (i.e.~about 400 calculations), for a specific EDF parametrisation. Consequently, their emulator can be applied for different EDF parameter values, while ours cannot. In theory, both ROM techniques could be combined to reduce the total computational cost even further.

The chapter is organised as follows. We first introduce ROM applied to the FAM equation in Sec.~\ref{sec:reduced-order-fam-equations}. In Sec.~\ref{subsec:galerkin-projection} two different projection methods are considered followed by a basis manipulation enhancing numerical stability. Attempts for smart snapshot selection are proposed in Sec.~\ref{sec:methods-for-snapshot-selection} and compared with straightforward ones. Additional comments regarding symmetry are discussed in Sec.~\ref{sec:strength-symmetry}. Throughout, the methodology is verified using examples of the ISGMR of deformed \ce{Mg-24} and spherical \ce{Pb-208} using the standard Skyrme parametrisation BSkG2.  The chapter ends with a benchmark and applications of the finalised emulator for strength functions for ISGMR of \ce{Mg-24} and \ce{Pb-208} in Sec.~\ref{sec:performance}, and its generalisability is illustrated for ISGMR and ISGQR excitations on other EDF forms and parametrisations ($t_4$-$t_5$: BSkG4, N2LO: BSkG5 and Gogny: D1M), and for an IVGDR excitation of a triaxially deformed nucleus \ce{Zr-100}. 

\section{Reduced order FAM equation}\label{sec:reduced-order-fam-equations}
Starting from a mean-field calculation, the FAM equation (Eq.~\eqref{eq:linear_response_QFAM}) can be expressed in the following matrix form:
\begin{equation}
    \mathcal{A}(\omega)\mathcal{X}(\omega)  \equiv (E-\omega \mathcal{M}+\mathcal{H})\mathcal{X} = -\mathcal{F} \;, \label{eq:FAM_equation}
\end{equation}
using 
\[
\begin{aligned}
\mathcal{X} &= \begin{pmatrix}
    X_{\mu\nu}&Y_{\mu\nu}
\end{pmatrix}^T && \in \mathbb{C}^{N\times 1} \;,\\
\mathcal{M} &= \begin{pmatrix}
    \unitmatrix &0\\0&-\unitmatrix
\end{pmatrix} && \in \mathbb{R}^{N\times N} \;,\\
\mathcal{F} &= \begin{pmatrix}
    F^{20}&F^{02}
\end{pmatrix}^T && \in \mathbb{C}^{N\times 1} \;,\\
\delta \mathcal{H} = \mathcal{H}\mathcal{X} &= \begin{pmatrix}
   \delta H^{20}_{\mu\nu}(\omega)&\delta H^{02}_{\mu\nu}(\omega)
\end{pmatrix}^T && \in \mathbb{C}^{N\times 1}\;,\\
E &= \begin{pmatrix}
    E_{\mu\nu}&0\\0&E_{\mu\nu}
\end{pmatrix} && \in \mathbb{R}^{N\times N} \;.
\end{aligned}
\]
and that $\delta \mathcal{H}$ depends linearly on the amplitudes $\mathcal{X}$. In this notation, $N$ represents the dimension of the configuration space from the mean-field solution, equal to \textbf{two times} the number of particle-hole pairs (in RPA) or two-quasiparticle pairs (in QRPA). 

Equation~\eqref{eq:FAM_equation} defines a relation between a given frequency $\omega$ and the response vector $\mathcal{X}$, which in turn relates to the nuclear response function (or strength) as discussed in Sec.~\ref{sec:strength-symmetry}. By evaluating the system at a set of frequencies $\{\omega_i\}_{1\leq i\leq N_s}$, we generate an  $N_s$-dimensional snapshot basis $\{\mathcal{X}(\omega_i)\}_{1\leq i\leq N_s}$, where $N_s \ll N$ and which span the reduced subspace. We will use a simplified notation $\mathcal{X}(\omega_i) \equiv \mathcal{X}_i$, but stress that each basis vector has a corresponding frequency $\omega_i$. In other words, for each basis vector we have that
\begin{equation}
    (E-\omega_i \mathcal{M}+\mathcal{H})\mathcal{X}_i = -\mathcal{F} \;.\label{eq:FAM_basis_solution}
\end{equation}
If we assume that this basis sufficiently spans the relevant phase space, any solution $\mathcal{X}$ for frequency $\omega (\neq \omega_i)$ can be approximated as a linear combination of the snapshots:
\begin{equation}
\mathcal{X(\omega)} \approx \sum_{i=1}^{N_s} \alpha_i(\omega) \mathcal{X}_i = \tilde{\mathcal{X}}(\omega) \;.\label{eq:X_approx}
\end{equation}
The frequency-dependent coefficients $\alpha_i(\omega)$ correspond to the projection of the full solution $\mathcal{X}(\omega)$ onto the reduced subspace (i.e.~the space spanned by the $N_s$ basis vectors). 

Substituting the expansion from Eq.~\eqref{eq:X_approx} into the FAM Eq.~\eqref{eq:FAM_equation} yields: 
\begin{align}
    \sum_{i=1}^{N_s}\left(E-\omega \mathcal{M}+\mathcal{H}\right)\alpha_i(\omega)\mathcal{X}_i & \approx  -\mathcal{F} \;. \label{eq:fam_dummy_label1}
\end{align}
By exploiting Eq.~\eqref{eq:FAM_basis_solution}, namely that each basis vector $\mathcal{X}_i$ is an exact solution to the FAM equation at frequency $\omega_i$, we can write
\begin{align}
    && (E-\omega_i \mathcal{M}+\mathcal{H})\mathcal{X}_i &= -\mathcal{F} \nonumber \\
    & \Leftrightarrow & (E+\mathcal{H})\mathcal{X}_i &= -\mathcal{F}+\omega_i \mathcal{M}\mathcal{X}_i \nonumber \\ & \Leftrightarrow & \qquad
    \mathcal{A}(\omega) \mathcal{X}_i &= -\mathcal{F}+(\omega_i-\omega)\mathcal{M}\mathcal{X}_i  \;,\label{eq:FAM_substitution} 
\end{align}
and substitute this back into the LHS of Eq.~\eqref{eq:fam_dummy_label1}:
\begin{align}
    \sum_{i=1}^{N_s}\left(-\mathcal{F} + (\omega_i -\omega)\mathcal{M}\mathcal{X}(\omega_i)\right)\alpha_i & \approx  -\mathcal{F}\;.\label{eq:overdetermined-approx}
\end{align}
We have shown that Eq.~\eqref{eq:fam_dummy_label1} can be expressed without explicitly referring to the induced fields $\delta H$, which are calculated using the FAM formulas~\cite{Avogadro2011}. Instead, in order to find the FAM amplitudes for a new frequency, we only need the (pre-calculated) snapshots and the external field $\mathcal{F}$. This bypasses the need for additional FAM calculations beyond those required to construct the snapshot basis, thereby potentially reducing computational cost significantly.

Note that putting an equality in Eq.~\eqref{eq:overdetermined-approx} does not hold in general. This leads to an overdetermined set of equations with $N_s$ unknown values ($\alpha_i$) and $N$ equations, yielding no solutions as typically $N_s\ll N$. For this reason, finding $\mathcal{X}$ changes into an optimisation problem where we seek an optimal approximation within the $N_s$-dimensional subspace that minimizes the discrepancy between the $N$ equations.

In this context, a solution is \textit{optimal} if it adheres to the constraints defining this subspace and it minimizes the residual of the approximation evaluated in the original equation, using a chosen metric (typically $L_2$ norm). Defining this subspace, or projection, however, is not unique and has a large impact on the quality of this optimal solution. In the next subsection, we discuss two different projection methods, namely the Galerkin and Least-Squares Petrov-Galerkin projection to make this choice.

Next, we define the snapshot matrix $\mathcal{V} \in \mathbb{R}^{N\times N_s}$ by stacking the basis vectors as follows:
\begin{equation}
   \mathcal{V} = \begin{pmatrix}
       | & | &  & | \\
        \mathcal{X}_1 & \mathcal{X}_2 & \ldots & \mathcal{X}_{N_s}\\
        |  & | &  & | 
   \end{pmatrix}\;. \label{eq:ROM_basis_matrix}
\end{equation}

\subsection{Reference spectra and RMS error}\label{sec:reference_data_and_RMS}
The solution $\mathcal{X}$, which contains $X_{\mu\nu}$ and $Y_{\mu\nu}$, determines the nuclear response function (strength), as defined in Eq.~\eqref{eq:S_XY}. To evaluate the the emulator's performance, we will compare the emulated strength against the reference spectra obtained through FAM calculations. These reference sets, $\mathbf{W}_\mathrm{ref}=\{\omega_r\}_{1\leq r \leq N_\mathrm{ref}}$, consist of $N_\mathrm{ref}>$300 frequencies. We obtained reference spectra for different Skyrme parametrisations and smearing values for \ce{Mg-24} and \ce{Pb-208}. 

Using the notation from Eq.~\eqref{eq:X_approx}, we quantify the emulator's accuracy using the absolute Root Mean Square (RMS) error on the strength $S$ as follows:
\begin{align}
    \varepsilon_\mathrm{RMS} = \sqrt{\frac{1}{N_{\mathrm{ref}}}\sum_{\omega \in \mathbf{W}_\mathrm{ref}} \Big(S(\mathcal{X}(\omega)) -S(\tilde{\mathcal{X}}(\omega))\Big)^2} \;. \label{eq:rms_absolute_error}
\end{align}
We prefer using an absolute error metric over a relative one for two reasons. First, the strength function is often near zero across significantly large frequency intervals. In these regions, relative errors can become disproportionately large even when the emulated strength is small. Moreover, most applications are not sensitive to this because the physical interpretation of the response is dominated by the resonance peaks. Second, an absolute metric effectively penalizes mislocated peaks.

Note, however, that this metric may not capture high frequency oscillations or numerical artefacts (e.g.~in Fig.~\ref{fig:PG_VS_G_Mg}) if they fall between the sampled reference frequencies. Another drawback of this metric is that it can be punishingly sensitive for a minor frequency shift in a sharp peak or a small difference in the strength. While an alternative error metric like the Mean Absolute Error is more robust against such outliers, it simultaneously lacks sensitivity to important features. Lastly, one could also check the energy-weighted sum rules to validate the strength function, however, since these only check the integrated strength value, they are less sensitive to the spectral shape. 

\section{Methods for reduced-basis projections}\label{sec:methods-for-reduced-basis-projections}
\subsection{Galerkin projection}\label{subsec:galerkin-projection}
The Galerkin projection is an intuitive way of reducing the model space. We try to find the coefficients $\alpha_i$ in Eq.~\eqref{eq:X_approx} such that $||\mathcal{X}(\omega) - \tilde{\mathcal{X}}(\omega)||$ is minimal, effectively finding the closest point in the space spanned by the snapshots to the actual solution $\mathcal{X}(\omega)$. 

Projecting the FAM equation on the $j$-th snapshot basis vector yields:
\begin{align}
    \mathcal{X}^{\dagger}_j(E-\omega \mathcal{M}+\mathcal{H})\left(\sum_{i=1}^{N_s}\alpha_i(\omega)\mathcal{X}_i\right) &= -\mathcal{X}^{\dagger}_j\mathcal{F} \nonumber \\
    \Leftrightarrow \qquad
    \sum_{i=1}^{N_s}\mathcal{X}^{\dagger}_j(-\omega \mathcal{M}-\mathcal{F}+\omega_i \mathcal{M})(\alpha_i(\omega)\mathcal{X}_i) &= -\mathcal{X}^{\dagger}_j\mathcal{F} \nonumber \\  \Leftrightarrow \qquad
    \sum_{i=1}^{N_s}\left(-\mathcal{X}^{\dagger}_j \mathcal{F} + (\omega_i -\omega)\mathcal{X}^{\dagger}_j \mathcal{M}\mathcal{X}_i\right)\alpha_i(\omega) &= -\mathcal{X}^{\dagger}_j\mathcal{F} \;. \label{eq:galerkin_basic}
\end{align}
We have used that $\mathcal{X}_i$ solves the FAM equation~\eqref{eq:FAM_substitution}. We thus obtain a linear system of the form 
\begin{equation}\sum_{i=1}^{N_s} A_{ji}\alpha_i= b_j \;,\label{eq:system_matrix_expression}\end{equation}
and refer to $A$ as the \emph{system-matrix} and $\kappa = \mathrm{cond}(A)= \tfrac{\sigma_\mathrm{max}}{\sigma_\mathrm{min}}$ the \emph{condition number} of the system matrix~\cite[Chap.~3.3]{demmel1997_singular}, which depends on the singular values $\sigma$ of $A$. In the literature, the condition number of square matrices is often defined (w.r.t. a different norm) as $\kappa = ||A^{-1}||\,||A||$ ~\cite[Chap.~2.2]{demmel1997_singular}. This is a measure of how sensitive a linear system is to small errors or perturbations, effectively indicating how close a matrix is to being singular~\cite[Thm.~2.1]{demmel1997_singular}. We now write the expression for the system matrix. Define
\[
\Omega  = \mathrm{diag}(\omega_i-\omega)_{N_s\times N_s}\qquad
\mathrm{and}\qquad
F = \begin{pmatrix}
| & | & & | \\
    \mathcal{F}&\mathcal{F}& \ldots &\mathcal{F}\\
| & | & & | 
\end{pmatrix}_{N \times N_s}\;,
\]
by copying the vector $\mathcal{F}$. We now obtain
\begin{align*}
    A_{ji}&= -\mathcal{X}^{\dagger}_j \mathcal{F} + (\omega_i -\omega)\mathcal{X}^{\dagger}_j \mathcal{M}\mathcal{X}_i\\
   \Leftrightarrow \qquad  A &= -\mathcal{V}^\dagger F + \mathcal{V}^\dagger \mathcal{M}\mathcal{V}\Omega \;,
\end{align*}
from which we can see that $A$ is not hermitian:
\begin{align*}
A^\dagger &= -F^\dagger \mathcal{V} + \Omega^\dagger \mathcal{V}^\dagger \mathcal{M}^\dagger \mathcal{V} = -F^\dagger \mathcal{V} + \Omega^* \mathcal{V}^\dagger \mathcal{M} \mathcal{V} \neq A\;.
\end{align*}

\subsection{Petrov-Galerkin Least-Squares projection}\label{subsec:petrov-galerkin}
Instead of using a Galerkin projection and projecting on the basis vectors, we can use a least-squares method on Eq.~\eqref{eq:FAM_equation}. More precisely, we define a residual function $r(\omega)$ which represents how much the approximation deviates with respect to the FAM equation:
\begin{equation}
r(\omega) = \mathcal{A}(\omega) \sum_{i=1}^{N_s}\alpha_i(\omega) \mathcal{X}_i + \mathcal{F} \;, \label{eq:residual_FAM}
\end{equation}
and minimise the $L^2$-norm with respect to the coordinates $\alpha_j(\omega)$: 
\begin{align}
0&= \frac{\partial}{\partial \alpha_j} || r(\omega) ||^2 \label{eq:deriv_alpha_start}\\
&=  \frac{\partial}{\partial \alpha_j} \left(\mathcal{A}(\omega) \sum_{i=1}^{N_s}\alpha_i(\omega) \mathcal{X}_i + \mathcal{F}\right)^\dagger\left(\mathcal{A}(\omega) \sum_{i=1}^{N_s}\alpha_i(\omega) \mathcal{X}_i + \mathcal{F}\right) \nonumber\\
&=  \frac{\partial}{\partial \alpha_j} \sum_{i,k=1}^{N_s} ( \alpha_i^*(\omega) \mathcal{X}^\dagger_i \mathcal{A}^\dagger(\omega)+ \mathcal{F}^\dagger)(\mathcal{A}(\omega) \alpha_k(\omega) \mathcal{X}_k + \mathcal{F}) \nonumber\\
&= \sum_{i=1}^{N_s} \left( \alpha_i^*(\omega) \mathcal{X}^\dagger_i \mathcal{A}^\dagger(\omega)+ \mathcal{F}^\dagger\right)\mathcal{A}(\omega) \mathcal{X}_j \nonumber \\
&=\sum_{i=1}^{N_s} \alpha_i^*(\omega) \mathcal{X}^\dagger_i \mathcal{A}^\dagger(\omega)\mathcal{A}(\omega) \mathcal{X}_j+ \mathcal{F}^\dagger \mathcal{A}(\omega) \mathcal{X}_j \;.
\end{align}

Using the same trick as before, by applying the FAM equation~\eqref{eq:FAM_substitution}, we replace $\mathcal{A}(\omega)\mathcal{X}_i$ with:
\begin{align*}
\begin{dcases}
     \mathcal{A}(\omega)\mathcal{X}_i &= -\mathcal{F} + (\omega_i - \omega)\mathcal{M}\mathcal{X}_i\\ 
    \mathcal{X}_i^\dagger \mathcal{A}(\omega)^\dagger &= -\mathcal{F}^\dagger + (\omega_i^* - \omega^*)\mathcal{X}^\dagger_i\mathcal{M}^\dagger \;,
\end{dcases}
\end{align*}
yielding:
\begin{align}
    0&=\sum_{i=1}^{N_s} \alpha_i^*(\omega) \left(-\mathcal{F}^\dagger + (\omega_i^* - \omega^*)\mathcal{X}^\dagger_i\mathcal{M}^\dagger\right)\Big(-\mathcal{F}+ (\omega_j - \omega)\mathcal{M}\mathcal{X}_j\Big) \nonumber \\&\qquad \qquad + \mathcal{F}^\dagger\left(-\mathcal{F} + (\omega_j - \omega_\alpha)\mathcal{M}\mathcal{X}_j\right) \nonumber \\
    &=\sum_{i=1}^{N_s} \alpha_i^*(\omega) \left(\mathcal{F}^\dagger\mathcal{F} - (\omega_i^* - \omega^*)\mathcal{X}^\dagger_i\mathcal{M}^\dagger\mathcal{F}\right) -\mathcal{F}^\dagger\mathcal{F} + (\omega_j - \omega)\mathcal{F}^\dagger\mathcal{M}\mathcal{X}_j \nonumber \\
    &\qquad \qquad +\sum_{i=1}^{N_s}\alpha_i^*(\omega)\left( -(\omega_j - \omega)\mathcal{F}^\dagger\mathcal{M}\mathcal{X}_j + (\omega_i^* - \omega^*)(\omega_j - \omega)\mathcal{X}^\dagger_i\mathcal{M}^\dagger\mathcal{M}\mathcal{X}_j\right) \;.\label{eq:rom_dummy2}
\end{align}
This expression is more involved compared to the Galerkin equivalent, but it is of the same linear form $A\alpha = b$. Equivalently, one could take the derivative w.r.t.~$\alpha^*$ instead in Eq.~\eqref{eq:deriv_alpha_start}. This would yield a system of equations in terms of $\alpha$ immediately (equal to the conjugate of Eq.~\eqref{eq:rom_dummy2}). Both methods have been implemented and yield identical output. 

Using $\alpha^\dagger A^\dagger = b^\dagger$ such that $\sum_{i=1}^{N_s}\alpha^*_i A^\dagger_{ij} = b^*_j$ and using the same expression for $F$ and $\Omega$ as earlier, we write the hermitian conjugate of the system matrix $A$ as follows:
\[
\begin{aligned}
A^\dagger_{ij} &=\mathcal{F}^\dagger\mathcal{F} - (\omega_i^* - \omega^*)\mathcal{X}^\dagger_i\mathcal{M}^\dagger\mathcal{F} -(\omega_j - \omega)\mathcal{F}^\dagger\mathcal{M}\mathcal{X}_j + (\omega_i^* - \omega^*)(\omega_j - \omega)\mathcal{X}^\dagger_i\mathcal{M}^\dagger\mathcal{M}\mathcal{X}_j \\
A^\dagger_{ij} &=\Big(\mathcal{F}^\dagger-(\omega_i^*-\omega^*)\mathcal{X}^\dagger_i\mathcal{M}^\dagger\Big)\Big(\mathcal{F}+(\omega_j-\omega)\mathcal{M}\mathcal{X}_j\Big)\\[0.5em]
\Leftrightarrow \qquad A^\dagger&= \left(F^\dagger-\Omega^\dagger\mathcal{V}^\dagger \mathcal{M}^\dagger\right)\left(F-\mathcal{M} \mathcal{V}\Omega\right)\;.
\end{aligned}
\]

In contrast with the Galerkin projection, the system matrix now is Hermitian:
\[
\begin{aligned}
A^\dagger &= \left(F^\dagger-\Omega^\dagger\mathcal{V}^\dagger \mathcal{M}^\dagger\right)\left(F-\mathcal{M} \mathcal{V}\Omega\right)\;,\\[1em]
A &= \left(F-\mathcal{M} \mathcal{V}\Omega\right)^\dagger\left(F-\Omega^\dagger\mathcal{V}^\dagger \mathcal{M}^\dagger\right)^\dagger\\
 &= \left(F^\dagger-\Omega^\dagger\mathcal{V}^\dagger \mathcal{M}^\dagger\right)\left(F-\mathcal{M} \mathcal{V}\Omega\right)=A^\dagger\;. \\
\end{aligned}
\]

Equation~\eqref{eq:rom_dummy2} can be further simplified by using that $\mathcal{M}^\dagger = \mathcal{M} $ and $\mathcal{M}^\dagger \mathcal{M} = 1$:
\[
\begin{aligned}
     \mathcal{F}^\dagger\mathcal{F} - (\omega_j - \omega)\mathcal{F}^\dagger\mathcal{M}\mathcal{X}_j &=\sum_{i=1}^{N_s}\alpha_i^*(\omega) \left[ \mathcal{F}^\dagger\mathcal{F}-(\omega_j - \omega)\mathcal{F}^\dagger\mathcal{M}\mathcal{X}_j \right]
     \\ &\quad + \sum_{i=1}^{N_s} \alpha_i^*(\omega) \left[(\omega_i^* - \omega^*)(\omega_j - \omega)\mathcal{X}^\dagger_i\mathcal{X}_j- (\omega_i^* - \omega^*)\mathcal{X}^\dagger_i\mathcal{M}^\dagger\mathcal{F}\right] \;.
\end{aligned}
\]

\begin{figure}
    \centering
    \includegraphics[width=0.435\linewidth]{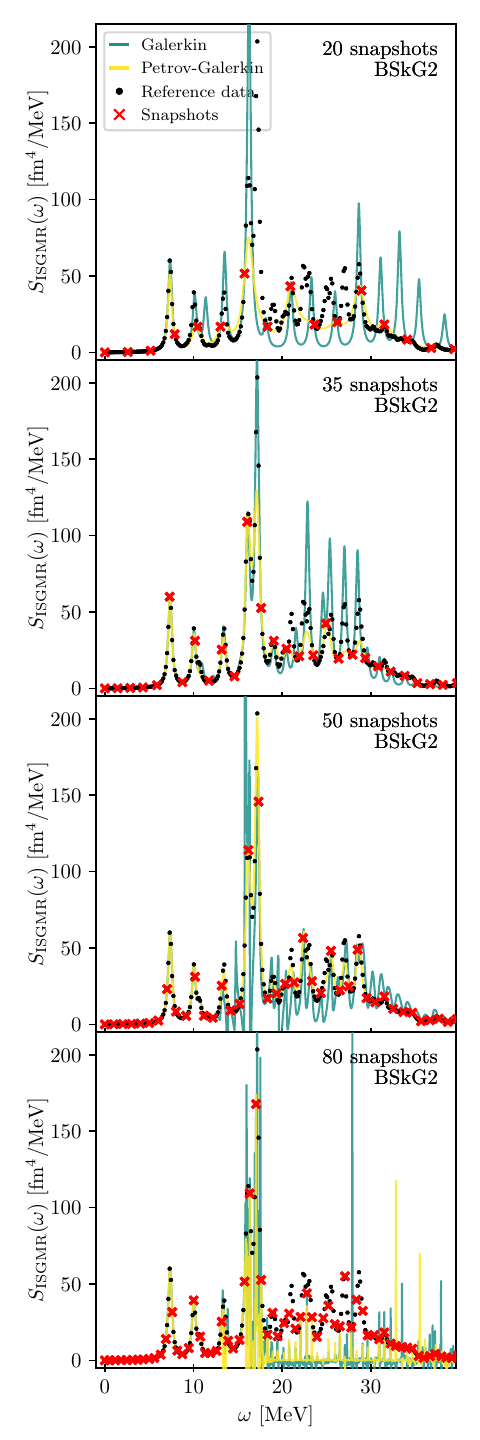}~\includegraphics[width=0.435\linewidth]{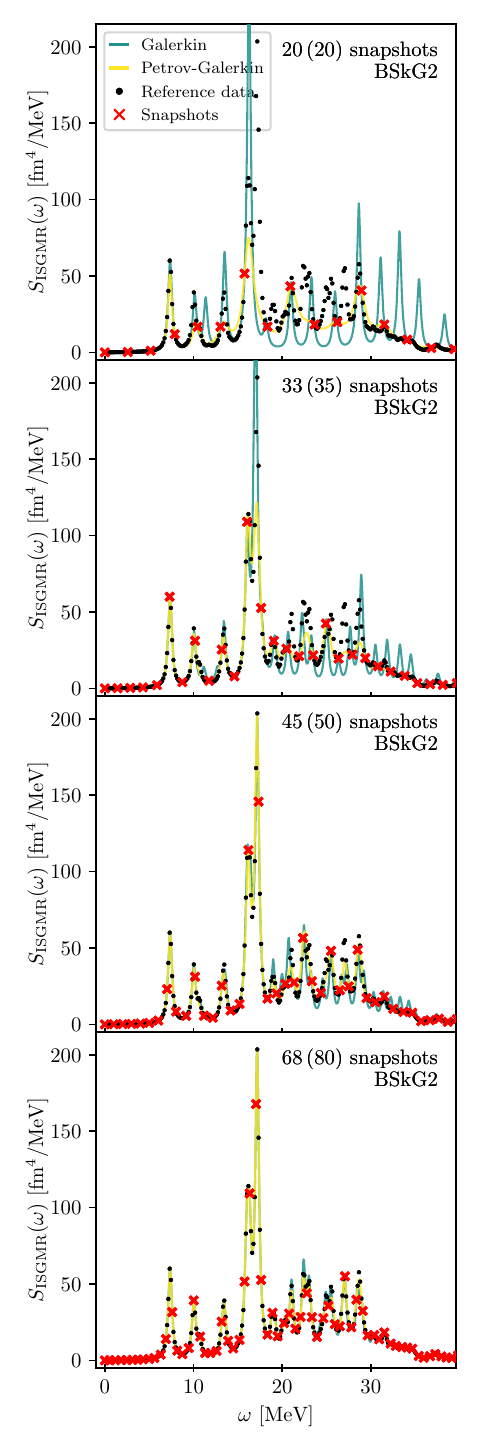}
    \caption{Strength of the ISGMR of \ce{Mg-24} with $\gamma=0.25$\,MeV calculated with FAM (black) and emulated using Galerkin (green) and Petrov-Galerkin (yellow) projection for a different number of (equidistantly-sampled) snapshots. The left column is without POD basis manipulations, while the right column uses a singular value cutoff of $0.001$. }
    \label{fig:PG_VS_G_Mg}
\end{figure}

To conclude this section, we compare the Galerkin and Petrov-Galerkin methods, as illustrated in the left column of Fig.~\ref{fig:PG_VS_G_Mg}. While the Galerkin method identifies peaks using fewer snapshots, such as at $N_s=20$, these peaks are often inaccurately positioned or absent in the reference spectrum. Conversely, the Petrov-Galerkin method is more conservative, requiring a higher number of snapshots for peaks to emerge. However, its predictions align more closely with the reference data at equal $N_s$.

Despite these differences, both methods encounter stability issues as the snapshot count increases. The Galerkin method exhibits instabilities as early as 50 snapshots, resulting in unphysical strength amplitudes. While more robust, the Petrov-Galerkin method eventually encounters similar instabilities, appearing at approximately 80 snapshots for these examples.

These results are counterintuitive, as one would expect the solution to converge toward the reference with increasing number of snapshots. In the following section, we treat this behaviour as numerical ill-conditioning  rather than a physical or methodological issue. Consequently, we try to stabilize the system by applying basis manipulations. 

\subsection{Basis manipulations}\label{subsec:basis-manipulations}
In the previous sections, we discussed different projection methods for model space reduction. We continue the analysis by considering basis vector operations which could improve numerical stability. While snapshot selection is also important for the emulator's performance, we only consider this later in Sec.~\ref{sec:methods-for-snapshot-selection}, and start from a given snapshot basis $\{\mathcal{X}_i\}_{0\leq i \leq N_s}$. 

An arbitrary snapshot basis is not, in general, linearly independent. This can be understood as follows. The FAM amplitudes $X(\omega)$ and $Y(\omega)$ are defined within the particle-hole space of finite dimension $N$, determined by the underlying mean-field solution. Since the frequency $\omega$ is a continuous parameter, selecting $N_s > N$ snapshots will mathematically guarantee linear dependence. Physically, this reflects the fact that nuclear response changes smoothly with $\omega$. Small shifts in the probing frequency yield highly correlated responses, with rapid variations only occurring near the resonance peaks. Consequently, if snapshots are chosen too densely or in redundant regions, the system matrix (Eq.~\eqref{eq:system_matrix_expression}) becomes ill-conditioned.

This can be a possible explanation for the unphysical approximated strength observed in Fig.~\ref{fig:PG_VS_G_Mg}, as the numerical solver struggles to represent $\mathcal{X}$ correctly. The condition number $\kappa$ plotted in Fig.~\ref{fig:compare_SVD_Mg_cond_nr} confirms this, showing values that grow exponentially with increasing number of snapshots.

The instabilities observed in our FAM emulator are reminiscent of so-called Kohn anomalies, typically encountered in variational scattering theory. In scattering emulators, these anomalies appear at discrete energies when the basis and the specific boundary conditions lead to a non-unique solution of the variational functional~\cite{Drischler2021, Maldonado2025, Giri2025}. Effectively, this manifests as near-singular system matrices, similar to our case. Moreover, it is shown that the Petrov-Galerkin method is less prone to these singularities compared to the Galerkin method~\cite{Maldonado2025}. Further, by applying a greedy sampling method (to be introduced in Sec.~\ref{subsubsec:greedy}), no such anomalies are encountered~\cite{Maldonado2025, Giri2025}.

To mitigate these instabilities, we employ a rather standard technique called Proper Orthogonal Decomposition (POD) to generate an orthonormal basis~\cite{Maldonado2025}~\cite[Sec. 3.2.1]{Hesthaven2016}. Consider a new basis $\{\mathcal{U}_i\}_{i\leq N_r}$ with $N_r<N_s$ vectors which we obtain by using a linear combination (and potentially truncation) of the original snapshot basis, such that $\mathcal{U} = \mathcal{V}U$ with $U \in \mathbb{R}^{N_s\times N_r}$ a transformation matrix and 
\[
   \mathcal{U} = \begin{pmatrix}
       | & | &  & | \\
        \mathcal{U}_1 & \mathcal{U}_2 & \ldots & \mathcal{U}_{N_r}\\
        |  & | &  & | 
   \end{pmatrix} \in \mathbb{C}^{N\times N_r}\;,
\]
the ROM matrix after transformation. The approximated $\mathcal{X}(\omega)$ can be expressed as follows:
\begin{align*}
\mathcal{X}(\omega) \approx \sum_{j=1}^{N_s}\alpha_j(\omega)\mathcal{X}_j \approx \sum_{j=1}^{N_r}\alpha_j(\omega)(\mathcal{U} U^{-1})_j &\equiv \sum_{i=1}^{N_r}\beta_i(\omega)\mathcal{U}_i  \;.
\end{align*}
Note that the first $\approx$ represents the projection within the reduced model space, while the second $\approx$ is due to possible further truncation of the model space, such that the equation holds in case $N_s = N_r$. We want to stress that, while we have used a simplified notation for $\mathcal{X}_i \equiv \mathcal{X}(\omega_i)$ which implies a bijection between $\omega_i$ and $\mathcal{X}_i$, there is no such relation between $\omega_i$ and $\mathcal{U}_i$ because it generally depends on all $\{\omega_i\}_{1\leq i \leq N_s}$.

For the remainder of this section, we focus on one specific method to apply a POD, namely the singular value decomposition (SVD) and analyse how it changes the projections considered in Sec.~\ref{subsec:galerkin-projection} and Sec.~\ref{subsec:petrov-galerkin}. Lastly, we point out that SVD may become expensive for large instances, due to the matrix diagonalization. While other well-known orthogonalisation methods exist and are often considered in the literature, e.g.~Gram-Schmidt procedure~\cite{Hesthaven2016}, we opt for the SVD as it preserves the structure of the FAM equations.

\subsection{Singular Value Decomposition}

\begin{wrapfigure}[18]{r}{0.45\linewidth}
    \vspace{-4.75em}
    \centering
    \includegraphics[width=\linewidth]{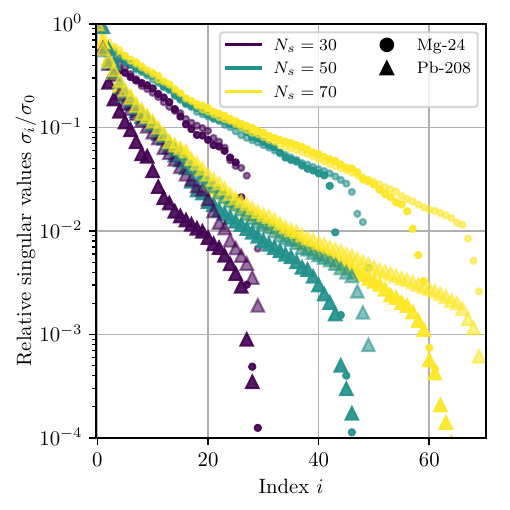}
    \vspace{-2em}
    \caption{Relative singular values of the snapshot basis $\mathcal{V} = [ \mathcal{X}_1, \mathcal{X}_2, \ldots]$ for different basis sizes $N_s$. To compare, the results using a greedy sampler are also added (semi-transparent).}
    \label{fig:singular_values}
\end{wrapfigure}
The transformation $U$ can be defined using a singular value decomposition (SVD) on the snapshot basis matrix $\mathcal{V}$. This means we can write
\[
\mathcal{V}= U_{S}SV^\dagger_{S}\;,
\]
where $V_S \in \mathbb{C}^{N_s \times N_s}$ is a complex unitary matrix, while $U_S \in \mathbb{C}^{N\times N_s}$. This assumes that the singular values of $\mathcal{V}$ are collected in descending order in the diagonal square matrix $S \in \mathbb{R}_{\geq 0}^{N_s \times N_s}$. The columns of $U_S$ and $V_S$ are orthonormal vectors and could be used as an alternative basis. We can thus transform the original snapshot as follows: 
\[
\mathcal{V}U \equiv \mathcal{V}(V_S^\dagger)^{T}S^{-1} = U_S \equiv \mathcal{U} \;.
\]
A truncated SVD reduces the dimensionality of the problem by removing basis vectors that correspond to small singular values. Fig.~\ref{fig:singular_values} shows the singular values relative to the largest one for different $N_s$, as a function of the index of the ordered singular values. One can observe an initial exponential decrease in these relative values (which appears linear on the figure) followed by a sudden jump around relative values of $10^{-1.5}$ and $10^{-2}$ for \ce{Mg-24} and around $10^{-3}$ for \ce{Pb-208}. This suggests that truncating any vectors that correspond to the singular values beyond this jump will not cause a significant loss of information. Moreover, this jump occurs later as the number of snapshots increases. This indicates that more information is captured in the space spanned by $\{\mathcal{X}_i \}$, and more relevant basis vectors can be constructed in $\{\mathcal{U}_i\}$. The same happens when applying the greedy sampling method, introduced later in Sec.~\ref{subsubsec:greedy}, indicating that this method selects the snapshots more effectively.

Henceforth we will use a relative cutoff on the singular values of $0.001$ and simply refer to it as a cutoff. Note that without imposing such a cutoff, the SVD transformation is of no use.

\subsubsection{Galerkin POD}
For the Galerkin projection, after applying a POD on the ROM basis, we obtain the following:
\begin{align*}
\mathcal{U}_j^\dagger(E-\omega \mathcal{M}+\mathcal{H})\left(\sum_{i=1}^{N_r}\beta_i(\omega)\mathcal{U}_i \right) &= -\mathcal{U}_j^\dagger\mathcal{F}\\
\sum_{i=1}^{N_r} (U^\dagger \mathcal{V}^\dagger)_j(E-\omega \mathcal{M}+\mathcal{H}) (\mathcal{V} U)_i\beta_i(\omega) &= -(U^\dagger \mathcal{V}^\dagger)_j\mathcal{F}\\
\sum_{i=1}^{N_r}\sum_{l,k=0}^{N_s}U^\dagger_{jk} \left(-\mathcal{X}_k^\dagger \mathcal{F} + \mathcal{X}_k^\dagger(\omega_l-\omega)\mathcal{M}\mathcal{X}_l \right) U_{li}\beta_i(\omega) &= -\sum_{k=0}^{N_s} U^\dagger_{jk} \mathcal{X}^\dagger_k \mathcal{F}\;.
\end{align*}
Similar to what we had in Eq.~\eqref{eq:galerkin_basic}, even after applying a POD, we can rewrite the equations such that no additional FAM calculations are required. In this case, we now solve for fewer $N_r < N_s$ coefficients. 

\subsubsection{Petrov-Galerkin POD}
We can do the same for the Petrov-Galerkin projection, this time optimising with respect to the coefficients in the transformed basis:
\begin{align*}
0&=  \frac{\partial}{\partial \beta_j} \left(\mathcal{A}(\omega) \sum_{i=1}^{N_r}\beta_i(\omega)\mathcal{U}_i + \mathcal{F}\right)^\dagger\left(\mathcal{A}(\omega) \sum_{k=1}^{N_r}\beta_k(\omega)\mathcal{U}_k + \mathcal{F}\right)\\
&=\sum_{i=1}^{N_r} \beta_i^*(\omega) \mathcal{U}^\dagger_i \mathcal{A}^\dagger(\omega)\mathcal{A}(\omega) \mathcal{U}_j+ \mathcal{F}^\dagger \mathcal{A}(\omega) \mathcal{U}_j\\
&=\sum_{i=1}^{N_r} \beta_i^*(\omega) (U^{\dagger}\mathcal{V}^\dagger)_i \mathcal{A}^\dagger(\omega)\mathcal{A}(\omega) (\mathcal{V}U)_j+ \mathcal{F}^\dagger \mathcal{A}(\omega) (\mathcal{V}U)_j\\
&=\sum_{i=1}^{N_r}\sum_{l,k=1}^{N_s}\beta_i^*(\omega) U^{\dagger}_{il}\mathcal{X}^\dagger_l \mathcal{A}^\dagger(\omega)\mathcal{A}(\omega) \mathcal{X}_kU_{kj}+\sum_{k=1}^{N_s} \mathcal{F}^\dagger \mathcal{A}(\omega) \mathcal{X}_kU_{kj} \;.
\end{align*}
Using the substitutions from Eq.~\eqref{eq:FAM_substitution}, we obtain:
\begin{align*}
0&=\sum_{i=1}^{N_r}\sum_{l,k=1}^{N_s} \beta_i^*(\omega) U^{\dagger}_{il} \Big(-\mathcal{F}^\dagger + (\omega_l^* - \omega^*)\mathcal{X}^\dagger_l\mathcal{M}^\dagger\Big)\Big(-\mathcal{F}+ (\omega_k - \omega)\mathcal{M}\mathcal{X}_k\Big)U_{kj} \\ &\qquad + \sum_{k=1}^{N_s}\mathcal{F}^\dagger\left(-\mathcal{F} + (\omega_k - \omega)\mathcal{M}\mathcal{X}_k\right)U_{kj} \\
0&=\sum_{i=1}^{N_r}\sum_{l,k=1}^{N_s}  \beta_i^*(\omega) \Big(\mathcal{F}^\dagger\mathcal{F}U^{\dagger}_{il}U_{kj} - (\omega_l^* - \omega^*)U^\dagger_{il}\mathcal{X}^\dagger_l\mathcal{M}^\dagger\mathcal{F}U_{kj} -(\omega_k - \omega)U^{\dagger}_{il}\mathcal{F}^\dagger\mathcal{M}\mathcal{X}_kU_{kj}\Big) \\ &\qquad +\sum_{i=1}^{N_r}\sum_{l,k=1}^{N_s} \beta_i^*(\omega)(\omega_l^* - \omega^*)(\omega_k - \omega)U^\dagger_{il}\mathcal{X}^\dagger_l\mathcal{M}^\dagger\mathcal{M}\mathcal{X}_k U_{kj} + \sum_{k=0}^{N_s}\Big((\omega_k - \omega)\mathcal{F}^\dagger\mathcal{M}\mathcal{X}_k U_{kj}  -\mathcal{F}^\dagger\mathcal{F}U_{kj} \Big) \;.
\end{align*}
Note that due to the presence of $\omega_k$ and $\omega_l$, both the untransformed basis $\{\mathcal{X}_i\}$ and the transformation matrix $U$ are needed to solve the system for $\beta$.

\begin{figure}
    \centering
    \begin{minipage}[t]{0.45\linewidth}
    \centering
    \includegraphics[width=\linewidth]{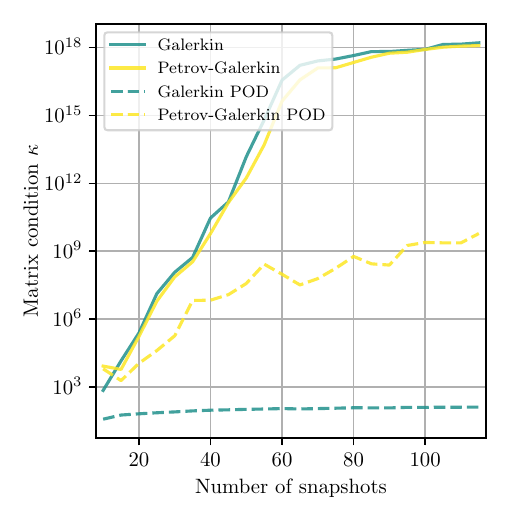}
    \caption{Condition number of the system matrix (applied for \ce{Mg-24} with $\gamma = 0.25$\,MeV) with different projection methods as a function of number of snapshots with (dashed) cutoff $0.001$ and without (full).}
    \label{fig:compare_SVD_Mg_cond_nr}
    \end{minipage}~
    \begin{minipage}[t]{0.45\linewidth}
    \centering
    \includegraphics[width=\linewidth]{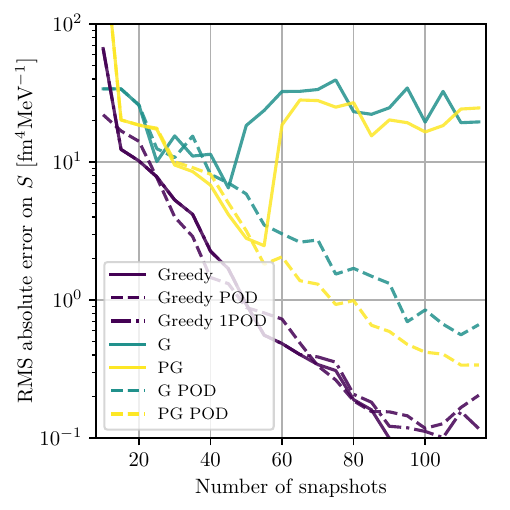}
    \caption{RMS absolute error on the \ce{Mg-24} ISGMR spectrum using $\gamma = 0.25$\,MeV. Different projection methods (Galerkin green, Petrov-Galerkin yellow) are shown using equidistant sampling, with and without POD. Greedy (purple, PG) POD uses an SVD every iteration, while greedy 1POD only uses an SVD at the end.}
    \label{fig:compare_SVD_Mg_rms}
    \end{minipage}
\end{figure}

\subsubsection{Normalisation}
Lastly, one could decide to normalise the basis vectors using $\mathcal{Y}_i = \mathcal{X}_i / || \mathcal{X}_i ||$. This would change the Galerkin projected equations as follows:
\begin{align*}
    \sum_{i=1}^{N_s}\left(-\mathcal{Y}^{\dagger}_j \frac{\mathcal{F}}{||\mathcal{X}_i ||} + (w_i -w)\mathcal{Y}^{\dagger}_j \mathcal{M}\mathcal{Y}_i\right)\alpha_i &= -\mathcal{Y}^{\dagger}_j\mathcal{F}\;.
\end{align*}
Since the first term on the LHS now depends on $\omega_i$, the equations are both more complicated and computationally demanding. While this could improve numerical conditioning of the system matrix, the effect of basis normalisation was only briefly investigated in this thesis and yielded no significant improvements. Moreover, by applying a POD transformation, the basis vectors are already normalised.

\subsection{Conclusions of projections}
We continue the discussion of Fig.~\ref{fig:PG_VS_G_Mg}. We have seen that, for a low amount of snapshots, the Galerkin method introduces peaks that are not in the reference data or incorrect peak locations, while the Petrov-Galerkin method is more conservative, requiring more snapshots to resolve the strength function accurately. Although both methods yield similar RMS errors on the spectrum, as show in Fig.~\ref{fig:compare_SVD_Mg_rms}, the Petrov-Galerkin approach is preferred as it avoids the generation of non-existing peaks. 

As the number of snapshots increases, both methods initially improve, but eventually break down, illustrated by the unphysical strength function in Fig.~\ref{fig:PG_VS_G_Mg} and sudden increase in error in Fig.~\ref{fig:compare_SVD_Mg_rms}. We have identified the cause as numerical ill-conditioning, which was evidenced by the high condition number in Fig.~\ref{fig:compare_SVD_Mg_cond_nr}, and have shown that this can be resolved by applying a proper orthogonal decomposition with a truncation of the reduced model space. As illustrated by the right column of Fig.~\ref{fig:PG_VS_G_Mg}, where the number of snapshots prior to SVD is mentioned in parenthesis, both stabilized methods converge towards the reference spectrum as expected.

Henceforth we will use the Petrov-Galerkin projection and apply an SVD at the end. We will continue the discussion after introducing different snapshot selection methods.

\clearpage
\section{Methods for snapshot selection}\label{sec:methods-for-snapshot-selection}

In FAM applied to (Q)RPA, the strength function is typically computed using a fixed smearing parameter. In this work, we develop a ROM emulator to reconstruct the same strength function while requiring fewer FAM calculations. We treat this strength function, at the target smearing parameter $\gamma_t$, as our target observable.

As a first step, we try different sampling methods choosing ROM snapshots at the target smearing. We work out the procedure of iterative snapshot selection. Subsequently, we investigate whether a ROM based on snapshots at the target smearing can predict the strength function at different smearing values. Since this evolves evaluating the model outside the parameter domain spanned by the ROM snapshots, we refer to this as extrapolation.  

To streamline the analysis, the iterative methods in this work sample exclusively from a dense, precomputed reference spectrum. While we provide a general implementation~\cite{code_github} that is capable of interfacing directly with the (Q)FAM solver to generate new snapshots on-the-fly, this constrained search significantly improves computational efficiency during the testing phase of the methodology. It should be noted that this approach introduces a conservative bias. Without the limitations of a discrete frequency grid, such approach would yield results that are equal to or superior to those presented here.

\subsection{Equidistant frequency sampling}
The equidistant frequency sampling is a uniform sampling scheme to select snapshot frequencies  within the domain $[0,\,\omega_\mathrm{max}]$. For $N_s$ snapshots, the frequency spacing is defined as \[\Delta\omega=\frac{\omega_\mathrm{max}}{N_s-1}\;.\] The resulting set of snapshot frequencies is given by $\{\omega_i\}=\{i\Delta w\}_{1\leq i\leq N_s}$. This ensures that the $N_s$ snapshots are evenly distributed across the specified range.

\subsection{Greedy frequency sampling}\label{subsubsec:greedy}
Following~\cite[Sec. 3.2.2]{Hesthaven2016} and~\cite{Maldonado2025}, we implement a greedy strategy for basis vector acquisition. Unlike the equidistant method discussed in the previous section, which generates the basis in one go, the greedy approach constructs the snapshot basis iteratively. This often produces a more compact basis, achieving convergence with fewer iterations. While the computational cost of these additional calculations is usually negligible compared to the FAM calculations, iterative samplers cannot be parallelised\footnote{It can, however, be parallelised in the sense that different emulators can be constructed at the same time, e.g.~for different nuclei and/or operators. This is often referred to as `embarrassingly parallel'.}. 

For the greedy algorithm, the availability of an error estimator to represent the loss induced by the model order reduction is crucial~\cite{Hesthaven2016}.  For practical implementation, the search space is generally limited to a finite set, e.g.~$\omega \in \mathbf{W}= \{\omega_i + \gamma_t |\, \omega_\mathrm{min} \leq \omega_i \leq \omega_\mathrm{max} \}$ with $\omega_i, \omega_\mathrm{min}, \omega_\mathrm{max} \in \mathbb{R}$, as opposed to an exhaustive search over the entire continuous domain. We adapt the general working algorithm from~\cite[Alg.: The greedy algorithm]{Hesthaven2016} slightly such that it can be applied to our case, i.e.~using the projected FAM equation and adding an intermediate FAM calculation for the new snapshot, as shown in Alg.~\ref{al:greedy}. 

\begin{algorithm}[H]
\SetKwInOut{Input}{Input}
\caption{Greedy basis method~\cite{Hesthaven2016}}\label{al:greedy}
\Input{Target smearing $\gamma_t$, range of $\omega$, max number of snapshots $N_\mathrm{max}$, initial number of snapshots $N_\mathrm{snap}$}
\BlankLine
Define $\mathbf{W}$ such that $\Im(\mathbf{W}) = \gamma_t$\;
Initialize ROM basis $\{\mathcal{X}_i\}$ for $i\leq N_\mathrm{snap}$\;
\BlankLine
\While{$N_\mathrm{snap} < N_\mathrm{max}$}{
\ForEach{$\omega \in \mathbf{W}$}{
    Solve projected FAM equation for $\alpha(\omega)$\;
    Evaluate error estimator $\eta(\omega)$\;
}
\BlankLine
Find $\omega_\mathrm{new}= \argmax_{\omega \in \mathbf{W}} \eta(\omega)$\;

Do FAM calculation to obtain $\mathcal{X}(\omega_\mathrm{new})$\;
Set $\{\mathcal{X}_i\}_{i\leq N_\mathrm{snap}} \mathrel{{+}{=}} \mathcal{X}(\omega_\mathrm{new})$ and $N_\mathrm{snap} \mathrel{{+}{=}} 1$
}
\Return ROM basis $\{\mathcal{X}_i\}_{i\leq N_\mathrm{max}}$
\end{algorithm}

In the next subsections, we discuss and compare some of the options for the error estimator $\eta$. The search space $\mathbf{W}$ is kept constant (at the target smearing), but variations of search space $\mathbf{W}$ are considered in Subsec.~\ref{subsubsec:greedy2D}.

\subsubsection{Greedy (naive)}
A simple example could be to start with a small basis (e.g.~2 vectors) and approximate the strength function $S(\omega) \approx \tilde{S}(\omega)$. A possible next sample is chosen at peaks of $\tilde{S}$, while ensuring the new samples are sufficiently far away (e.g.~more than a threshold frequency $d\omega$) from the ones already included in the basis. We define the error estimator $\eta(\omega)$ by:
\begin{align}
\eta(\omega) =
    \begin{cases}
        \tilde{S}(\omega)\quad \mathrm{for} \;\omega \notin \bigcup_{i\in \{0\ldots N_s\}} [\omega_i - d\omega, \omega_i + d\omega ]  \\
        0 \quad \mathrm{elsewhere\;.}\\
    \end{cases}\label{eq:cost_naive_greedy}
\end{align}
This requires solving the linear system (denoted $A\mathcal{\alpha} = b$ earlier) every time a new snapshot needs to be selected instead of once at the end. The size of the linear system increases by one every iteration. We additionally constrain the error estimator such that only snapshots near the peak of the emulated spectra can be selected. 

\begin{figure}
    \centering
    \includegraphics[width=0.45\linewidth]{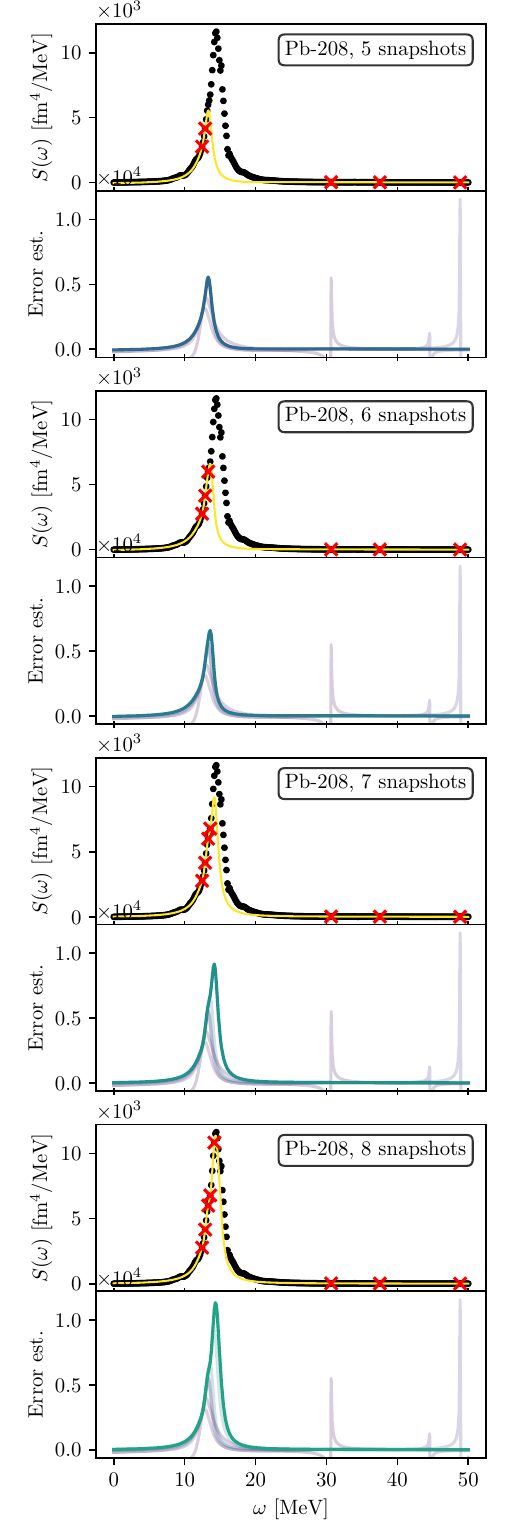}~
    \includegraphics[width=0.45\linewidth]{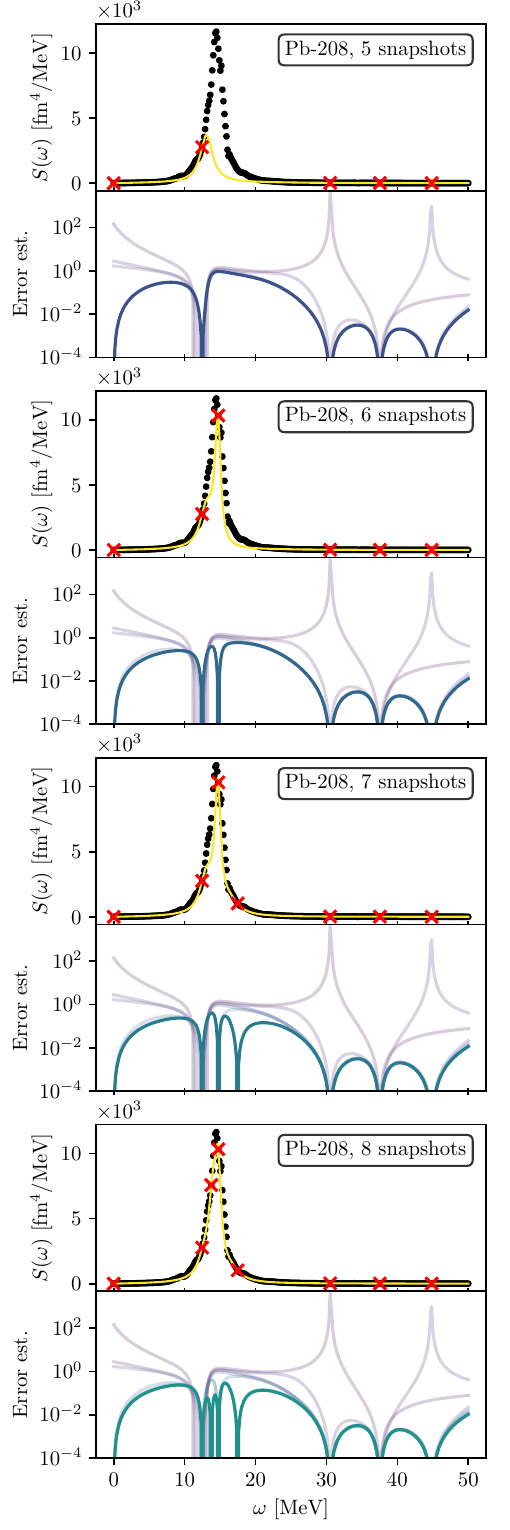}
    \caption{(left) Illustration of the naive greedy sampling Eq.~\eqref{eq:cost_naive_greedy}, which uses the emulator's output to find the maximum of the emulated strength to select the next snapshot. Initial snapshots are chosen at 25$\%$ and 75$\%$ of the domain. The error estimator is in units of fm$^4$\,MeV$^{-1}$. (right) Equivalent simulation but using the greedy sampler from Eq.~\eqref{eq:cost_greedy}. Reference spectrum (black) is from \ce{Pb-208} with $\gamma = $0.5\,MeV. The semi-transparent curves represent the error estimator of previous iterations.}
    \label{fig:greedy_naive}
\end{figure}

\begin{figure}
    \centering
    \includegraphics[width=0.45\linewidth]{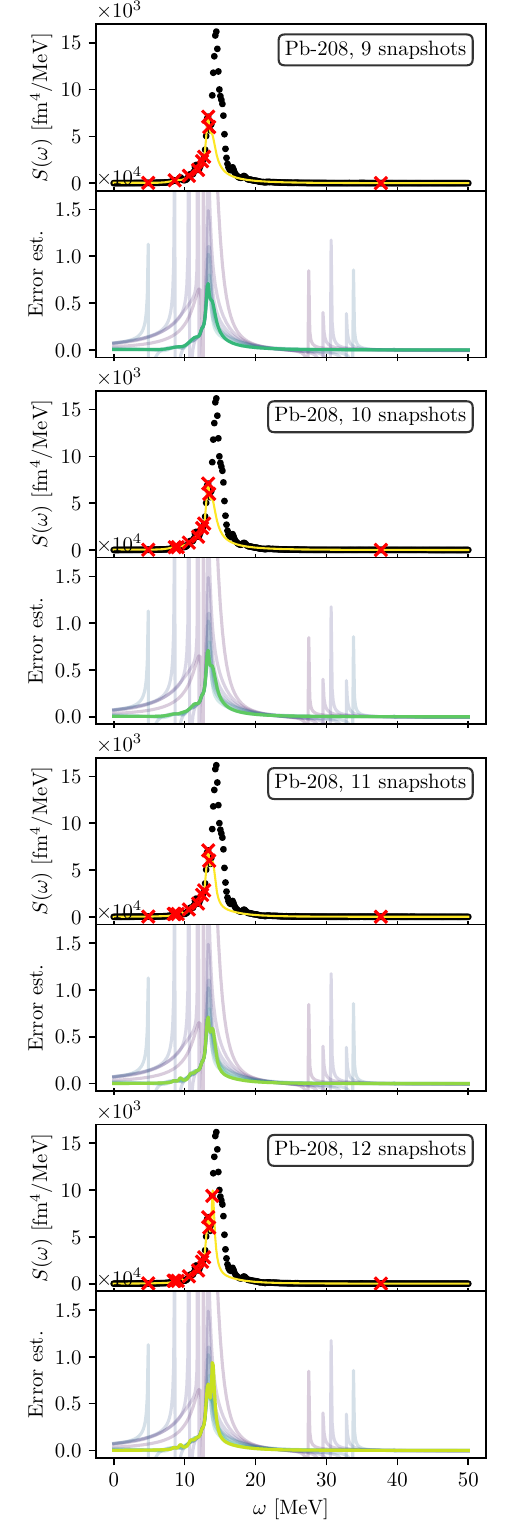}~
    \includegraphics[width=0.45\linewidth]{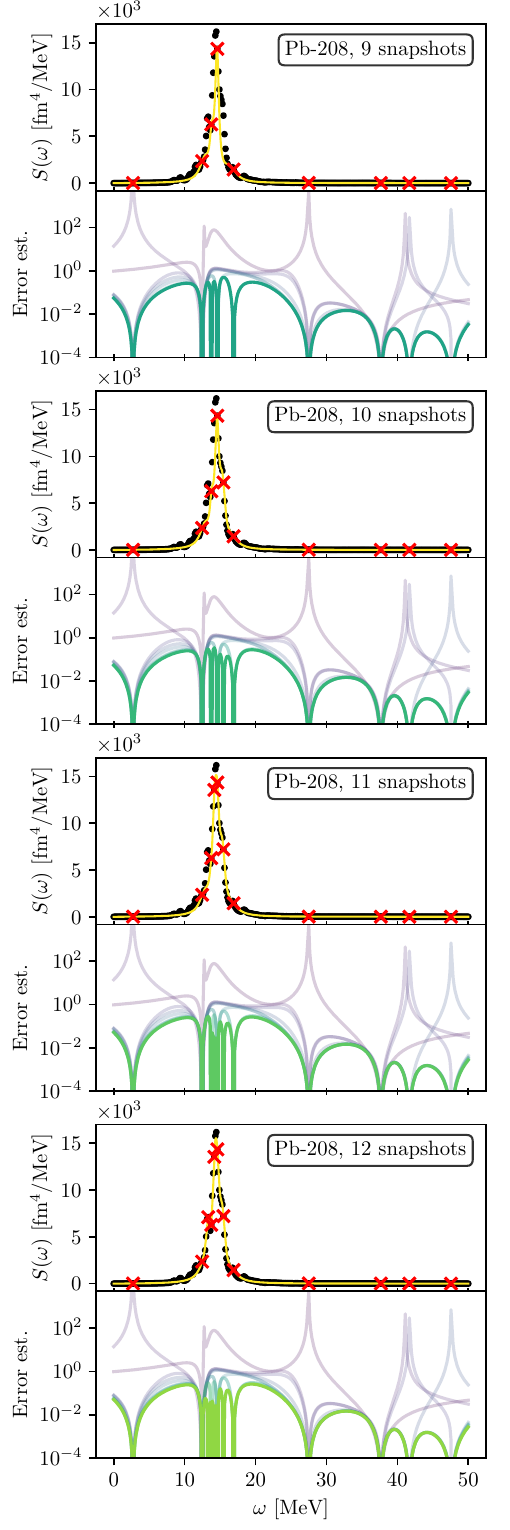}
    \caption{Same figure as in Fig.~\ref{fig:greedy_naive}, but with smearing $\gamma = $0.25\,MeV.}
    \label{fig:greedy_naive_0.25}
\end{figure}

This method is illustrated for \ce{Pb-208} in the left columns of Figs.~\ref{fig:greedy_naive} and~\ref{fig:greedy_naive_0.25} and we will discuss some drawbacks. The method often requires many snapshots close to the peak before determining the correct position. This can be helped by choosing a larger threshold, forcing the snapshots to be reasonably far apart. Note, however, that a large threshold value may prevent the method from resolving smaller peaks completely. When the spectrum exhibits multiple peaks, the method may require many samples near one peak to either start predicting a second peak or fail to predict a second peak completely (not shown here). 

\begin{figure}
    \centering
    \includegraphics[width=0.5\linewidth]{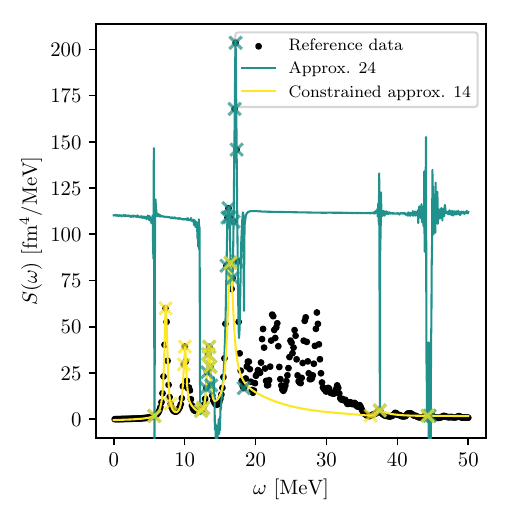}
    \caption{Comparing the naive greedy method with and without the 'peak' constraint. See text for details. }
    \label{fig:explosion}
\end{figure}

\begin{figure}
    \centering
    \includegraphics[width=0.5\linewidth]{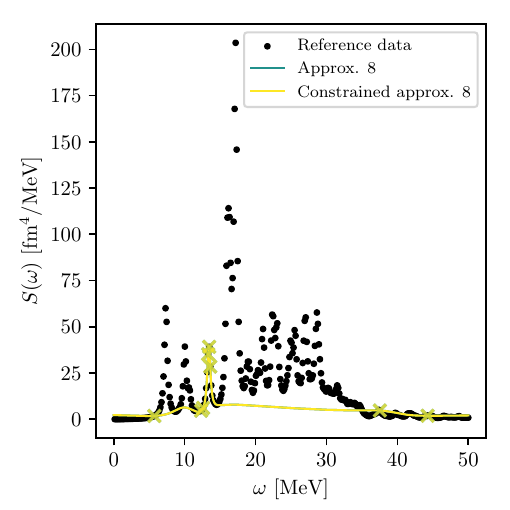}~
    \includegraphics[width=0.5\linewidth]{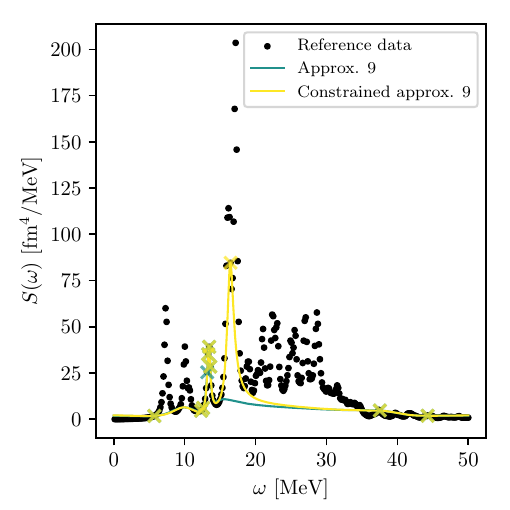}
    \caption{Illustration of the naive greedy method with and without constraint. When going from 8 to 9 snapshots, the unconstrained method (green) takes a snapshot at the maximum of the strength  (ignoring the neighbourhood of other snapshots), while the constrained method (yellow) searches for peaks. The first peak around 13\,MeV is checked, is not selected as it is in the neighbourhood of another snapshot. The new snapshot is selected at the second highest peak (17\,MeV).}
    \label{fig:before_explosion}
\end{figure}

Removing the requirement that new snapshots be located near the peaks of the emulated spectra may cause numerical instabilities, as illustrated in Fig.~\ref{fig:explosion}. The constrained method stops once no more peaks are detected; in this case, it yields a maximum number of snapshots of 14 and fails to predict peaks in the $[20-30]$\;MeV range. In contrast, while the unconstrained method provides a similar approximation for a small number of snapshots as shown in Fig.~\ref{fig:before_explosion}, it eventually adds snapshots causing the system to become ill-conditioned and fails to produce accurate strength functions.

We conclude that this method will not work in general due to the flaws mentioned above, but may be considered for predicting peak positions of giant resonances where only one peak is expected. In the next section, we propose a different error estimator which yields a significantly better method.

\subsubsection{Greedy}
A second alternative is to look at the residuals, as defined in Eq.~\eqref{eq:residual_FAM}. Every iteration, the norm of the residual needs to be calculated for every $\omega$, and the next snapshot is chosen according to Alg.~\ref{al:greedy}. More precisely,
\begin{align*}
   ||r(\omega)||^2 &=  \sum_{i,j=0}^{N_s} \Big( \alpha_i^*(\omega) \mathcal{X}^\dagger_i\mathcal{A}^\dagger(\omega)+ \mathcal{F}^\dagger\Big)\Big(\mathcal{A}(\omega) \alpha_j(\omega) \mathcal{X}_j + \mathcal{F}\Big) \\
    &= \sum_{i,j=0}^{N_s} \Big( \alpha_i^*(\omega) (-\mathcal{F}^\dagger + (\omega_i^* - \omega^*)\mathcal{X}^\dagger_i\mathcal{M}^\dagger)+ \mathcal{F}^\dagger\Big)\\ &\qquad \cdot \Big(\alpha_j(\omega)(-\mathcal{F} + (\omega_j - \omega)\mathcal{M}\mathcal{X}_j) + \mathcal{F}\Big) \;,
\end{align*}
where we have used Eq.~\eqref{eq:FAM_substitution} to simplify $\mathcal{A}$. We note that the physical units of $||r(\omega)||^2$ are the same as $[\mathcal{F}]^2$ (for instance, fm$^4$ for monopole transitions). To ensure that the error estimator is generally interpretable and independent of the transition type, it is preferable to have a dimensionless quantity. Therefore, we normalise it using the squared norm of $\mathcal{F}$\footnote{This also removes the $A$-dependence from the error estimate, making it more suitable as a convergence criteria which can be used for all nuclei.}:
\begin{equation}
 \eta(\omega) =    ||r(\omega)||^2  / ||\mathcal{F}||^2 \;. \label{eq:cost_greedy}
\end{equation}
 The previous expression can be evaluated for all $\omega \in \mathbf{W}$, given that the coefficients $\alpha$ are known. Hence, for every iteration, we need to solve the projected FAM equation for all $\omega \in \mathbf{W}$, and evaluate the expression above. Although the matrix product containing $\mathcal{X}$ changes every iteration, evaluation of $\eta$ in $\omega \in \mathbf{W}$ is fast since we can precompute the matrix products once each iteration. This yields an additional computational cost compared to the equidistant sampling, but this is negligible compared to running even a single extra FAM calculation. Consequently, this greedy search is advantageous if it yields the same accuracy with fewer FAM evaluations. 

\begin{figure}
    \centering
    \includegraphics[width=0.45\linewidth]{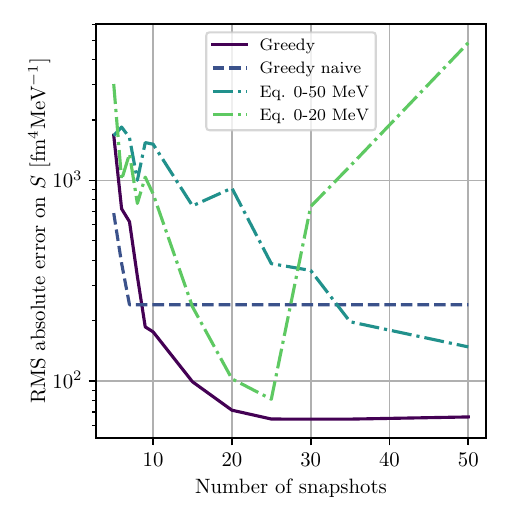}~
    \includegraphics[width=0.45\linewidth]{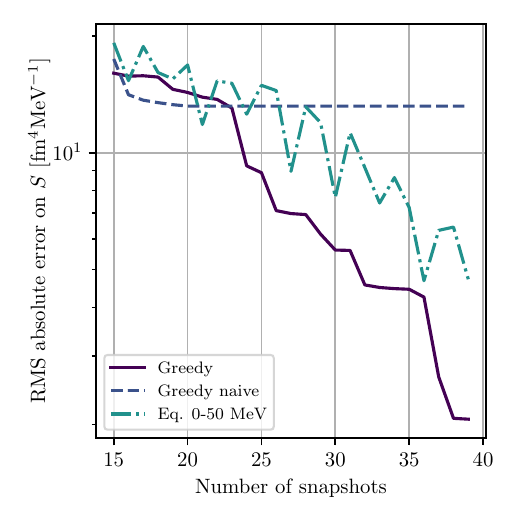}
    \caption{Comparing the performance of the greedy (and naive greedy) with the equidistant sampling method, on \ce{Pb-208} with $\gamma = 0.5$\;MeV (left) and \ce{Mg-24} with $\gamma=0.25$\;MeV (right). All methods are applied to the frequency domain $[0, 50]$\;MeV, except for the curve in green, which is applied to $[0,20]$\;MeV.}
    \label{fig:pLaCeHoLdEr}
\end{figure}

Figure~\ref{fig:pLaCeHoLdEr} illustrates the RMS error comparison between the greedy and equidistant samplers. Notably, restricting the equidistant sampling to a narrower domain ([0-20]\,MeV instead of [0-50]\,MeV) accelerates convergence for up to 25 snapshots, after which numerical instabilities arise. While this is effective for \ce{Pb-208}, it lacks generalizability as it requires a priori knowledge of the resonance peak's location. Furthermore, the staggered pattern observed in the RMS error curve is an artifact of the equidistant sampling, due to the global shifting of the snapshot positions as the sample size increases. Lastly, while the naive greedy performs slightly better than the greedy sampler for the initial snapshots, its accuracy is limited due to the necessary `peak' constraint discussed earlier. 

In addition,~\cite{Hesthaven2016} suggests to orthonormalise the basis vectors to increase numerical stability. This can be achieved by applying a POD as discussed in Sec.~\ref{subsec:basis-manipulations}. Doing so, one needs to update the error estimator from Eq.~\eqref{eq:cost_greedy} as follows:
\begin{align*}
   ||r(\omega)||^2 &=  \sum_{i,j=0}^{N_r} \Big( \beta_i^*(\omega) \mathcal{U}^\dagger_i \mathcal{A}^\dagger(\omega)+ \mathcal{F}^\dagger\Big)\Big(\mathcal{A}(\omega) \beta_j(\omega) \mathcal{U}_j + \mathcal{F}\Big) \\
    &=  \sum_{i,j=0}^{N_r} \sum_{k,l=0}^{N_s} \Big( \beta_i^*(\omega) U^\dagger_{ik}\mathcal{X}^\dagger_k \mathcal{A}^\dagger(\omega)+ \mathcal{F}^\dagger\Big)\Big(\mathcal{A}(\omega) \beta_j(\omega) \mathcal{X}_lU_{lj} + \mathcal{F}\Big) \\
    &=\sum_{i,j=0}^{N_r} \sum_{k,l=0}^{N_s}  \Big( \beta_i^*(\omega) (-U_{ik}^\dagger\mathcal{F}^\dagger + (\omega_k^* - \omega^*)U_{ik}^\dagger\mathcal{X}^\dagger_k\mathcal{M}^\dagger)+ \mathcal{F}^\dagger\Big)\\ &\qquad \cdot \Big(\beta_j(\omega)(-\mathcal{F}U_{lj} + (\omega_l - \omega)\mathcal{M}\mathcal{X}_lU_{lj}) + \mathcal{F}\Big) \;,
\end{align*}

There are two problems with this method. One is that the greedy method is no longer robust against selecting the same snapshot twice, when applying a POD with a nonzero cutoff value. This typically occurs once a significant number of snapshots have been incorporated into the basis such that the estimator exhibits a flat profile, yielding no clear choice for the next snapshot. This however prevents the model from further converging to lower RMS values. One could circumvent this by dynamically lowering the cutoff or removing the snapshot from $\mathbf{W}$. Either way, this approach is not elegant.

A second drawback is the (non-)scalability of this method. While the greedy search without POD requires fast evaluation of the error estimator every iteration, this version demands an SVD-decomposition of the snapshot matrix at every iteration! We stress that these additional operations can become noticeable due to diagonalization of large matrices and memory constraints, especially on smaller machines (like a laptop), yielding it potentially infeasible for modelling heavy nuclei (e.g.~\ce{Pb-208}) and/or with many basis vectors. 

The accuracy of the method can be examined and compared to those from Sec.~\ref{sec:methods-for-reduced-basis-projections}. As illustrated by Fig.~\ref{fig:compare_SVD_Mg_rms}, the greedy method without POD reaches an absolute RMS error below 0.1\;fm$^4$\,MeV$^{-1}$ using 90 snapshots, already one order of magnitude lower compared to the equidistant sampling using POD reaching 1\;fm$^4$\,MeV$^{-1}$.  Note that the numerical instability does not yet occur for the greedy method. Hence, when the SVD is applied in addition to the greedy method, the error seems to stagnate just above the 0.1 mark, so within this range of snapshots the SVD does not yield additional benefits. The error can be lowered by decreasing the threshold value. Note that the error can only be lower than those of the full model when numerical instabilities arise. 

Based on the analysis in this section, we decide to apply the greedy search without SVD and, optionally, only apply it once at the end to avoid possible numerical issues. 
 
\newpage

\subsection{Effect of the smearing parameter}
\begin{minipage}{0.5\linewidth}
\subsubsection{Performance at different smearing}\label{sec:performance-at-different-smearing}
Fig.~\ref{fig:RMS_different_smearing} compares the performance of the emulator at different smearing parameter values, without extrapolation, for \ce{Mg-24}. This means that the snapshots are taken at the target smearing. Similarly, we have added a simple spline interpolation method to compare to and see how the smearing parameter can influence the performance. 
\\
\\
First, we note that the RMS error decreases with an increasing number of snapshots. For the spline interpolation, this means that there are more interpolation points in the strength spectra. Next, we note that the RMS error is generally lower for larger smearing, which can be expected intuitively since this corresponds to a smoother curve in the strength spectra. The ROM model outperforms a simple spline interpolation. Note that, given that the number of snapshots is sufficiently high, the ROM accuracy compared to the interpolation can vary from $\times 100$ higher for large smearing and $\times 10$ higher for low smearing. 
\end{minipage}
\hfill
\begin{minipage}{0.45\linewidth}
    \centering
    \includegraphics[width=\linewidth]{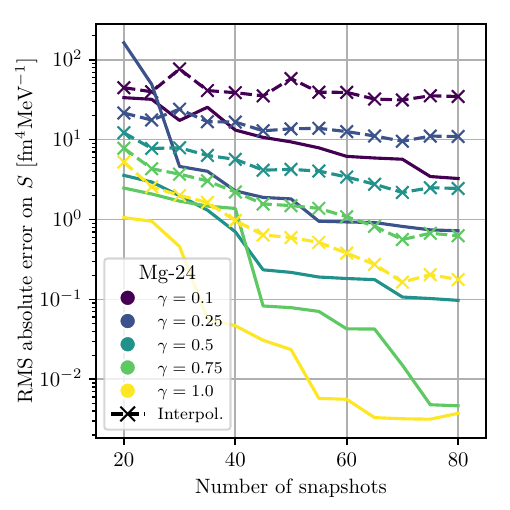}
    \captionof{figure}{RMS error for different smearing as function of number of snapshots. We compare the ROM model using a greedy sampling method with a spectrum obtained by cubic interpolation between snapshot points, i.e.~without using the ROM framework. }
    \label{fig:RMS_different_smearing}
\end{minipage}

\subsubsection{Performance by extrapolation}
\begin{figure}
    \centering
    \includegraphics[width=0.45\linewidth]{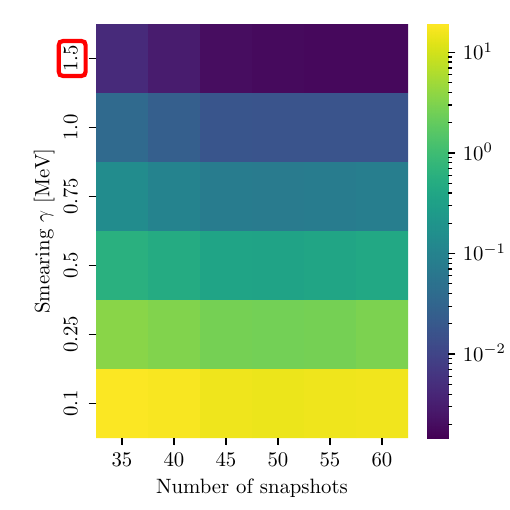}~\includegraphics[width=0.45\linewidth]{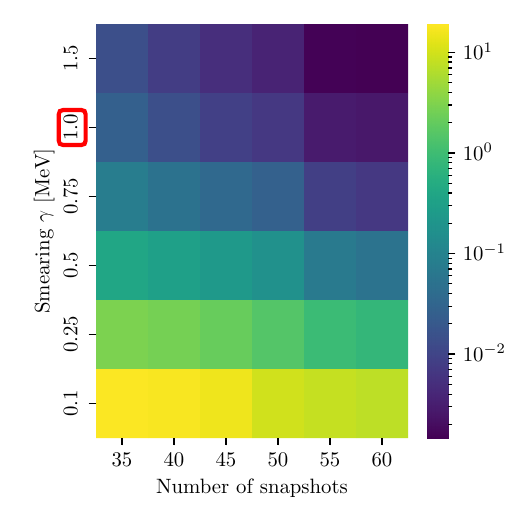}

    \includegraphics[width=0.45\linewidth]{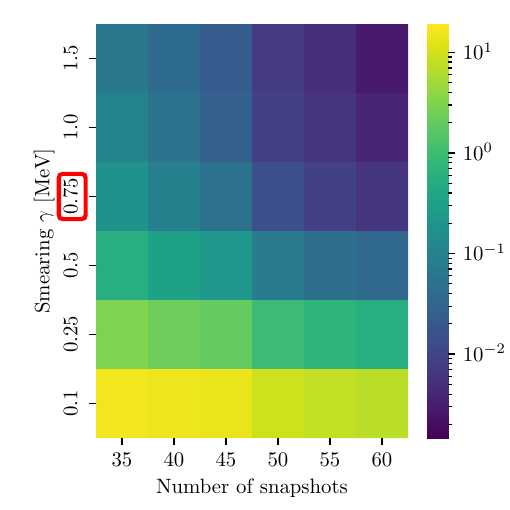}~\includegraphics[width=0.45\linewidth]{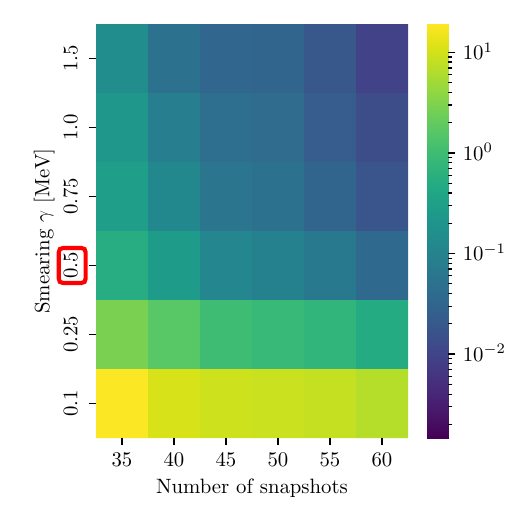}

    \includegraphics[width=0.45\linewidth]{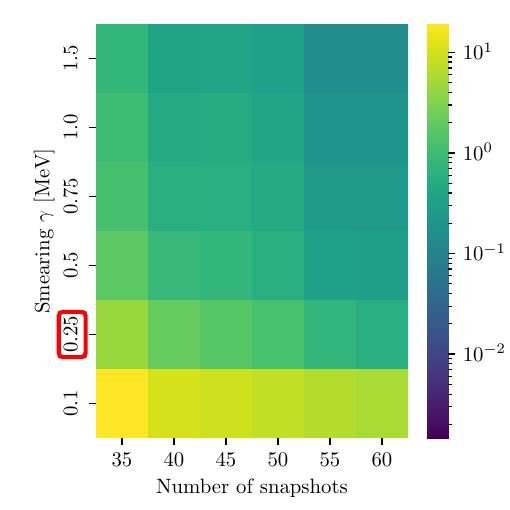}~\includegraphics[width=0.45\linewidth]{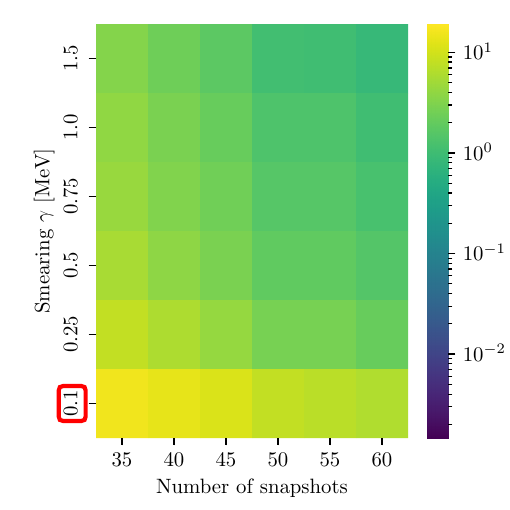}
    \caption{RMS absolute error visualised in a heatmap  (Eq.~\eqref{eq:rms_absolute_error}) to show the emulator's performance when extrapolating. A greedy sampler is used and this is for \ce{Mg-24}. The y-axis represents the target smearing, and the red indication shows at what smearing the snapshots are selected. }
    \label{fig:extrapolation_MG}
\end{figure}

Next, we evaluate the emulator's performance when extrapolating across different smearing, as summarized in Fig.~\ref{fig:extrapolation_MG} for \ce{Mg-24}. Consistent with Sec.~\ref{sec:performance-at-different-smearing}, when looking at the smearing marked in red, the RMS error decreases with an increasing number of snapshots. When comparing the different figures, the RMS error is lower for larger smearing. Looking at an individual plot, we see the same trend for extrapolation, namely that towards larger smearing values a lower RMS error is obtained, while towards smaller smearing values the RMS error increases. 

However, one can observe an asymmetry. Extrapolating from a small smearing ($\gamma = 0.1$\,MeV) to a larger one ($1.0$\,MeV) results in a high RMS error. Conversely, an emulator with snapshots at $\gamma=1.0$\,MeV can predict the spectrum at $\gamma=0.1$\,MeV with an error of magnitude comparable to an emulator trained specifically at that lower value. This suggests that snapshots taken at higher smearing values capture more global information. Intuitively, broader peaks allow a single snapshot to resolve features across a wider frequency range.  

Remarkably, the emulator correctly predicts that strength function peaks narrow as $\gamma$ decreases. While such extrapolation is mathematically possible, its effectiveness is not a trivial observation. This raises the question whether it could be beneficial for the emulator to include snapshots at different smearing.

\clearpage
\subsection{Greedy pseudo-2D sampling}\label{subsubsec:greedy2D}
\begin{wrapfigure}[20]{r}{0.45\linewidth}
    \vspace{-3.5em}
    \centering
    \includegraphics[width=\linewidth]{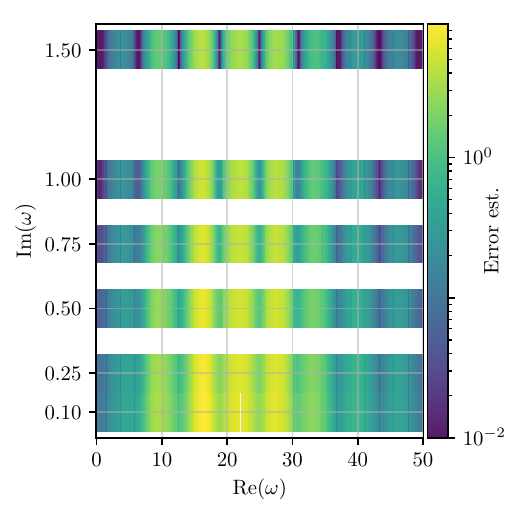}
    \caption{Error estimate after initialisation with 11 vectors at $\gamma=1.5$\,MeV. Note that the error estimate is higher for smaller smearing, which seems to be generally true (although not shown here). As a result, the next snapshots will inevitably be selected at $\gamma=0.1$\,MeV.}
    \label{fig:error_est_example_heatmap_full_spectrum}
\end{wrapfigure}

The greedy method can be extended to search across the complex plane by discretizing the parameter space $\mathbf{W}$ as a 2D grid, allowing for more variation of the model space. However, this approach introduces two problems (selection and optimality). As illustrated in Fig.~\ref{fig:error_est_example_heatmap_full_spectrum}, the error estimator tends to yield higher values for lower smearing. Consequently, a straightforward 2D greedy implementation typically selects snapshots at the lowest smearing levels, which are often the most computationally demanding. Furthermore, these snapshots do not necessarily ensure the most rapid convergence. Because the error estimator only predicts the inaccuracy of $\mathcal{X}$ at a given $\omega$, it does not guarantee that adding the corresponding true snapshot to the basis is the optimal choice. One could propose to use a different error estimator.

We discuss a few options to address the aforementioned problems. A first, more straightforward approach, is to simply add a penalty to the error estimator to take into account computational cost indirectly. Specifically, the error estimator can be weighted by a smearing-factor $\gamma^k$ with $k$ a free parameter. However, choosing $k$ to counterweight the actual error estimator (such that it is not only determined by the $\gamma$ dependence), proved to be difficult and not universal. Additionally, this approach could lead to overcorrection; namely leading to the selection of snapshots at excessively high smearing, which is also undesirable. This approach was not further investigated.

Instead of changing the error estimator, we can also address the second issue by selecting sub-optimal snapshots. We therefore propose a 2-step greedy method (or pseudo-2D greedy). 

\begin{figure}[hb]
    \centering
    \definecolor{yellow}{RGB}{253, 231, 37}
\definecolor{green}{RGB}{94, 201, 98}
\definecolor{bluegreen}{RGB}{33,145,140}
\definecolor{blue}{RGB}{59,82,139}
\definecolor{purpur}{RGB}{68,1,84}

\begin{tikzpicture}[>=Stealth, scale=0.8]
    \tikzstyle{snapshot} = [purpur, draw, cross out, minimum size=6pt, inner sep=0pt, thick, ]
    \tikzstyle{newpoint} = [circle, fill=purpur, inner sep=1.5pt]
    \tikzstyle{region} = [green, very thick, {Bar[width=4pt]}-{Bar[width=4pt]}]

    \def\xr{2.5}
    \draw[thick]
plot[smooth, tension=0.8] coordinates {
(0.55,3.0)
(0.6,1.8)
(0.7,1.3)    
(0.9,1.8)
(1.2,2.8)
(1.5,3.3)
(1.8,2.5)
(1.95,2.0)
(2.1,1.4)    
(2.2,2.0)
(2.3,3.5)
(2.4,4.2)
(2.5,4.6)    
(2.6,4.1)
(2.7,3.4)
(3.0,2.9)
(3.3,2.6)
(3.6,2.8)
(3.8,1.6)
(3.9,1.35)   
(4.05,1.9)
(4.15,2.4)
(4.3,2.8)
(4.8,3.2)
(5.0,3.3)
};
    
    \draw[region] (0.5, 0) -- (5.1, 0) node[below left] {$\mathbf{W}$};
    \draw[->, thick] (0, -0.25) -- (0, 5.1) node[below left] {$\eta(\omega)$};
    \draw[white] (-0.25, 0) -- (-0.1, 0) node[] {}; 
    \node[snapshot, label={[text=purpur]above:$\omega_1$}] at (0.7, 0) {};
    \node[snapshot, label={[text=purpur]above:$\omega_2$}] at (2.1, 0) {};
    \node[snapshot, label={[text=purpur]above:$\omega_3$}] at (3.9, 0) {};

    \node[snapshot] (wnext) at (\xr, 4.6) {};
    \node[right, purpur] at (wnext) {$\omega_\mathrm{new}$};

\end{tikzpicture}
    ~
    \definecolor{yellow}{RGB}{253, 231, 37}
\definecolor{green}{RGB}{94, 201, 98}
\definecolor{bluegreen}{RGB}{33,145,140}
\definecolor{blue}{RGB}{59,82,139}
\definecolor{purpur}{RGB}{68,1,84}

\begin{tikzpicture}[>=Stealth, scale=0.8]
    \tikzstyle{snapshot} = [purpur, draw, cross out, minimum size=6pt, inner sep=0pt, thick, ]
    \tikzstyle{newpoint} = [circle, fill=purpur, inner sep=1.5pt]
    \tikzstyle{region} = [green, very thick, {Bar[width=4pt]}-{Bar[width=4pt]}]

    \def\xr{2.5}

    \draw[->, thick] (-0.25, 0) -- (5.1, 0) node[below left] {$\Re(\omega)$};
    \draw[->, thick] (0, -0.25) -- (0, 5.1) node[below left] {$\Im(\omega)$};

    \draw[purpur] (0.0, 4) node[left, black] {$10\,\gamma_t$} -- (4.8, 4) ;
    \draw[purpur] (0.0, 1) node[left, black] {$\gamma_t$}-- (4.8, 1) ;

    \draw[region] (0.25, 1) -- (3.8, 1);
    \node[green, above right] at (0.25, 1) {$\mathbf{W}_r$};
    
    \draw[region] (\xr, 1) -- (\xr, 4.6);
    \node[green, right] at (\xr, 3.5) {$\mathbf{W}_i$};

    \node[snapshot, label={[text=purpur]above:$\omega_1$}] at (0.5, 4) {};
    \node[snapshot, label={[text=purpur]above:$\omega_2$}] at (2.1, 4) {};
    \node[snapshot, label={[text=purpur]above:$\omega_3$}] at (3.7, 4) {};

    \node[newpoint] (wr) at (\xr, 1) {};
    \node[below, purpur] at (wr) {$\omega_r+i\gamma_t$};

    \node[snapshot] (wnext) at (\xr, 2.5) {};
    \node[right, purpur] at (wnext) {$\omega_\mathrm{new}$};

\end{tikzpicture}
    \caption{(left) Simplified illustration of the greedy method. Given a basis $\{\mathcal{X}(\omega_i)\}_{i=1,2,3}$, the error estimator is evaluated on the set $\mathbf{W}$. The next basis vector $\mathcal{X}(\omega_\mathrm{new})$ is calculated where the error estimate is maximal. (right) Simplified illustration of the pseudo-2D greedy method. Given a basis $\{\mathcal{X}(\omega_i)\}_{i=1,2,3}$, the error estimator is evaluated on $\mathbf{W}_r$. Using $\omega$ with the maximum error, a new set $\mathbf{W}_i$ is constructed. The next snapshot location is found by evaluating the error estimator on $\mathbf{W}_i$ with a threshold. }
    \label{fig:greedy2D_illustration}
\end{figure}
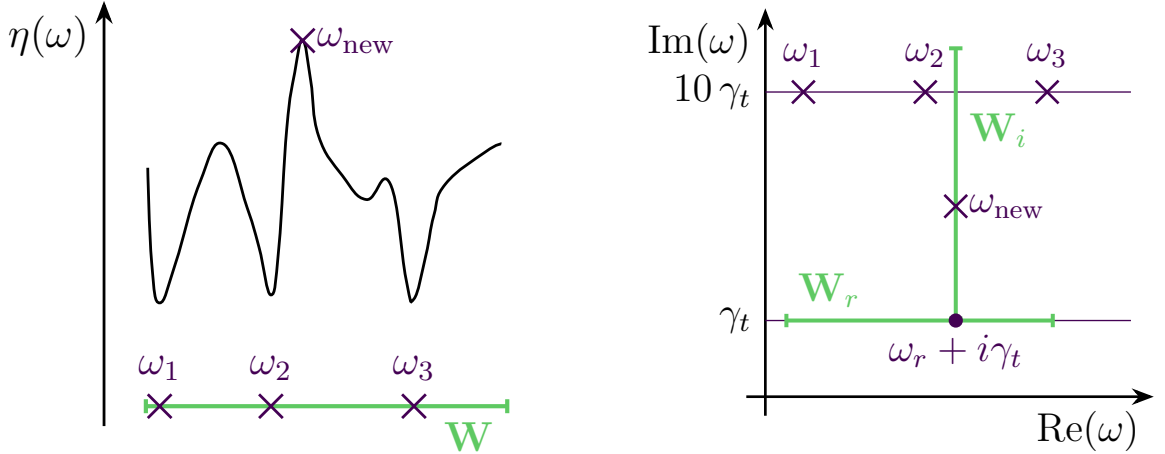

\clearpage
\subsubsection{Method}

The proposed pseudo-2D greedy method is explained below, and the pseudo-code can be found in Alg.~\ref{al:greedy2D}.

We first define a target smearing value $\gamma_t$, representing the intended smearing for the emulator. In most applications, $\gamma_t$ is small, corresponding to a high resolution in the strength spectrum. The ROM basis is initialised using a small number of snapshots at a larger smearing, e.g.~$\{\mathcal{X}_1, \mathcal{X}_2, \mathcal{X}_3\}$ for frequencies at $\gamma = 10\gamma_t$ as shown in Fig.~\ref{fig:greedy2D_illustration}. Using these snapshots to form a ROM basis, the error estimator $\eta(\omega)$ is evaluated on a predefined set of frequencies with smearing $\gamma_t$, called $\mathbf{W}_r$. We now identify the frequency $\omega_r+i\gamma_t$ where the error estimator is maximal. This corresponds to the frequency at which the current ROM basis fails to capture the strength. This is effectively using the exact same approach as for the 1D greedy in Sec.~\ref{subsubsec:greedy} except for initialisation. 

Next, instead of immediately adding snapshots at $\omega_r+i\gamma_t$, we prefer to select a snapshot at $\omega_r + i\gamma$ where $\gamma > \gamma_t$, on the condition that this snapshot still captures most of the same features as the snapshot at smearing $\gamma_t$. This means that we should calculate both snapshots and compare the performance of the extended ROM basis where one has the optimal snapshot and the other has the suboptimal snapshot. This, of course, defeats the purpose since we want to avoid the snapshot calculation. 

In order quantify the performance of the addition of a suboptimal snapshot, the error estimator is used again to approximate this. Defining a new set of frequencies $\mathbf{W}_i$, where only the complex part is varied and thus $\forall \omega \in \mathbf{W}_i: \mathrm{Re}(\omega) = \omega_r$ as mentioned earlier. The error estimate $\eta(\omega)$ along $\mathbf{W}_i$ typically shows a descending curve, as presented by Fig.~\ref{fig:H_scan}. We claim that the steeper the curve, the faster important information is lost when sampling at larger smearing. We therefore define a threshold $\tau$ to select the next snapshot. The choice for this threshold is analysed next.  

\begin{algorithm}[H]
\SetKwInOut{Input}{Input}
\caption{Greedy Basis Method: pseudo-2D}\label{al:greedy2D}
\Input{Target smearing $\gamma_t$, threshold $\tau$, range of $\omega$, max number of snapshots $N_\mathrm{max}$, initial number of snapshots $N_\mathrm{snap}$}
\BlankLine
Define $\mathbf{W}_r$ such that $\Im(\mathbf{W}_r) = \gamma_t$\;
Initialize ROM basis $\{\mathcal{X}_i\}$ for $i\leq N_\mathrm{snap}$ where $\Im(\omega_i)=10\gamma_t$ \;
\BlankLine
\While{$N_\mathrm{snap} < N_\mathrm{max}$}{
\ForEach{$\omega \in \mathbf{W}_r$}{
    Solve projected FAM equation for $\alpha(\omega)$\;
    Evaluate error estimator $\eta(\omega)$\;
}
\BlankLine
Find $\omega_r = \Re\left( \argmax_{\omega \in \mathbf{W}_r} \eta(\omega) \right)$\;
Set $\mathbf{W}_i$ such that $\Re(\mathbf{W}_i) = \omega_r$\;
\BlankLine
\ForEach{$\omega \in \mathbf{W}_i$}{
    Solve projected FAM equation for $\alpha(\omega)$\;
    Evaluate error estimator $\eta(\omega)$\;
    \If{$\eta(\omega) > \tau \eta(\omega_r)$}{
        Set $\eta(\omega) = 0$
    }
}
Do FAM calculation to obtain $\mathcal{X}(\omega_\mathrm{new})$ at $\omega_\mathrm{new}= \argmax_{\omega \in \mathbf{W}_i} \eta(\omega)$\;
Set $\{\mathcal{X}_i\}_{i\leq N_\mathrm{snap}} \mathrel{{+}{=}} \mathcal{X}(\omega_\mathrm{new})$ and $N_\mathrm{snap} \mathrel{{+}{=}} 1$
}
\Return ROM basis $\{\mathcal{X}_i\}_{i\leq N_\mathrm{max}}$
\end{algorithm}

\clearpage
\subsubsection{Threshold}\label{sec:therhold_2D}~
\begin{wrapfigure}[29]{r}{0.45\linewidth}
    \centering
    \vspace{-4em}
    \includegraphics[width=\linewidth]{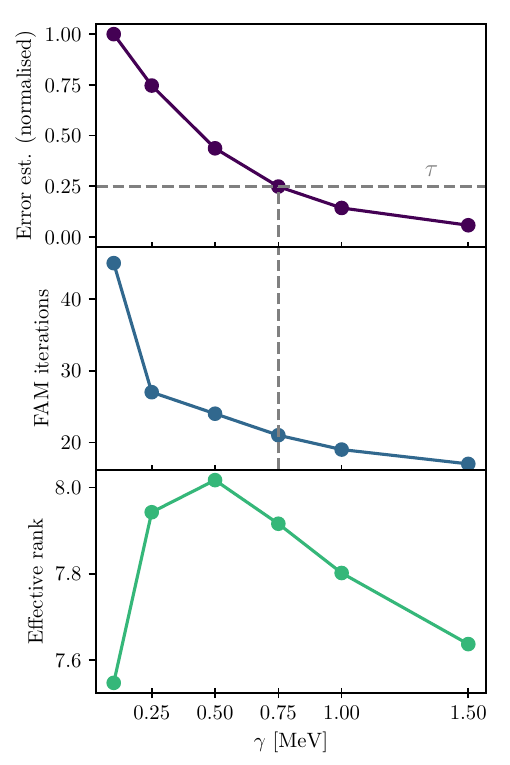}
    \caption{Intermediate step of the pseudo-2D greedy algorithm when selecting the 16th snapshot along the complex axis (marked as $\mathbf{W}_i$ in Fig.~\ref{fig:greedy2D_illustration}). (top) Relative error estimate and a threshold value $\tau=0.25$ in gray. (middle) Computation time (in CPU s) required for calculating the new snapshot. (bottom) Effective rank~\cite{Roy_Effective_Rank} of the basis including the new snapshot.}
    \label{fig:H_scan}
\end{wrapfigure}
To select suboptimal snapshots effectively, determining an appropriate threshold value $\tau$ is crucial. As observed in Fig.~\ref{fig:error_est_example_heatmap_full_spectrum}, the error estimator generally increases as the smearing value decreases. Figure~\ref{fig:H_scan} provides an alternative representation where the error estimator value  (along $\mathbf{W}_i$) is scaled w.r.t. the error at $\omega_r + i \gamma_t$. A steep curve indicates that the error estimator strongly favours snapshots at small smearing. Conversely, a broader curve suggests that snapshots at larger smearing may already provide sufficient information. As demonstrated in the middle panel of Fig.~\ref{fig:H_scan}, selecting a suboptimal snapshot allows for a reduction in computation time. 

These observations motivate the following definition: the threshold $\tau \in [0,1] $ defines the fraction of $\eta(\omega_r + i \gamma_t)$, such that the new snapshot is selected at the maximum below this threshold. This results in the expected behaviour: selection at smaller smearing for steep curves and selection at larger smearing for broad curves. In this context, a threshold value of $\tau=1$ corresponds to the standard greedy selection, i.e.~at the smallest smearing, while lower threshold values allow for a more relaxed selection, favouring larger smearing. 

\begin{figure}
    \centering
    \begin{subfigure}{0.48\textwidth}
    \includegraphics[width=\linewidth]{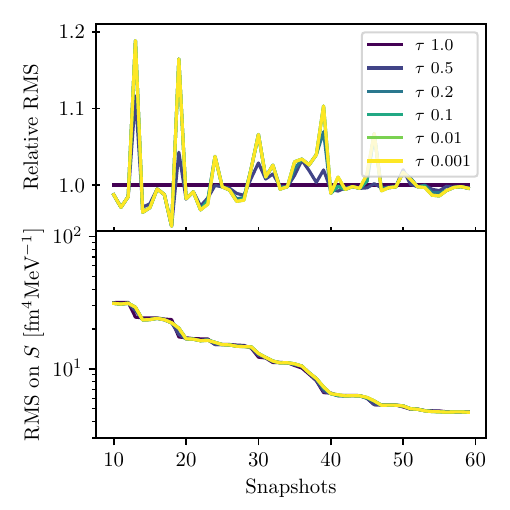}
    \caption{}\label{fig:thresholdsssA}
    \end{subfigure}~
    \begin{subfigure}{0.48\textwidth}
    \includegraphics[width=\linewidth]{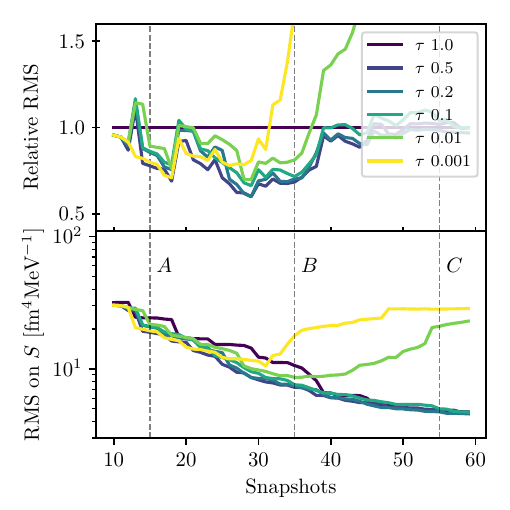}
    \caption{}\label{fig:thresholdsssB}
    \end{subfigure}
    
    \begin{subfigure}{\textwidth}
    \includegraphics[width=\linewidth]{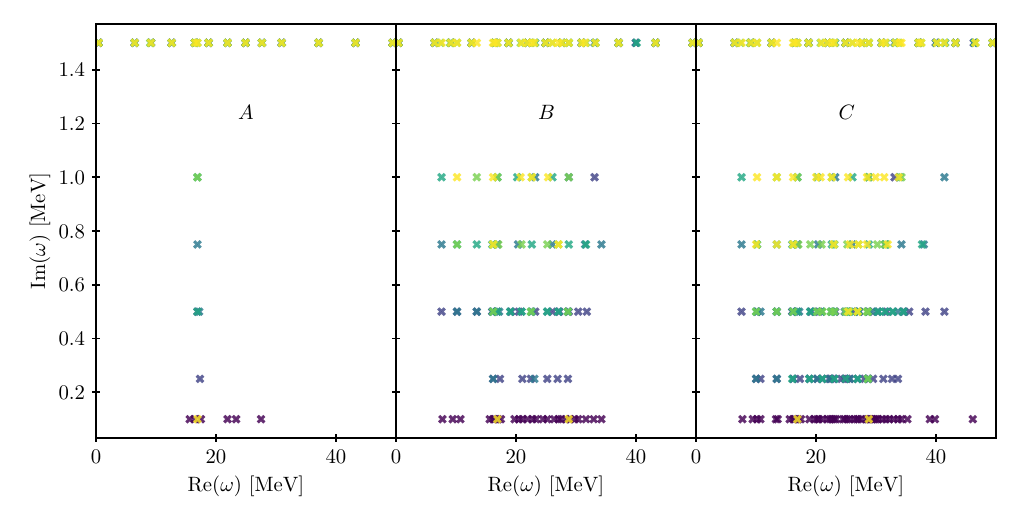}
    \caption{} \label{fig:threshold_omega}
    \end{subfigure}
    \caption{RMS values as a function of number of snapshots for different threshold values $\tau$. Relative RMS is with respect to $\tau = 1$, which means the method samples at the highest error estimate. (\ref{fig:thresholdsssA}) To illustrate the effect of adding one suboptimal snapshot, only the newest snapshot is taken at the $\tau$ from the legend and all previous snapshots were taken at the maximum error estimate (using $\tau = 1$). (\ref{fig:thresholdsssB}) Use case where all snapshots are selected using $\tau$ from the legend. (\ref{fig:threshold_omega}) Snapshot location in the complex plane of the frequency $\omega$ for different threshold values $\tau$.}
    \label{fig:thresholdsss}
\end{figure}

The impact of choosing a suboptimal threshold can be analysed from different perspectives. First, we examine the effect of adding a single suboptimal snapshot to a basis with optimal snapshots ($\tau = 1$), as shown in Fig.~\ref{fig:thresholdsssA}. Overall, the impact on the RMS value is reasonably small and, as expected, decreases with increasing number of snapshots. More pronounced differences are visible for small values of $\tau$.

Second, we compare cases where all snapshots are selected using a consistent threshold value, as shown in Fig.~\ref{fig:thresholdsssB}. Remarkably, for smaller $\tau$, the RMS error initially decreases rapidly as more snapshots are added. This suggests that choosing sub-optimal snapshots improves performance. However, when including even more snapshots, the RMS error using very low threshold values (e.g.~$\tau \leq 0.01$) may increase.

While we have tested extremely small values for $\tau$, it is important to note that these results are influenced by the boundaries of the snapshot search space. For instance, if the relative error estimate at $\gamma = 1.5$\,MeV is larger than $\tau$, the greedy sampler is forced to sample at the boundary of 1.5\,MeV, even if the true ideal location is at a higher smearing value. This introduces a bias to our analysis. 

To assess the impact of this bias, Fig.~\ref{fig:threshold_omega} visualises the snapshot locations of the bases denoted with $A, B$ and $C$ from Fig.~\ref{fig:thresholdsssB}. There is a clear correlation between the threshold and snapshot location: for $\tau = 1$, snapshots are sampled at $\gamma = 0.1$\;MeV, while smaller $\tau$ values allow for selection at larger smearing. Even at the smallest $\tau$, certain snapshots (at least 1 for $A$ and 2 for $B$ and $C$) are located at $\gamma = 0.1$. Similarly, for $\tau=0.2$ and $\tau =0.5$, some snapshots are selected at large smearing as well. We conclude that the bias is more pronounced for the smallest $\tau$ values as most snapshot locations are restricted by the limited sampling space in this test, but the effect remains small for the other $\tau$. Ultimately, this sample-space bias will only lead to an underestimation of the emulator's performance. Based on this analysis, we have selected a threshold value of $\tau=0.1$ to find a balance between low thresholds (to reduce computation time) and selecting relevant snapshots.

\begin{wrapfigure}[20]{r}{.45\linewidth}
    \centering
    \vspace{-2em}
    \includegraphics[width=\linewidth]{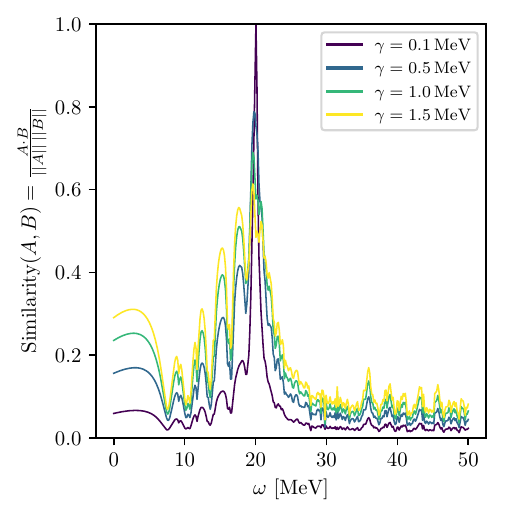}
    \caption{Similarity between a snapshot vector at $\omega~=~20$\,MeV for different smearing (colour) and the vector for frequencies at the target smearing $\gamma~=~0.1$\,MeV. }
    \label{fig:similarity_smearing}
\end{wrapfigure}

To visualise how a larger smearing can be more informative, we calculated the overlap (or cosine similarity) of a snapshot at $\omega = 20$\;MeV with different smearing values and other frequencies at the target smearing $\gamma=0.1$\,MeV. This is shown in Fig.~\ref{fig:similarity_smearing}, where the purple curve is trivially equal to $1$ at $20$\;MeV. While the overlap at the same frequency decreases with increasing smearing, the overlap with surrounding frequencies increases. This implies a vector with higher smearing contains more information regarding its surroundings compared to a vector at lower smearing.

One way to quantify this extra information is by calculating the Shannon entropy (or Effective Rank) of the snapshot matrix~\cite{Roy_Effective_Rank}. Looking at Fig.~\ref{fig:H_scan}, this reveals that snapshots at larger smearing can have a larger entropy. As a consequence, selecting one of these snapshots should yield a better RMS, which matches our findings. A logical next step would then be to use entropy as an error estimator for basis selection (e.g.~in~\cite{Coifman_Entropy_cost}). However, this requires an SVD for every potential new snapshot, which quickly becomes computationally prohibitive for large model spaces. In contrast, the proposed pseudo-2D greedy search is much faster as it only requires an evaluation of the error estimator from Eq.~\eqref{eq:cost_greedy}.

\subsection{Conclusions on snapshot selection}

In this subsection, we explored various snapshot selection procedures. We implemented an iterative greedy search approach and compared it to a naive alternative and equidistant sampling methods. We found that the 1D-greedy search achieves similar accuracy with significantly fewer snapshots. When increasing the number of snapshots, lower RMS errors can be obtained, and the instability discussed in Sec.~\ref{subsec:basis-manipulations} seems to have vanished within a reasonable number of snapshots. Therefore, only applying an SVD at the end as a safety measure is recommended. 

Moreover, we noticed that, while the emulator is able to extrapolate towards different smearing values, it is also beneficial to include snapshots at different smearing to form a mixed basis. While a straightforward approach did not work, a pseudo-2D greedy search compromising computational speed and sub-optimal snapshot selection appears to be feasible. This method is equivalent in terms of accuracy and number of snapshots to the 1D-greedy alternative, while it benefits from faster FAM convergence. This yields a significant decrease in the total computational cost to evaluate a strength function. The methods are benchmarked and compared in Sec.~\ref{sec:performance}.

\section{Strength function symmetry}\label{sec:strength-symmetry}

In the previous sections, we mainly focussed on developing the emulator. However, one can also check the symmetries following from the theory, as shown in Eq.~\eqref{eq:FAM_symmetries}, especially since we limited our snapshot selection to only search for $\mathrm{Re}(\omega)>0$ and $\mathrm{Im}(\omega)>0$. For instance, $\Im(\omega)$ is directly relevant for finite-temperature (FT)-QRPA~\cite{ravlic2025_FTQRPA}.

\begin{figure}
    \centering
    \includegraphics[width=0.85\linewidth]{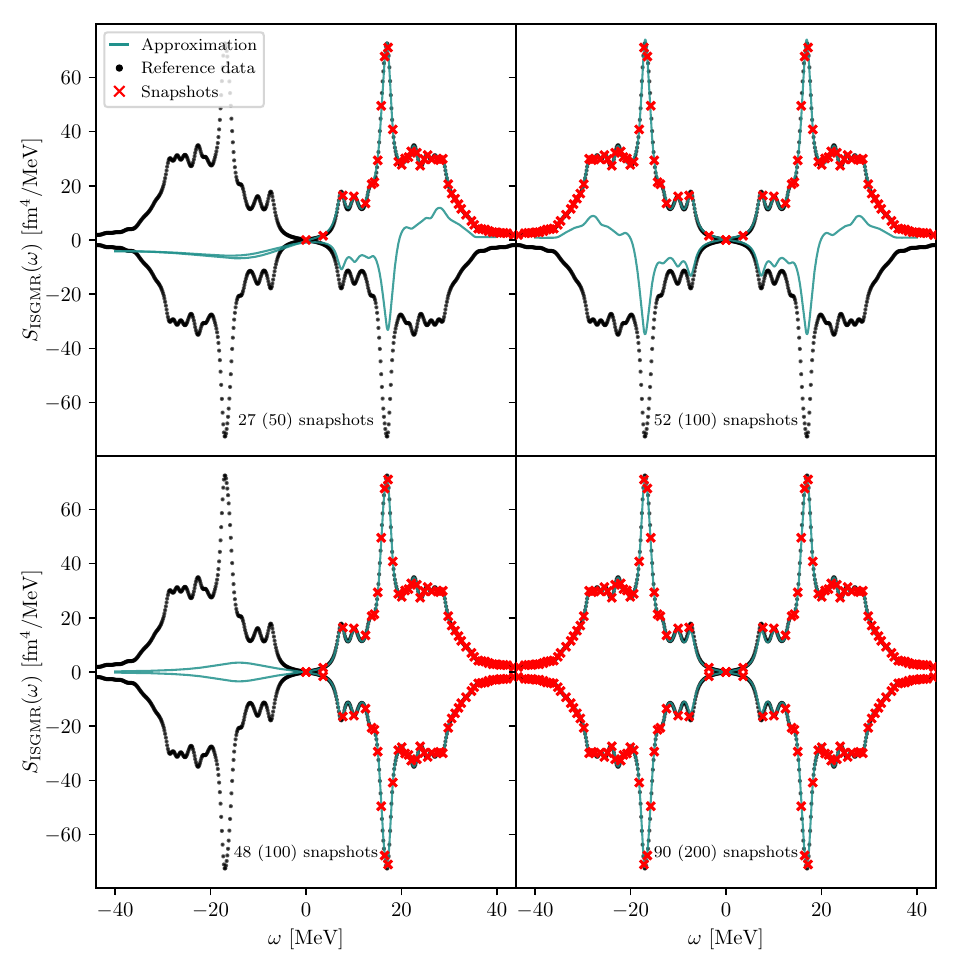}
    \caption{Emulated strength for \ce{Mg-24} and $\gamma=1.0$\,MeV using a greedy sampler and SVD with cutoff $0.01$.}
    \label{fig:symmetry}
\end{figure}

While these symmetries follow from the theory, Fig.~\ref{fig:symmetry} illustrates that the ROM model does not inherently preserve them! Luckily, the ROM basis can be easily extended to include these symmetries with minimal computational cost. More precisely, a FAM calculation at frequency $\omega$, yielding $X(\omega)$ and $Y(\omega)$ can be used to add a snapshot at frequency $-\omega$ by setting $X(-\omega)=Y(\omega)$ and $Y(-\omega)=X(\omega)$. This corresponds to the top-right panel in Fig.~\ref{fig:symmetry}, where now the symmetry $S(-\omega) = -S(-\omega^*)$ is conserved\footnote{Although the strength at $S(-\omega^*)$ is incorrect!}. One can also add a snapshot at $\omega^*$, namely with $X(\omega^*) = X^*(\omega)$ and $Y(\omega^*) = Y^*(\omega)$, yielding the bottom-left panel and securing the $S(\omega^*) = - S(\omega)$. 

We find that both symmetries must be explicitly included to the basis to accurately reproduce the expected behaviour. Furthermore, the SVD reveals that the number of significant basis vectors increases approximately linearly with the addition of these symmetries, suggesting that the snapshot spaces for different frequency domains is largely linearly independent. Similarly, adding these symmetries does not improve the emulation quality in the first quadrant.

\newpage
\section{Emulator performance}\label{sec:performance}
\subsection{Benchmark}
\begin{table}
    \captionof{table}{Benchmark of ROM models and a simple interpolation (using cubic interpolation on $S$ obtained by FAM), separated in emulator creation and evaluation. Snapshots were increased to obtain an RMS lower than the target RMS. Reported values are number of snapshots, the number of total FAM iterations, the CPU calculation time (on Lucia) and the total speed-up compared to simple interpolation. Model evaluation time is separated into SVD and calculation time of the strength in 10\,000 frequencies averaged 10 times. Since these times are obtained on a laptop with limited performance, they can merely be used to compare qualitatively.  Indications imply that (\textbf{*}) it is assumed the spectrum is contained in one peak and (\textbf{**}) the error convergence below the threshold of the method is not guaranteed.}
    \label{tab:greedy_comp_speed}
    \centering
    
    \centering
    \renewcommand{\arraystretch}{1.1}
    \begin{tabular}{l S[table-format=3.0] S[table-format=5.0]  S[table-format=7.0] S[table-format=2.1]  S[table-format=1.2] S[table-format=1.3]}
    \toprule
     & \multicolumn{4}{c}{Creation} & \multicolumn{2}{c}{Evaluation} \\
    \cmidrule(lr){2-5} \cmidrule(lr){6-7}
     Method & {\# snap.} & {\# iter.}&{CPU [s]} & {Speed-up}&{SVD [s]} & {Evaluation [s]} \\
    \midrule
    \rowcolor{gray!40} \multicolumn{7}{l}{\textbf{\ce{Mg-24}} ($\gamma = 0.1$\,MeV, target RMS $<10$\,fm$^4$MeV$^{-1}$)} \\
    Interpolation & 258& 6029& 36542& 1.0&{-} & 0.001 \\
    Equidistant G + POD& 73 &1732&10447&3.5&0.05& 0.417\\
    Equidistant PG + POD& 62& 1461& 8809&4.1&0.04& 0.345 \\
    Greedy 1D & 46&1168& 6989&5.2&{-}& 0.256\\
    Greedy 1D + 1POD & 47 &1193&7152& 5.1& 0.04& 0.311 \\
    \rowcolor{gray!20}Greedy 2D & 30&485&2787&13.1&{-}&0.144 \\  \\[-6pt]
 \rowcolor{gray!40} \multicolumn{7}{l}{\textbf{\ce{Pb-208}} ($\gamma = 0.1$\,MeV, target RMS $<250$\,fm$^4$MeV$^{-1}$)} \\
    Interpolation & 200&10070&4010337&1.0&{-}&0.045\\
    Equidistant G + POD  & 190&9605& 3825412 &1.1& 3.75& 3.703 \\
    Equidistant PG + POD & 122 & 6404& 2524672 &1.6& 3.45& 3.482 \\
    Greedy 1D & 23&1645&568077&7.1& {-}& 0.188\\
    Greedy 1D + 1POD & 23&1645&568077&7.1& 0.13&0.227\\
    \rowcolor{gray!20}  Greedy 2D &20& 417&194944&20.6&{-}&0.201 \\
    \\[-6pt]
     \rowcolor{gray!40} \multicolumn{7}{l}{\textbf{\ce{Pb-208}} ($\gamma = 1.0$\,MeV, target RMS $<75$\,fm$^4$MeV$^{-1}$)} \\
    Interpolation & 32& 541& 286379&1.0 &{-}& 0.009\\
    Lorentzian fit\textbf{*}&23& 386&205513& 1.4&{-}& <0.001 \\
    Equidistant G\textbf{**}&21 & 354 &184809& 1.5&{-}& 0.195\\
    Equidistant PG\textbf{**}&22& 368&195094& 1.5&{-}& 0.239\\
    Equidistant G + POD& 22& 368&195094& 1.5&0.09& 0.512 \\
    Equidistant PG + POD& 22 & 368&195094& 1.5&0.05& 0.371 \\
    Greedy 1D& 7&130&67806& 4.2&{-}& 0.166 \\
    Greedy 1D + 1POD& 7&130&67806&4.2&0.02& 0.263 \\
    \bottomrule
    \end{tabular}

\end{table}

In this section, we discuss the benchmark of the developed methodologies. As previously described in Sec.~\ref{sec:reference_data_and_RMS}, reference data were generated for \ce{Mg-24} and \ce{Pb-208} using FAM calculations across a dense set of equidistantly distributed frequencies. This slightly compromises the emulator's performance, as it restricts the basis vectors to be part of the discrete values within the reference set (see Sec.~\ref{sec:therhold_2D}).

For a given element and target smearing, we define a target RMS error to represent the desired precision. For each method, the number of snapshots was gradually increased until the resulting strength spectrum matched the reference spectrum with an RMS error lower than the target RMS error. In practice, this convergence criterion cannot be applied directly, as it requires knowledge of the exact FAM spectrum. In such cases, one could instead define a convergence criterion based on the error estimation provided by the greedy sampling methods. 

We report both the emulator creation and emulator evaluation times in Tab.~\ref{tab:greedy_comp_speed}. The emulator creation accounts for the majority of the total computational cost and is dominated by the computationally expensive FAM calculations. While the computation time scales with the number of snapshots, not all snapshots are equally expensive (this is the crux of the pseudo-2D greedy method). Consequently, we have included the total number of FAM iterations and the corresponding CPU time required for generating the snapshot basis. 

Once the emulator is constructed, it can be used to rapidly compute strength functions. In a sense, one could argue that the effective speed-up of the emulator is set by its evaluation only (as is reported in~\cite{Jin2025}). However, reporting only this as `the speed-up' would be misleading, as a significant cost lies in the emulator construction, which has to be performed for every element and operator. Therefore, while we note that the emulator evaluation and (single) SVD costs are negligibly small, our analysis focuses on the emulator creation time.

We report results for the equidistant sampling methods using Galerkin (G) and Petrov-Galerkin (PG) projections with an SVD, as discussed in Sec.~\ref{sec:methods-for-reduced-basis-projections}. We also evaluate the 1D greedy method from Sec.~\ref{subsubsec:greedy} (using PG) and verify that similar results are obtained with and without the SVD. Finally, the pseudo-2D greedy method is employed for a smearing of $0.1$\,MeV to avoid the sampling-space bias discussed in Sec.~\ref{subsubsec:greedy2D}. All of this is compared against a method denoted as `Interpolation', applying a basic cubic spline to the strength values obtained by FAM. This serves as a natural baseline that obtains a continuous spectrum from discrete data points.

The first two examples involve the ISGMR of \ce{Mg-24} and \ce{Pb-208} with small smearing. This choice is motivated by the fact that low RMS errors are more difficult to achieve at smaller smearing. Briefly, we conclude that the equidistant samplers (G and PG) yield either no speed-up or a modest improvement (up to $\times$4), depending on the element. The 1D-greedy sampler yields a speed-up of $\times$5 to $\times$7 due to the reduction in required number of snapshots. While the pseudo-2D greedy allows for another minor reduction in number of snapshots, resulting in a speed-up between $\times 13$ and $\times 20$.

The final example considers \ce{Pb-208} with a larger smearing parameter. This example can be useful when the primary interest does not require high resolution on the strength, for instance when determining the centroid position of the resonance. We observe that lower RMS errors are achieved with fewer snapshots compared to the small smearing examples. Furthermore, the speed-up trend across the methods remains consistent with the previous examples. 

Because sufficient precision can be obtained with fewer snapshots, we can verify\footnote{Otherwise, the G and PG method would not converge due to numerical issues.} that the G and PG without SVD perform almost identically compared to the SVD variants. Additionally, for strength function characterised by a single resonant peak, one could employ a Lorentzian fit instead of interpolation. While this approach is not universally applicable and fails for lower smearing values, it allows for a reduction in snapshots, thereby performing equally good as the equidistant samplers. Nevertheless, the greedy sampler still yields the largest speed-up. 

To further reduce computational costs, the partially-constructed emulator could be used to warmstart MOCCa~\cite{RyssensMOCCa}. When starting a new FAM calculation, the emulator can already provide an improved initial guess for $\mathcal{X}(\omega)$, thereby accelerating convergence and reducing the total number of iterations required. This feature has not yet been implemented.

\subsection{Generalisability}
So far, the emulator was tested for the ISGMR spectra for \ce{Mg-24} and \ce{Pb-208}, based on mean-field solutions with the BSkG2 parametrisation. It is essential to verify whether the methodology remains valid beyond these assumptions. 

To validate the approach, reference data (as described in Sec.~\ref{sec:reference_data_and_RMS}) were also generated using BSkG4 and BSkG5. As discussed in Sec.~\ref{sec:edf-parameters}, these parametrisations are different: while BSkG2 is a standard Skyrme parametrisation with specific fit corrections, BSkG4 extends this form with additional $t_4$ and $t_5$ terms. BSkG5, by contrast, uses a different functional form based on N2LO Skyrme interactions. 

For brevity, only the pseudo-2D greedy search was applied to assess convergence and accuracy. The emulated strength functions, generated using the same protocol as for BSkG2, are shown in Fig.~\ref{fig:BSkG4-5}. These results demonstrate that the emulator's performance is consistent across the BSkG parametrisations.

Furthermore, we tested the emulator using the Gogny D1M interaction~\cite{goriely2009_D1Ma, goriely2016_D1Mb}. This interaction differs significantly from the BSkG family: while the latter is based on zero-range Skyrme forces, the Gogny force incorporates finite-range interactions using Gaussian functions. Their numerical implementations also differ significantly: BSkG calculations are performed on a 3D coordinate-space grid, whereas the D1M results rely on an expansion in a harmonic oscillator basis. Consequently, the D1M reference data were generated using the PAN@CEA code~\cite{panaCEA}, rather than the MOCCa~\cite{RyssensMOCCa} code used for the BSkG data.

\begin{figure}
    \centering
    \includegraphics[width=0.45\linewidth]{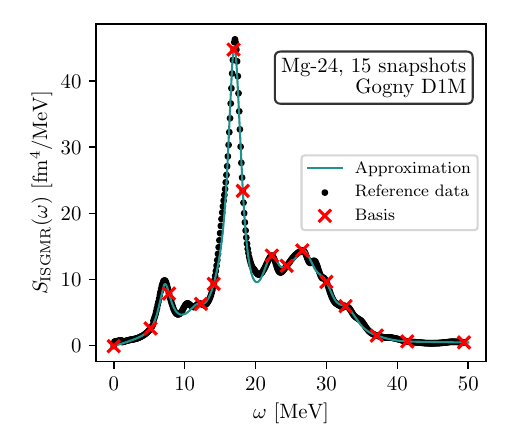}~    \includegraphics[width=0.45\linewidth]{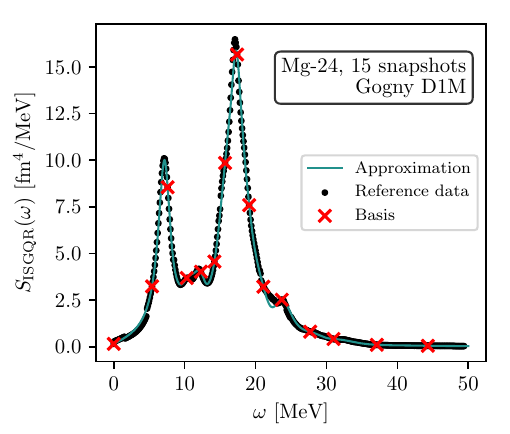}
    \caption{The ROM emulator using Greedy-1D sampler applied to a Gogny D1M interaction, for \ce{Mg-24} monopole and quadrupole strengths ($Q_{20}$ component) with $\gamma = 1$\,MeV.}
    \label{fig:Gogny_isgmr-isgqr}
\end{figure}

\begin{figure}
    \centering
    \includegraphics[width=0.45\linewidth]{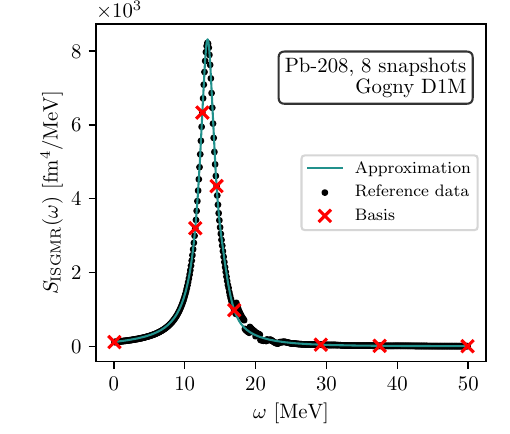}~    \includegraphics[width=0.45\linewidth]{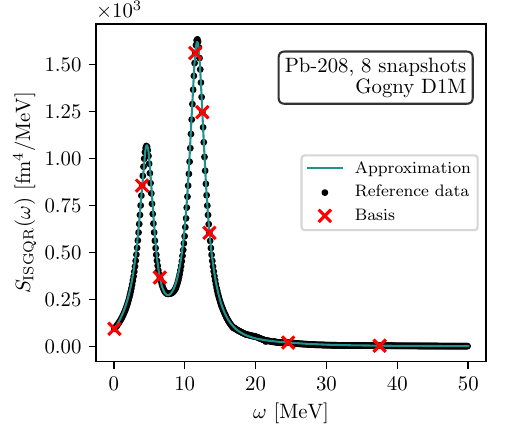}
    \caption{The ROM emulator using Greedy-1D sampler applied to a Gogny D1M interaction, for \ce{Pb-208} monopole and quadrupole strengths ($Q_{20}$ component) with $\gamma=1$\,MeV.}
    \label{fig:Gogny_isgmr-isgqr_pb}
\end{figure}

The results for \ce{Mg-24} and \ce{Pb-208} using the 1D greedy sampling method are shown in Fig.~\ref{fig:Gogny_isgmr-isgqr} and Fig.~\ref{fig:Gogny_isgmr-isgqr_pb}, respectively. In addition to the monopole resonance, we have verified the emulator for the isoscalar giant quadrupole resonance (ISGQR). 

Lastly, we have verified the approach to emulate the IVGDR for the triaxially deformed nucleus \ce{Zr-100}. The reference data is obtained with the QRPA implementation in MOCCa~\cite{RyssensMOCCa} using BSkG2. The $x$, $y$ and $z$ component denote along which axis the excitation operator is applied, and their sum yields the total response. The emulator is built for each component separately using a greedy sampler and 10 snapshots.

These results indicate that the ROM emulator is robust and can be applied to various nuclear parametrisations, functional forms, multipole operators, mean-field basis, deformed nuclei and with pairing effects.

\newpage
\begin{subappendices}
\section{Additional figures}
    \centering
    \includegraphics[width=0.5\linewidth]{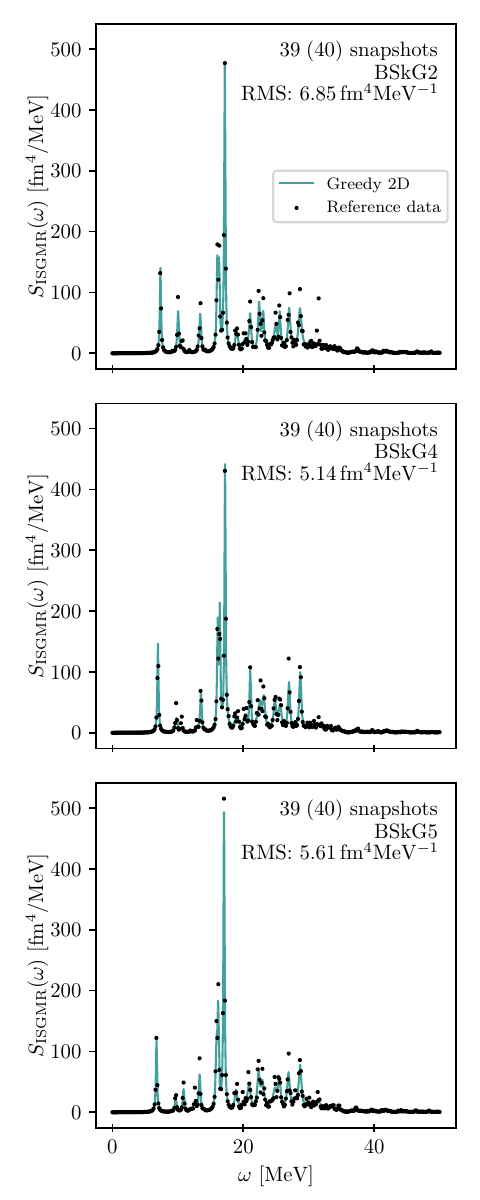}~        \includegraphics[width=0.5\linewidth]{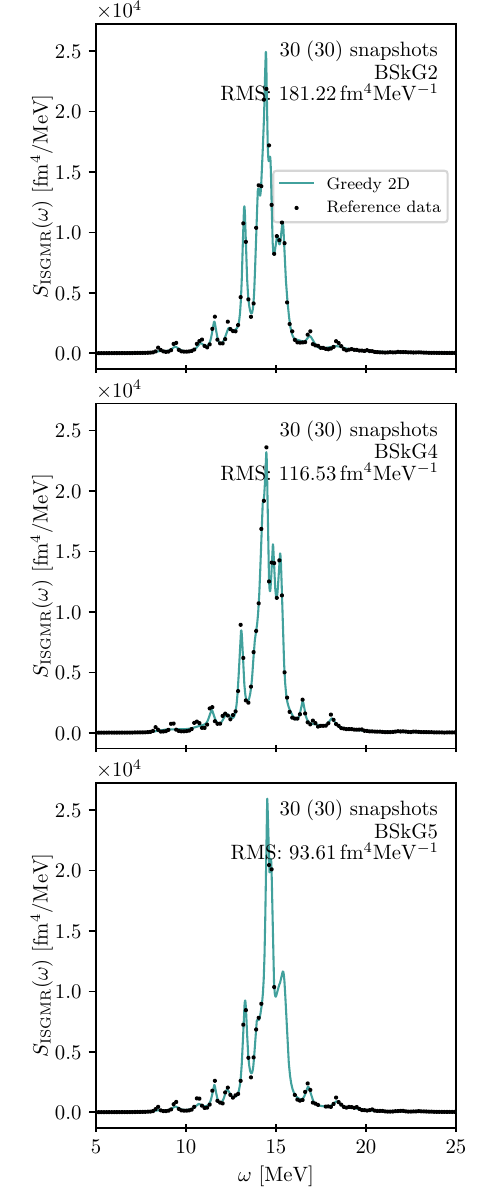}
    \captionof{figure}{Emulated strength functions for the ISGMR for \ce{Mg-24} (left) and \ce{Pb-208} (right) obtained by the pseudo-2D greedy sampling method using the same precision as in Tab.~\ref{tab:greedy_comp_speed} for $\gamma=0.1$\,MeV.}
    \label{fig:BSkG4-5}
    
\begin{figure}
    \centering
    \includegraphics[width=0.9\linewidth]{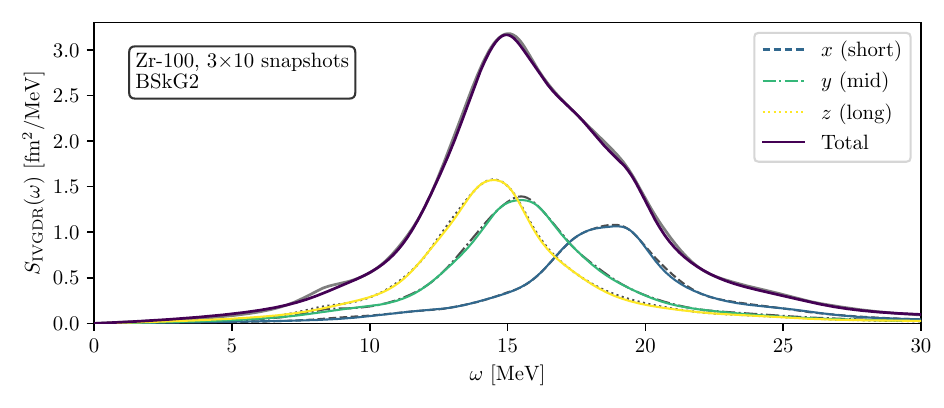}
    \caption{The ROM emulator using Greedy-1D sampler to calculate for the BSkG2 Skryme interaction to find dipole resonance of \ce{Zr-100}, which has an axial deformed ground
state. The strength along the $x$,$y$ and $z$ directions are obtained separately (each with 10 snapshots) and their sum yields the total strength. Reference values are marked in gray.}
    \label{fig:Zr100}
\end{figure}
\end{subappendices}
\chapter{Nuclear incompressibility}\label{chap:4}
\section{Introduction}
The nuclear Equation of State (EoS) relates the energy density, pressure and matter density of nuclear systems. The EoS is a fundamental input for deriving astrophysical properties, such as masses and radii of neutron stars. Furthermore, it is essential for simulating and understanding astrophysical events, including core-collapse supernova explosions~\cite{Garg2018}, astrophysical signals emitted from neutron stars and gravitational waveforms following from the merger of two dense stars~\cite{Margueron2026}.

A key parameter is the nuclear incompressibility $K_\infty$, which characterises the stiffness of the EoS near the saturation density $\rho_0$. Determining $K_\infty$ is challenging because infinite nuclear matter is not directly observable. 

Experimentally, one must instead measure the incompressibility of finite nuclei, $K_A$, typically extracted from giant monopole resonances (see for example~\cite{Garg2018}). These observed resonances are influenced by surface, Coulomb, and asymmetry effects that should not appear in $K_\infty$. While extracting $K_\infty$ from finite nuclei is not straightforward, it can be established via theoretical models.

Beside its importance in the EoS, $K_\infty$ also appears in the fitting procedure for the EDFs. Because this value can be calculated directly from the EDF parameters, as we will show in this chapter, adding a constraint on its value is crucial to reduce the degrees of freedom and improve the fitting process. 

Attempts to determine $K_\infty$ appear inconsistent. Most (non-relativistic) Skyrme- and Gogny-type models yield values in the range of 210-250\,MeV, whereas relativistic models yield larger values, 260-270\,MeV (see~\cite[Tab.~1]{stone} for a summary). This difference may partially be explained by including density dependence in the symmetry energy~\cite{Zamora2024}. Nevertheless, these results have large uncertainties of 10\%-20\% ~\cite{Colo2004, Zamora2024}.

Conventionally, $K_\infty$ is taken around 240$\pm$20\,MeV~\cite{Colo2004, stone,  Garg2018, Margueron2026}. As a result, such EDF parametrisations are built to correctly predict ISGMR centroid positions for \ce{Pb-208} and \ce{Zr-90}. For \ce{Pb-208}, the centroid is experimentally determined to be 14.17$\pm$0.28\,MeV~\cite{Colo2004}, though other studies report slightly different values, e.g.~13.96$\pm$0.20\,MeV~\cite{Youngblood2004}. To consistently describe the ISGMR in both Sn and Pb isotopes within a single framework, the inclusion of quasiparticle-vibration coupling is necessary \cite{Li2023}.

The BSk and BSkG parametrisations are constrained by $K_\infty =$ 240$\pm$10\,MeV~\cite{goriely2016b}. However, the correlation between the \ce{Pb-208} ISGMR centroid and $K_\infty$ has not been verified for these functionals. This raises the possibility that the EDF parameters are over-constrained, and that an unconstrained parametrisation might reproduce a correct \ce{Pb-208} centroid while yielding a  $K_\infty$ value outside of the conventional range. 

Recent studies suggest that the relationship between $K_\infty$ and the ISGMR is not as straightforward as previously thought, proposing the use of models where $K_\infty$ are more loosely constrained~\cite{Khan2012, Khan2013, Margueron2026}.

In this chapter, we aim to verify this correlation for the BSk and BSkG to determine whether or not its use in the fitting procedure is justified. In Sec.~\ref{sec:K_theory}, we briefly introduce the required theoretical background. Next, the $K_\infty$ value for the relevant EDF-types are derived in Sec.~\ref{sec:K_derive}. The ISGMR of \ce{Pb-208} is obtained using the FAM-QRPA code, and the correlation between the $K_\infty$ and centroid position is discussed in Sec.~\ref{sec:K_conclusion}.

\clearpage
\section{Theoretical description}\label{sec:K_theory}
\begin{wrapfigure}[15]{r}{0.45\linewidth}
    \vspace{-3.5em}
    \centering
    \includegraphics[width=.9\linewidth]{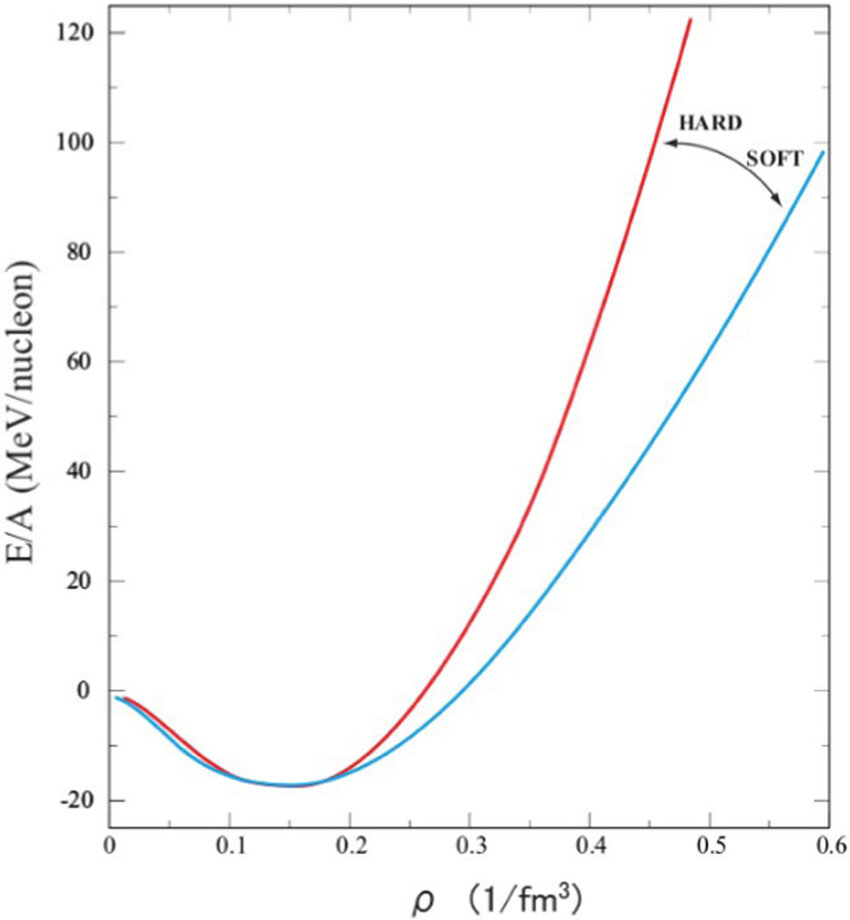}
    \vspace{-0.3em}
    \caption{Binding energy per nucleon as function of nuclear density, from~\cite[Fig.~4]{Harakeh2025_GR_fig}.}
    \label{fig:energy_per_nucleon}
\end{wrapfigure}
This section is devoted to introducing the required concepts for the rest of the chapter. We consider homogeneous, isotropic, infinite nuclear matter, where the number of protons and neutrons are equal and the Coulomb force is neglected to prevent energy divergence. Spatial or current-dependent properties can be neglected as they would average out. 

The energy per nucleon $E/A(\rho)$ curve has a minimum at the saturation density $\rho_0 \approx 0.166$\,fm$^{-3}$~\cite{Garg2018}. The pressure $P$ can be defined using $\rho=A/V$ as: 
\begin{align*}
P = -\left.\frac{\partial E}{\partial V}\right|_{A}
= -\frac{d E}{d \rho}\frac{d \rho}{d V}
=\frac{\rho}{V}\frac{d E}{d \rho} =\rho^2\frac{d E/A}{d \rho}\;.
\end{align*}
The compression modulus $\chi$ is related to the incompressibility of the system. 

For a finite nucleus, $K_A$ is defined as~\cite{chabanat1997}:
\begin{align}
\chi &= -\frac{1}{V}\left.\frac{dV}{dP}\right|_{A} = \frac{1}{\rho}\left(\frac{dP}{d\rho}\right)^{-1}\nonumber \;,\\
    K_A &=  \frac{9}{\rho\chi} =18\frac{P}{\rho} + 9\rho^2 \frac{\partial^2E/A}{\partial \rho ^2} \label{eq:compressibility_K_A} \;.
\end{align}
The infinite matter incompressibility, $K_\infty$, can be found by evaluating this at the saturation point $\rho=\rho_*$\footnote{Note that $\rho_0$ is used to denote $\rho_*$ in literature, though this leads to an ambiguous definition of $\rho=\rho_0=\rho_p+\rho_n$.}. 

An alternative way to find $K_\infty$ is by using the curvature of the energy per particle around the saturation density as shown in Fig.~\ref{fig:energy_per_nucleon}. We Taylor expand around the minimum $\rho=\rho_*$ as follows:
\begin{align*}
E/A(\rho) &= E/A(\rho_*) + \left.\frac{d E/A}{d \rho}\right|_{\rho_*}\left(\rho-\rho_*\right) +\frac{1}{2}\left.\frac{d^2 E/A}{d \rho}\right|_{\rho_*^2}\left(\rho-\rho_*\right)^2+\ldots \\
&= E/A(\rho_*) +\frac{1}{2} K_\infty \left(\frac{\rho-\rho_*}{9\rho_*}\right)^2+\ldots \;.
\end{align*}
The first derivative evaluated at $\rho_*$ is equal to zero by construction, meaning that the pressure is zero at $\rho_*$. Now, $K_\infty$ is related to the curvature of $E/A$ with respect to $\rho$, such that:
\begin{equation}
K_\infty = 9 \rho_*^2 \left.\frac{d^2(E/A)}{d\rho^2}\right|_{\rho=\rho_*}\;.\label{eq:compressibility_K_inf}
\end{equation}
Equivalently, it can also be expressed as a function of the pressure\footnote{Recall that the first derivative vanishes in $\rho=\rho_*$.}:
\[
K_\infty = 9 \left.\frac{dP}{d\rho}\right|_{\rho=\rho_*}\;.
\]
This relation allows for finding a $K_\infty$ value associated to a specific EDF. However, to construct such an EDF we need to deduce $K_\infty$ from experimental data, which requires a different approach.

The incompressibility of a finite nucleus is related to the ISGMR centroid energy via~\cite{Colo2004, Garg2018, Blaizot1980, stone}:
\begin{equation*}
K_A = \frac{m \langle r^2\rangle_0 E^2_\mathrm{ISGMR}}{\hbar^2} \;,
\end{equation*}
where $\langle r^2 \rangle_0$ is the ground-state mean-square radius and $m$ is the reduced nucleon mass. Using the sum rules from Sec.~\ref{sec:sumrules}, we can rewrite this to:
\begin{align}
   K_A &=  \frac{m \langle r^2 \rangle_0}{\hbar^2} \frac{m_1}{m_{-1}} \;.\label{eq:compressibility_K_A_sumrule}
\end{align}
This means that we can calculate $K_A$ from the ISGMR mean energy obtained by the FAM calculations, provided that the charge radius of a nucleus is known with sufficient accuracy from experiments.

Now that we have obtained a way to define $K_A$, we consider a few options to relate this with $K_\infty$. The first option is inspired by the liquid-drop model to expand $K_A$ as follows:
\begin{equation}
K_A = K_\mathrm{vol} + K_\mathrm{surf}A^{-1/3}+K_\mathrm{symm}\left(\frac{N-Z}{A}\right)^2+K_\mathrm{Coul}Z^2A^{-1/3}\;,\label{eq:liquiddrip}
\end{equation}
including volume, surface, symmetry and Coulomb contributions. As discussed in~\cite{Garg2018}, a first approach is to believe this liquid drop model and assume the infinite nuclear incompressibility can be identified with the volume constant: $K_\infty \equiv K_\mathrm{vol}$. Since Eq.~\eqref{eq:liquiddrip} contains multiple unknown liquid-drop parameters, these parameters must be fitted simultaneously to data from different isotopes. A second approach is to assume a linear dependency of $K_\infty$ and $K_A$, and fit~\cite{Colo2004, Garg2018}:
\[
K_A = a K_\infty + b \qquad \mathrm{or}\qquad E_\mathrm{ISGMR}= a' \sqrt{K_\infty} + b'\;.
\]
Some recent papers opt to abandon this relation and try to obtain $K_\infty$ via a third approach, by defining a density-dependent incompressibility~\cite{Khan2012, Khan2013, Margueron2026}.

In the following sections, we derive $K_\infty$ directly from a known EDF form. 

\section{Deriving the nuclear incompressibility from an EDF}\label{sec:K_derive}
\subsection{Densities for INM}
To simplify the expressions, we can relate the matter density $\rho$ to the kinetic density $\tau$ in INM. We derive it in two different ways. First, we note that single-particle waves are plane waves and can be represented by $\psi_k(r) = \tfrac{1}{(2\pi)^{3/2}}e^{-ik\cdot r}$~\cite{Grams2025_notes}. After adding an index to distinguish protons/neutrons, we can write the densities as:
\begin{align*}
    \rho_q &= \frac{N}{V} = \int \mathrm{d}k^3 n_{k,q} \psi_{k,q}^\dagger(r)\psi_{k,q}(r)=\tfrac{1}{3\pi^2}k_{F,q}^3\;,\\
    \tau_q &= \int \mathrm{d}k^3 n_{k,q} \nabla \psi_{k,q}^\dagger(r)\nabla \psi_{k,q}(r)= \tfrac{1}{5\pi^2}k_{F,q}^5 \\&= \tfrac{1}{5\pi^2}\rho_q^{5/3}(3\pi^2)^{5/3}=\tfrac{3^{5/3}\pi^{4/3}}{5}\rho_q^{5/3}=\tfrac{3}{5}(3\pi^2)^{2/3}\rho_q^{5/3}\;.
\end{align*}

The same relation can also be derived from Fermi-Dirac statistics. The number of states $n$ (also known as occupation) can be written as
\[
n = \frac{g}{h^3}\int f \mathrm{d}^3p\;,
\]
where $h$ is Planck's constant and $h^3$ represents the volume of a phase space cell. The function $g(p)$ represents the degeneracy of a state and $f(p)$ is the Fermi function. At $T=0$\,K, the fermi function $f(p)$ is a Heaviside function: $f(p) = 1$ for $|p| \leq p_F$ and $f(p) = 0$ elsewhere. For protons and neutrons the degeneracy is equal to $2$ due to the spin. We obtain:
\begin{equation*}
    \rho \equiv n=\frac{8\pi }{h^3}\int_0^{p_F}p^2\mathrm{d}p=\frac{8\pi }{3h^3}p_F^3 = \frac{8\pi }{3\hbar^3 (2\pi)^3}(\hbar k_F)^3= \frac{1}{3\pi^2}k_F^3\;.
\end{equation*}
Note that this only holds for protons or neutrons, and if both are considered, we should distinguish them with an index: $\rho_q = \frac{1}{3\pi^2}k_{Fq}^3$.
The kinetic density $\tau$ is related to the energy by $\varepsilon = \frac{\hbar^2}{2m}\tau$, and since 
\[
    \varepsilon = \frac{g}{2m h^3}\int p^2 \mathrm{d}^3p =  \frac{4\pi g}{h^3}\int p^4 \mathrm{d}p \;.
\]
We obtain that $\tau = \frac{1}{5\pi^2}k^5_F$. Now $\rho_q$ and $\tau_q$ can be related to each other:
\begin{align}
\tau_q &=  \rho_q^{5/3}(3\pi^2)^{5/3}\frac{1}{5\pi^2} = \rho_q^{5/3} \frac{3}{5}(3\pi^2)^{2/3}\label{eq:tau_ifv_rho} \;,
\end{align}
which is the same as above.

\subsection{NLO}
We aim to derive the expression for $K_\infty$ for the standard Skyrme form.

\paragraph{Symmetric nuclear matter}
First, we will limit the calculations to symmetric nuclear matter, hence $\rho_p = \rho_n$ and, by convention, $\rho = \rho_p +\rho_n$. Starting with $\mathcal{E}_\mathrm{Sky}$ from Eq.~\eqref{eq:energy_density_skyrme_parameters}, and simplifying the terms leads to:
\begin{align*}
    \mathcal{E}_\mathrm{K}&=\tfrac{\hbar^2}{2m}\tau \;,\\
    \mathcal{E}_0 &= \tfrac{1}{4} t_0((2+x_0)\rho^2 -(2x_0+1)(\rho_p^2+\rho_n^2))= \tfrac{3}{8} t_0 \rho^2\;,\\
    \mathcal{E}_3 &= \tfrac{1}{24} t_3\rho^\gamma ((2+x_3)\rho^2-(2x_3+1)(\rho_p^2+\rho_n^2)) = \tfrac{1}{16}t_3 \rho^{\gamma+2}\;,\\
    \mathcal{E}_\mathrm{eff}&=\tfrac{1}{8}(t_1(2+x_1)+t_2(2+x_2))\tau\rho +\tfrac{1}{8}(t_2(2x_2+1)-t_1(2x_1+1))(\tau_p\rho_p +\tau_n\rho_n)\;.
\end{align*}
Using Eq.~\eqref{eq:tau_ifv_rho} and noting that we assume symmetric nuclear matter:
\[
\tau =\tfrac{3}{5}(3\pi^2)^{2/3} ((\tfrac{1}{2}\rho)^{5/3} + (\tfrac{1}{2}\rho)^{5/3}) = \tfrac{3}{5}(\tfrac{3\pi^2}{2})^{2/3} \rho^{5/3}\;,
\]
the first and last equation can be simplified as follows:
\begin{align*}
\mathcal{E}_\mathrm{K}&=\tfrac{\hbar^2}{2m}\tfrac{3}{5}(\tfrac{3\pi^2}{2})^{2/3}\rho^{5/3}\;,\\
\mathcal{E}_\mathrm{eff}&=\tfrac{1}{8}(t_1(2+x_1)+t_2(2+x_2))\tau\rho +\tfrac{1}{8}(t_2(2x_2+1)-t_1(2x_1+1))(\tau_p\rho_p +\tau_n\rho_n)\\
&=\tfrac{1}{8}(t_1(2+x_1)+t_2(2+x_2))\tau\rho +\tfrac{1}{8}(t_2(2x_2+1)-t_1(2x_1+1))\tfrac{3}{5}(3\pi^2)^{2/3}(\rho_p^{8/3} +\rho_n^{8/3}) \\
&=\tfrac{1}{8}(t_1(2+x_1)+t_2(2+x_2))(\tfrac{3}{5}(\tfrac{3\pi^2}{2})^{2/3})\rho^{8/3} +\tfrac{1}{8}(t_2(2x_2+1)-t_1(2x_1+1))\tfrac{3}{5}(\tfrac{3\pi^2}{2})^{2/3}2^{2/3}2(\tfrac{1}{2}\rho)^{8/3}\\
&=\tfrac{3}{40}(\tfrac{3\pi^2}{2})^{2/3}\rho^{8/3}[(t_1(2+x_1)+t_2(2+x_2)) +\tfrac{1}{2}(t_2(2x_2+1)-t_1(2x_1+1))]\\
&=\tfrac{3}{80}(\tfrac{3\pi^2}{2})^{2/3}\rho^{8/3}[3t_1+5t_2+4t_2x_2]\;.
\end{align*}
We calculate the following relations appearing in Eqs.~\eqref{eq:compressibility_K_A} and~\eqref{eq:compressibility_K_inf}:
\begin{align*}
    \frac{\mathcal{E}}{\rho} &= \tfrac{\hbar^2}{2m}\tfrac{3}{5}(\tfrac{3\pi^2}{2})^{2/3}\rho^{2/3} + \tfrac{3}{8}t_0\rho+\tfrac{1}{16}t_3\rho^{\gamma+1}+ \tfrac{3}{80}(\tfrac{3\pi^2}{2})^{2/3}\rho^{5/3}[3t_1+5t_2+4t_2x_2]\;,\\
    \frac{\partial}{\partial \rho}\frac{\mathcal{E}}{\rho} &= \tfrac{\hbar^2}{2m}\tfrac{2}{5}(\tfrac{3\pi^2}{2})^{2/3}\rho^{-1/3}+\tfrac{3}{8}t_0 + \tfrac{1}{16}(\gamma+1)t_3\rho^{\gamma}+\tfrac{1}{16}(\tfrac{3\pi^2}{2})^{2/3}\rho^{2/3}[3t_1+5t_2+4t_2x_2]\;,\\
     \frac{\partial^2}{\partial \rho^2}\frac{\mathcal{E}}{\rho} &= -\tfrac{\hbar^2}{2m}\tfrac{2}{15}(\tfrac{3\pi^2}{2})^{2/3}\rho^{-4/3} +\tfrac{1}{16}(\gamma(\gamma+1)t_3\rho^{\gamma-1}+\tfrac{1}{24}(\tfrac{3\pi^2}{2})^{2/3}\rho^{-1/3}[3t_1+5t_2+4t_2x_2]\;,
\end{align*}
leading to
\[
K_{\infty} =  -\tfrac{\hbar^2}{2m}\tfrac{6}{5}(\tfrac{3\pi^2}{2})^{2/3}\rho_*^{2/3}+\tfrac{9}{16}\gamma(\gamma+1)t_3\rho_*^{\gamma+1}+\tfrac{3}{8}(\tfrac{3\pi^2}{2})^{2/3}\rho_*^{5/3}[3t_1+5t_2+4t_2x_2] \;.
\]

The critical density $\rho_*$ is defined as the minimum of $\tfrac{\mathcal{E}}{\rho}$, thus it solves:
\begin{equation*}
    0= \tfrac{\hbar^2}{2m}\tfrac{2}{5}(\tfrac{3\pi^2}{2})^{2/3}\rho_*^{-1/3}+\tfrac{3}{8}t_0 + \tfrac{1}{16}(\gamma+1)t_3\rho_*^{\gamma}+\tfrac{1}{16}(\tfrac{3\pi^2}{2})^{2/3}\rho_*^{2/3}[3t_1+5t_2+4t_2x_2]\;.
\end{equation*}
For a given parametrisation, a minimum is found numerically near $\rho_* = 0.16$\,fm$^{-3}$. This expression is equivalent to those in~\cite{Bartel1982, Chamel2009}, despite the different terms. This is because they have included the first term from Eq.~\eqref{eq:compressibility_K_A} to approximate $K_\infty$, which vanishes when evaluating the expression in $\rho_*$. 

If the EDF is of a different form, the above expression for $K_\infty$ no longer holds. From here, we could derive the $K_A$, but we will derive a more general form in the next section.

\paragraph{Arbitrary asymmetry} In this case, we no longer assume equal amounts of proton and neutrons. Following the conventions in the literature, we define $I= \tfrac{N-Z}{A} = \tfrac{(N-Z)V}{(N+Z)V} = \eta$ and
\begin{align}
F_m (I) &= \tfrac{1}{2}((1+I)^m+(1-I)^m) \notag \\
    &= \tfrac{1}{2\rho_0^m}((\rho_0+\rho_1)^m+(\rho_0-\rho_1)^m)\notag\\
    &= \tfrac{1}{2\rho^m}((2\rho_n)^m+(2\rho_p)^m) =  2^{m-1}\rho^{-m}(\rho_n^m+\rho_p^m)\;,\label{eq:F_ma} \\[1em]
\tilde{F}_m(I) &= \tfrac{1}{2}((1+I)^m-(1-I)^m)=2^{m-1}\rho^{-m}(\rho_n^m-\rho_p^m)\;.\label{eq:F_mb}
\end{align}
This leads to the following relation:
\begin{align*}
    \tau_0 &= \tfrac{3}{5}(\tfrac{3\pi}{2})^{2/3}F_{5/3}\rho^{5/3}\;,\\
    \tau_1 &=  \tfrac{3}{5}(\tfrac{3\pi}{2})^{2/3}\tilde{F}_{5/3}\rho^{5/3}\;.
\end{align*}
Similar to the previous calculation, we rewrite each term of $\mathcal{E}$ separately:
\begin{align*}
    \mathcal{E}_K &= \tfrac{\hbar^2}{2m}\tfrac{3}{5}(\tfrac{3\pi}{2})^{2/3}F_{5/3}\rho^{5/3}\\[1em]
    \mathcal{E}_0 &= \tfrac{1}{8}t_0 \rho^2(2(x_0+2)-(2x_0+1)F_2)\;,\\[0.5em]
    \mathcal{E}_3 &= \tfrac{1}{48}t_3\rho^{\gamma+2}(2(2+x_3) - (2x_3+1)F_2)\;,\\[0.5em]
    \mathcal{E}_\mathrm{eff}&=\tfrac{1}{8}(t_1(2+x_1)+t_2(2+x_2))\tfrac{3}{5}(3\pi^2)^{2/3}(\rho_p^{5/3}+\rho_n^{5/3})\rho\\&\quad+ \tfrac{1}{8}(t_2(2 x_2+1)-t_1(2x_1+1))\tfrac{3}{5}(3\pi^2)^{2/3}{\rho_p^{8/3}+\rho_n^{8/3}} \\[0.5em]
    &=\tfrac{1}{8}(t_1(2+x_1)+t_2(2+x_2))\tfrac{3}{5}(3\pi^2)^{2/3} F_{5/3}\rho^{5/3}2^{-2/3}\rho\\&\quad+ \tfrac{1}{8}(t_2(2x_2+1)-t_1(2x_1+1))\tfrac{3}{5}(3\pi^2)^{2/3}F_{8/3}\rho^{8/3}2^{-5/3}\\[0.5em]
    &=\tfrac{3}{40}(\tfrac{3\pi^2}{2})^{2/3}\rho^{8/3}\left[(t_1(2+x_1)+t_2(2+x_2)) F_{5/3}+ \tfrac{1}{2}(t_2(2x_2+1)-t_1(2x_1+1))F_{8/3}\right]\;.
\end{align*}
Adding these together and dividing by $\rho$ leads to the same $\mathcal{E}/\rho$ as in~\cite[Eq.~3.18]{chabanat1997}. The expression for $K_\infty$ reads:
\begin{align*}
K_{\infty,\,\mathrm{NLO}} \approx \left.9\rho^2\frac{\partial^2 \mathcal{E}/\rho }{\partial \rho^2}\right|_{\rho=\rho_*} &= -\tfrac{6}{5}\tfrac{\hbar^2}{2m}(\tfrac{3\pi^2}{2})^{2/3}\rho^{2/3}_*F_{5/3} + {\tfrac{3}{16}}\gamma(\gamma+1)t_3\rho_*^{\gamma+1}(2(x_3+2)-(2x_3+1)F_2) \\ &\quad + \tfrac{3}{4}(\tfrac{3\pi^2}{2})^{2/3}\rho_*^{5/3}\left[(t_1(2+x_1)+t_2(2+x_2)) F_{5/3}\right.\\ &\hspace{3.5cm}+ \left.\tfrac{1}{2}(t_2(2x_2+1)-t_1(2x_1+1))F_{8/3}\right]\;.
\end{align*}
This value is fully determined by the EDF parameters, and also depends on the $p$-$n$ asymmetry $I$ appearing in the $F_m$ terms. Finally, we derive an expression for the finite nuclear incompressibility $K_A$ using Eq.~\eqref{eq:compressibility_K_A} (omitting some intermediate steps):
\begin{align*}
P &=\tfrac{2}{5}\tfrac{\hbar^2}{2m}(\tfrac{3\pi^2}{2})^{2/3}\rho^{5/3}F_{5/3} + \tfrac{1}{8}t_0\rho^2(2(x_0+2)-(2x_0+1)F_2) |\\&\qquad+ \tfrac{1}{48}(\gamma+1)t_3\rho^{\gamma+2}(2(2+x_3)-(2x_3+1)F_2))\\ &\qquad+ 
\tfrac{1}{8}(\tfrac{3\pi^2}{2})^{2/3}\rho^{8/3}[t_1(2+x_1)+t_2(2+x_2))F_{5/3}+\tfrac{1}{2}(t_2(2x_2+1)-t_1(2x_1+1))F_{8/3}]\;,\\[1em]
K_{A,\,\mathrm{NLO}} &= 6\tfrac{\hbar^2}{2m}(\tfrac{3\pi^2}{2})^{2/3}\rho^{2/3} F_{5/3} + \tfrac{9}{4}t_0\rho(2(x_0+2)-(2x_0+1)F_2) \\&\qquad+ \tfrac{3}{16}(\gamma+2)(\gamma+1)t_3\rho^{\gamma+1}(2(x_3+2)-(2x_3+1)F_2) \\ &\qquad + 3(\tfrac{3\pi^2}{2})^{2/3}\rho^{5/3}\left[(t_1(2+x_1)+t_2(2+x_2)) F_{5/3}\right.+ \left.\tfrac{1}{2}(t_2(2x_2+1)-t_1(2x_1+1))F_{8/3}\right]\;.
\end{align*}

\subsection{Skyrme EDF with $t_4$ and $t_5$ extension}
Another type of EDF function is the generalized Skyrme, containing additional terms in the Skyrme force~\cite{Chamel2009}. As discussed in Sec.~\ref{sec:t4t5_skyrme} the energy density can be written as the previous one plus the additional contributions. Consequently, the new $K_\infty$ value will be the of the same form with additional terms. 
Eq.~\eqref{eq:energy_density_skyrme_parameters_t4t5} yields (and dropping the INM notation):
\begin{align*}
    \mathcal{E}'_\mathrm{Sky} &= \mathcal{E}_\mathrm{Sky} + \mathcal{E}_\mathrm{4} + \mathcal{E}_\mathrm{5}\\ &=\mathcal{E}_\mathrm{Sky} + \tfrac{1}{8}t_4((2+x_4)\rho\tau - (1+2x_4)(\rho_p\tau_p+\rho_n\tau_n))\rho^\beta \\&\qquad\quad + \tfrac{1}{8}t_5((2+x_5)\rho\tau+(1+2x_5)(\rho_p\tau_p+\rho_n\tau_n)) \rho^\gamma\;.
\end{align*}
The derivation of the $K_\infty$ parameter is provided in the appendix of~\cite{Chamel2009}, which indeed has the first few terms identical to what we have before. To verify this, we write down the new contributions:
\begin{align*}
    \mathcal{E}_4&=\tfrac{1}{8}t_4[(2+x_4)\tfrac{3}{5}(3\pi^2)^{2/3}F_{5/3}\rho^{5/3+1}2^{-2/3} - (1+2x_4)\tfrac{3}{5}(3\pi^2)^{2/3}F_{8/3}\rho^{8/3}2^{-5/3}]\rho^\beta\\
    &=\tfrac{3}{40}(\tfrac{3\pi^2}{2})^{2/3}t_4\rho^{\beta+8/3}[(2+x_4)F_{5/3} - (\tfrac{1}{2}+x_4)F_{8/3}]\;,
\end{align*}
and then $9\rho_0$ times second order derivative, yields:
\[
K_\infty = \ldots + 
\tfrac{3}{40}(\tfrac{3\pi^2}{2})^{2/3}t_4(3\beta+5)(3\beta+2)\rho_*^{\beta+5/3}[(2+x_4)F_{5/3} - (\tfrac{1}{2}+x_4)F_{8/3}]\;.
\]
The $t_5$ contribution yields\footnote{Note the error in~\cite[Eq.~(A.13)]{Chamel2009}: $k_F$ instead of $k_F^2$.}:
\[ K_\infty = \ldots + 
\tfrac{3}{40}(\tfrac{3\pi^2}{2})^{2/3}t_5(3\gamma+5)(3\gamma+2)\rho_*^{\gamma+5/3}[(2+x_5)F_{5/3} + (\tfrac{1}{2}+x_5)F_{8/3}]\;.
\]

This can be compared to~\cite{Chamel2009} if we consider symmetric nuclear matter ($F_m=1$), to obtain:
\[
K_{\infty,\,45} = K_{\infty,\,\mathrm{NLO}} +  \tfrac{9}{80}(\tfrac{3\pi^2}{2})^{2/3}t_4(3\beta+5)(3\beta+2)\rho_*^{\beta+5/3}
+  \tfrac{3}{80}(\tfrac{3\pi^2}{2})^{2/3}t_5(3\gamma+5)(3\gamma+2)(4x_5 +5)\rho_*^{\gamma+5/3} \;.
\]

Again, recall that $\rho_*$ is the minimum of $\mathcal{E}/\rho$, such that:
\begin{align*}
0 &= \tfrac{\hbar^2}{2m}\tfrac{2}{5}(\tfrac{3\pi^2}{2})^{2/3}\rho_*^{-1/3}+\tfrac{3}{8}t_0 + \tfrac{1}{16}(\alpha+1)t_3\rho_*^{\alpha}+\tfrac{1}{16}(\tfrac{3\pi^2}{2})^{2/3}\rho_*^{2/3}[3t_1+5t_2+4t_2x_2]\\
&\qquad + \tfrac{3}{40}(\tfrac{3\pi^2}{2})^{2/3}t_4(\beta+\tfrac{5}{3})\rho_*^{\beta+2/3}((2+x_4)F_{5/3}-(\tfrac{1}{2}+x_4)F_{8/3})\\&\qquad+ \tfrac{3}{40}(\tfrac{3\pi^2}{2})^{2/3}t_5(\gamma+\tfrac{5}{3})\rho_*^{\gamma+2/3}((2+x_5)F_{5/3}+(\tfrac{1}{2}+x_5)F_{8/3})\;.
\end{align*}

Then similarly, for $K_A$ we find:
\begin{align*}
	P = P_\mathrm{NLO} &+ \tfrac{1}{3}(3\beta+5)\tfrac{3}{40}(\tfrac{3\pi^2}{2})^{2/3}t_4\rho^{\beta+8/3}[(2+x_4)F_{5/3} - (\tfrac{1}{2}+x_4)F_{8/3}]\\
	&+(\gamma+\tfrac{5}{3})\tfrac{3}{40}(\tfrac{3\pi^2}{2})^{2/3}t_5\rho^{\gamma+8/3}[(2+x_5)F_{5/3} + (\tfrac{1}{2}+x_5)F_{8/3}]\;,
\end{align*}%
\begin{align*}
K_{A,\,45} = K_{A, \mathrm{NLO}} &+ \tfrac{3}{40}(\tfrac{3\pi^2}{2})^{2/3}t_4(3\beta+5)\{3\beta+8\}\rho^{\beta+5/3}[(2+x_4)F_{5/3} - (\tfrac{1}{2}+x_4)F_{8/3}]\\ & + \tfrac{3}{40}(\tfrac{3\pi^2}{2})^{2/3}t_5(3\gamma+5)\{3\gamma+8\}\rho^{\gamma+5/3}[(2+x_5)F_{5/3} + (\tfrac{1}{2}+x_5)F_{8/3}] \;.
\end{align*}

\subsection{N2LO extension}
This type of functional has additional contributions from the $\mathcal{E}^{(4)}_e$ term, as shown in Subsec.~\ref{subsec:N2LO}. We will take the second-order derivative of $\mathcal{E}^{(4)}_e/\rho$ w.r.t.~$\rho$, which will yield additional contributions to $K_\infty$.

The higher-order kinetic density term $Q$ occurring in Subsec.~\ref{subsec:N2LO} can be written as follows~\cite{Grams2025_notes}: $Q_q~=~\tfrac{1}{7\pi^2}k^7_{F,q}$, with $k_{F,q}~=~(3\pi^2\rho_q)^{1/3}$, leading to $Q_q~=~\tfrac{1}{7\pi^2}(3\pi^2\rho_q)^{7/3}$. Using Eqs.~\eqref{eq:F_ma} and~\eqref{eq:F_mb}, one finds:
\begin{align*}
Q_0 &= Q_n+Q_p = \tfrac{1}{7\pi^2}(3\pi^2)^{7/3} 2^{-4/3}F_{7/3}\rho_0^{7/3} = \tfrac{3}{7}(\tfrac{3\pi^2}{2})^{4/3} F_{7/3}\rho_0^{7/3} \;,\\ 
Q_1 &= Q_n-Q_p = \tfrac{3}{7}(\tfrac{3\pi^2}{2})^{4/3} \tilde{F}_{7/3}\rho_0^{7/3}\;.
\end{align*}
Together with $\tau_{q,\mu\mu} = \tfrac{1}{3}\tau_q$, the additional N2LO contribution to the EDF simplifies as follows:
\begin{align*}
    \mathcal{E}^{(4)}_e &= C^{\tau\tau}_0(\tau^2_0 +2\sum_{\mu\mu}(\tfrac{1}{3}\tau_{0})^2+C^{\tau\tau}_1(\tau^2_1+2\sum_{\mu\mu}(\tfrac{1}{3}\tau_{1})^2) + C^{\rho Q}_0 \rho_0Q_0 + C^{\rho Q}_1 \rho_1Q_1\\
    &= \tfrac{5}{3}C^{\tau\tau}_0\tau^2_0+\tfrac{5}{3}C^{\tau\tau}_1\tau^2_1 + C^{\rho Q}_0 \rho_0Q_0 + C^{\rho Q}_1 \rho_1Q_1\\
    &=\tfrac{5}{3}C^{\tau\tau}_0\tfrac{9}{25}(\tfrac{3\pi}{2})^{4/3}F_{5/3}^2\rho_0^{10/3}+\tfrac{5}{3}C^{\tau\tau}_1\tfrac{9}{25}(\tfrac{3\pi}{2})^{4/3}\tilde{F}_{5/3}^2\rho_0^{10/3} \\&\qquad+ C^{\rho Q}_0 \tfrac{3}{7}(\tfrac{3\pi^2}{2})^{4/3} F_{7/3}\rho_0^{10/3} + C^{\rho Q}_1 \tfrac{3}{7}(\tfrac{3\pi^2}{2})^{4/3} \tilde{F}_{7/3}\rho_0^{7/3}\rho_1\\[0.5em]
    &= \tfrac{3}{5}(\tfrac{3\pi}{2})^{4/3}\rho_0^{10/3}(C_0^{\tau\tau}F^2_{5/3}+C_1^{\tau\tau}\tilde{F}^2_{5/3})+\tfrac{3}{7}(\tfrac{3\pi^2}{2})^{4/3}\rho_0^{10/3}(C_0^{\rho Q} F_{7/3}+ C^{\rho Q}_1 \tilde{F}_{7/3}\eta)\\
    &= \tfrac{3}{5}(\tfrac{3\pi}{2})^{4/3}\rho_0^{10/3}(C_0^{\tau\tau}F^2_{5/3}+C_1^{\tau\tau}\tilde{F}^2_{5/3})+\tfrac{3}{7}(\tfrac{3\pi^2}{2})^{4/3}(\rho_0^{10/3}C_0^{\rho Q} F_{7/3}+ \rho_0^{7/3}C^{\rho Q}_1 \tilde{F}_{7/3}\rho_1)\;.
\end{align*}
We emphasize that the derivative is taken assuming $\rho_1 / \rho_0 = \eta$ is a constant. Replacing $\rho_0$ by $\rho$, we obtain for the derivatives:
\begin{align*}
    \frac{\partial \mathcal{E}^{(4)}_e/\rho}{\partial \rho} &= \tfrac{7}{5}(\tfrac{3\pi}{2})^{4/3}\rho^{4/3}(C_0^{\tau\tau}F^2_{5/3}+C_1^{\tau\tau}\tilde{F}^2_{5/3})+(\tfrac{3\pi^2}{2})^{4/3}\rho^{4/3}(C_0^{\rho Q} F_{7/3}+C_1^{\rho Q}\eta\tilde{F}_{7/3})\;,\\
    \frac{\partial^2 \mathcal{E}^{(4)}_e/\rho}{\partial \rho^2} &= \tfrac{28}{15}(\tfrac{3\pi}{2})^{4/3}\rho^{1/3}(C_0^{\tau\tau}F^2_{5/3}+C_1^{\tau\tau}\tilde{F}^2_{5/3})+\tfrac{4}{3}(\tfrac{3\pi^2}{2})^{4/3}\rho^{1/3}(C_0^{\rho Q} F_{7/3}+C_1^{\rho Q}\eta\tilde{F}_{7/3})\;,\\[1em]
    K_{\infty,\,\mathrm{N2LO}} &=  K_{\infty,\,\mathrm{NLO}} + \tfrac{84}{5}(\tfrac{3\pi}{2})^{4/3}\rho_*^{7/3}(C_0^{\tau\tau}F^2_{5/3}+C_1^{\tau\tau}\tilde{F}^2_{5/3})+12(\tfrac{3\pi^2}{2})^{4/3}\rho_*^{7/3}(C_0^{\rho Q} F_{7/3}+C_1^{\rho Q}\eta\tilde{F}_{7/3})\;, 
\end{align*}

\begin{equation*}
    P = \rho_0^2\frac{\partial \mathcal{E}^{(4)}_e/\rho}{\partial \rho} = \tfrac{7}{5}(\tfrac{3\pi}{2})^{4/3}\rho_0^{10/3}(C_0^{\tau\tau}F^2_{5/3}+C_1^{\tau\tau}\tilde{F}^2_{5/3})+(\tfrac{3\pi^2}{2})^{4/3}\rho_0^{10/3}(C_0^{\rho Q} F_{7/3}+C_1^{\rho Q}\eta\tilde{F}_{7/3})\;,
\end{equation*}

\begin{align*}
    K_{A,\,\mathrm{N2LO}} &=  K_{A,\,\mathrm{NLO}}+  \tfrac{18*7}{5}(\tfrac{3\pi}{2})^{4/3}\rho^{7/3}(C_0^{\tau\tau}F^2_{5/3}+C_1^{\tau\tau}\tilde{F}^2_{5/3})+18(\tfrac{3\pi^2}{2})^{4/3}\rho^{7/3}(C_0^{\rho Q} F_{7/3}+C_1^{\rho Q}\eta\tilde{F}_{7/3}) \\
    &\qquad+\tfrac{84}{5}(\tfrac{3\pi}{2})^{4/3}\rho^{7/3}(C_0^{\tau\tau}F^2_{5/3}+C_1^{\tau\tau}\tilde{F}^2_{5/3})+12(\tfrac{3\pi^2}{2})^{4/3}\rho^{7/3}(C_0^{\rho Q} F_{7/3}+C_1^{\rho Q}\eta\tilde{F}_{7/3}) \\
    &= K_{A,\,\mathrm{NLO}} +42(\tfrac{3\pi}{2})^{4/3}\rho^{7/3}(C_0^{\tau\tau}F^2_{5/3}+C_1^{\tau\tau}\tilde{F}^2_{5/3})+30(\tfrac{3\pi^2}{2})^{4/3}\rho^{7/3}(C_0^{\rho Q} F_{7/3}+C_1^{\rho Q}\eta\tilde{F}_{7/3})\;.
\end{align*}
We obtain the same result as in~\cite{Grams2025_notes}.

\subsection{Verification}
We have verified our expressions for $K_\infty$ and $K_A(\rho=\rho^*)$ by evaluating them and comparing with the values reported in literature.  This was done for various parametrisations, more specifically SLy4, BSk, BSkG and those presented in~\cite{Colo2004}. These parametrisations cover the different EDF forms, including NLO, N2LO and the $t_4$-$t_5$ extended Skyrme.

\section{Discussion and conclusion}\label{sec:K_conclusion} 
We proceed to verify the correlation discussed in~\cite{Colo2004} which was also confirmed in~\cite{Zamora2024, Sagawa2019} for Gogny and relativistic Skyrme models. Using the expressions derived in the previous section, we calculate $K_\infty$ for each Skyrme functional and parametrisation. 

Next, we calculate the mean-field solutions (including those from~\cite{Colo2004}) of \ce{Pb-208} and apply the FAM code to obtain ISGMR spectrum. From this, the centroid energy of the resonance is extracted using the formulas in Sec.~\ref{sec:K_theory}. Following~\cite{Colo2004, Sagawa2019}, we report $\sqrt{\tfrac{m_1}{m_{-1}}}$ as the centroid energy.

\begin{figure}
    \centering
    \includegraphics[width=0.49\linewidth]{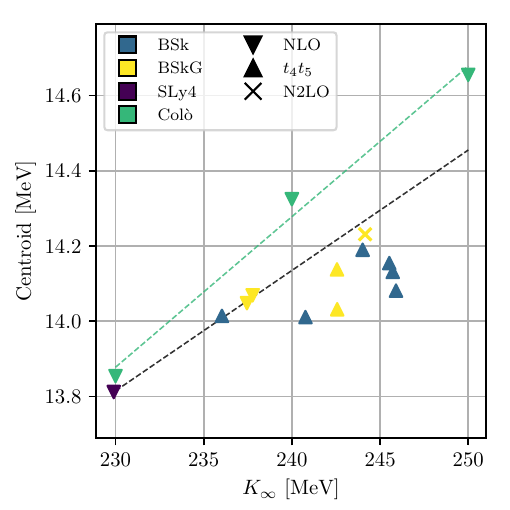}
    \caption{Centroid position of \ce{Pb-208} for different parametrisations (grouped, color) and EDF types (symbols). A linear fit is included to indicate a correlation for the parametrisation from~\cite{Colo2004}. The other linear fit uses the standard Skyrmes: BSkG1, BSkG2 and SLy4.}
    \label{fig:K_inf_trend}
\end{figure}

The results are summarised in Fig.~\ref{fig:K_inf_trend}. We first consider three models from~\cite{Colo2004} that share the same density dependence and symmetry energy $J$, which is also used for the BSk(G) models ($\alpha=1/6, J=32$\,MeV). These results are marked in green on the figure. We confirm that the correlation is present for these models, although our calculated centroid positions are slightly larger than those originally reported.

In contrast, the centroids obtained with the $t_4$-$t_5$ and N2LO EDFs (i.e.~BSk and BSkG3-5 parametrisations) show little to no correlation. While a correlation might be argued if one considers only the standard Skyrme EDFs, it appears weaker than the one from~\cite{Colo2004}. Given that we only have three data points, this conclusion must be treated with caution. 

The analysis in~\cite{Colo2004} concluded that maintaining correct centroid energies at higher $K_\infty$ requires a larger symmetry energy. Because this can lead to physically unrealistic symmetry energy values, an upper bound for $K_\infty$ is justified. However, we conclude that this correlation no longer holds for the non-standard EDF forms. This implies that we can maintain the same symmetry energy while increasing $K_\infty$, without compromising the centroid energy. Consequently, the assumption that $K_\infty$ must be constrained within 240$\pm$10\;MeV can be questioned.

Additionally, we previously noted that the $K_\infty$ of the extended Skyrme EDFs take the form $K_\mathrm{NLO} + K_\mathrm{rest}$. These contributions are summarised in Tab.~\ref{tab:K_inf_tab}. Recall that the parameters are fitted such that $K_\mathrm{NLO}+K_\mathrm{rest}$ is close to 240\,MeV. Remarkably, the $t_4$-$t_5$ models except BSk31, BSkG2 and BSkG3 show large negative and large positive values for both $K_\mathrm{NLO}$ and $K_\mathrm{rest}$ respectively, and precisely compensate to satisfy the constraint of summing to roughly 230\,MeV.  Consequently, the inclusion of $t_4$-$t_5$ terms allows for a significantly broader exploration of the parameter space, whereas other models effectively behave as standard NLO forms with minor perturbations, at least as far as $K_\infty$ is concerned.

Admittedly, this analysis is not exhaustive and should be interpreted carefully. These are only indications that some of the model assumptions are not well understood. We note that our findings are in line with~\cite{Margueron2026}. To better understand this issue, one should probe the effect of imposing different values of $K_\infty$  directly on the fitting procedure, rather than after as we have done here. However, fitting an EDF is a highly involved process and is out of the scope of this thesis.

Novel techniques to accurately measure ISGMR of unstable nuclei~\cite{Zamora2024} could improve the understanding of $K_\infty$. Regarding the fitting procedure of EDF parameters, it may suffice to omit $K_\infty$ as an explicit constraint and instead use the centroid positions of known resonances. While this has not been done previously due to the high computational effort to calculate the strength function, the emulator developed in Chap.~\ref{chap:emulator}, or applying the complex integration method from~\cite{Hinohara2015}  may make such an approach feasible. 

\begin{table}[]
    \caption{Values for $K_\infty$ separated in $K_\mathrm{NLO}$ and the $K_\mathrm{rest}$ part for the different parametrisations and EDF types. To verify the mean-field properties, the Fermi level and highest represented single-particle energies are reported, both (neutron, proton).}
    \label{tab:K_inf_tab}
    \centering
    \begin{tabular}{l l r r r r}
    \toprule
    Name& Type & \multicolumn{1}{c}{Fermi level} & \multicolumn{1}{c}{Highest spe} & $K_\mathrm{NLO}$ & $K_\mathrm{rest}$ \\
                && (MeV, MeV)                      & (MeV, MeV)                                  &                (MeV)  &  (MeV)                \\
    \midrule 
        K230~\cite{Colo2004}& NLO & (-5.51, -6.45) & (33.4, 34.5) & 230.00 & -\\
        K240~\cite{Colo2004}&NLO& (-5.35, -6.32) & (33.8, 34.7) & 240.00 & -\\
        K250~\cite{Colo2004}&NLO & (-5.20, -6.37) & (34.1, 35.0) & 250.00 & -\\
      \arrayrulecolor{black!40}\midrule
        SLy4&  NLO& (-5.60, -6.31) & (33.5, 34.6) & 229.90 & -\\
      \arrayrulecolor{black!40}\midrule
        BSk22& $t_4$-$t_5$ & (-5.69, -5.74) & (32.5, 34.3) & -1941.70&2187.60\\
        BSk23& $t_4$-$t_5$ & (-5.67, -5.74) & (32.5, 34.3) & -1942.00& 2187.72\\
        BSk24& $t_4$-$t_5$ & (-5.52, -5.91) & (32.6, 34.2) & -1942.13& 2187.65\\
        BSk25& $t_4$-$t_5$ & (-5.70, -5.70) & (32.6, 34.3) & -1895.55& 2131.59\\
        BSk26& $t_4$-$t_5$ & (-5.80, -5.72) & (32.5, 34.4) & -1483.51& 1724.28\\
        BSk31& $t_4$-$t_5$ & (-5.65, -6.05) & (32.5, 34.3) & 255.76&-11.74\\
      \arrayrulecolor{black!40}\midrule
        BSkG1&  NLO& (-5.67, -6.03) & (32.6, 34.0) & 237.78 & - \\
        BSkG2&  NLO& (-5.69, -6.04) & (32.7, 33.9) & 237.45 & -\\
        BSkG3&  $t_4$-$t_5$& (-5.60, -6.04) & (32.3, 33.9) & 240.51 & 2.05\\
        BSkG4&  $t_4$-$t_5$& (-5.61, -5.99) & (32.3, 33.9) & 220.79 & 21.78\\
        BSkG5&  N2LO& (-5.68, -6.05) & (32.7, 33.9) & 191.24&52.91\\
    \arrayrulecolor{black}\bottomrule
    \end{tabular}
\end{table}
\chapter{Conclusions and outlooks}\label{chap:5}
\section{Emulator}
The main goal of this thesis was the development of a ROM emulator for FAM-QRPA~\cite{code_github}, which can be applied to obtain the full strength function for a fixed nucleus, EDF and excitation operator. 

We successfully worked out a formalism to apply ROM directly to the FAM amplitudes of the linear response equations. To the best of our knowledge, this approach has not been done before. We discovered that the approximation completely removes the necessity for additional FAM-QRPA calculations beyond those for constructing emulator itself. We compared the Galerkin to the Least-Squares Petrov-Galerkin projection method, which yielded similar performance. The main difference is that the Petrov-Galerkin avoids generation of non-existing peaks, and is therefore preferred. To resolve numerical instabilities, applying a singular value decomposition to the basis vectors is effective. A crucial step in the approximation of new FAM amplitudes is that the necessity of additional FAM-QRPA calculations beyond the emulator basis can be avoided by exploiting the FAM equation. 

We realised that the method for snapshot selection strongly affects the emulator's performance. We found that a straightforward generalisation of a 1D-greedy approach does not work. Instead, we have developed a pseudo-2D greedy search for snapshot selection, allowing the emulator to combine snapshots taken in the complex-frequency plane. This strategy effectively manages the trade-off between computational speed and information gain. The benchmark indicates a speed-up of at least a factor 20 compared to applying FAM-QRPA. We note that the benchmark underestimates the emulator's performance and that a thorough examination will be done in the near future. 

While the emulator does not recover symmetries of the strength function, these can be recovered by adding new  snapshots without additional cost. Additionally, the results show that the methodology is robust for different multipole operators, functional forms, mean field basis, inclusion of pairing effects and for deformed nuclei.

\paragraph{Outlook}
The emulator allows for a considerable speed-up for the calculations of strength functions, making large-scale calculations more feasible. 

A first step to potentially further improve the emulator is to apply it for the FAM initialisation. More precisely, since the FAM procedure is an iterative method, the number of iterations could be reduced by improving the initial guess. It is not clear how much this can reduce the computational cost, though, it seems to be rather easy to implement for the MOCCa code~\cite{RyssensMOCCa}. 

The next step of this work is the application of the emulator to charge-exchange operators, which is necessary to model $\beta$-decay strength functions. We expect the methodology to work without major issues; however, this remains to be verified. 

Furthermore, to examine the effect of varying EDF parameters, the methodology proposed in~\cite{Jin2025} can be improved by combining it with our approach to obtain full strength functions with fewer FAM-QRPA calculations. 

Finally, due to its mathematical flexibility and generalizability, we believe this emulator is not limited to nuclear physics. Linear response and RPA theory also appear in quantum chemistry and condensed matter physics~\cite{chen, Ziegler, Ren_2012}. Therefore, a straightforward application of our emulator should be explored. 

\section{Infinite matter incompressibility}
In the second part of the thesis, we re-examined the correlation between the infinite nuclear incompressibility $K_\infty$ and the ISGMR centroid energies of \ce{Pb-208}. While this correlation is found for standard and relativistic Skyrmes~\cite{Colo2004}, we found that it longer holds for extended Skyrme functionals like $t_4$-$t_5$ and N2LO. This suggests that the traditional constraint on $K_\infty$ may be questioned for these functionals, and implies that we could allow different $K_\infty$ values without compromising the ISGMR centroid position. We also noticed that for most of the BSk models, the $t_4$-$t_5$ terms allowed for a broader exploration in the EDF parameter space, which was evidenced by the different contributions to $K_\infty$ from the $K_\mathrm{NLO}$ and $K_\mathrm{rest}$. 

\paragraph{Outlook}
A first idea is to remove the $K_\infty$ constraint from the fitting procedure, and to replace it with the ISGMR centroid of \ce{Pb-208} instead. However, as this centroid position cannot be computed directly, the FAM-QRPA equations need to be solved. This adds a significant computational cost to the fitting procedure. Even with the emulator or with the integration method of~\cite{Hinohara2015}, this will most likely not be feasible in the foreseeable future. 

A more thorough analysis of $K_\infty$ and its effect on the EDF fit is required. Recent efforts are made to improve our understanding of $K_\infty$ on both the experimental and theoretical parts. For example, novel techniques for measuring ISGMR of unstable nuclei~\cite{Zamora2024} are developed.  Recent studies propose a different relation between $K_\infty$ and the ISGMR~\cite{Khan2012, Khan2013, Margueron2026}, which would effectively lead to a looser constraint in the EDF fitting procedure. 


\backmatter

\printbibliography

\restoregeometry
\newpage
\thispagestyle{empty}
\sffamily
\begin{textblock}{191}(113,-11)
{\color{blueline}\rule{160pt}{5.5pt}}
\end{textblock}
\begin{textblock}{191}(168,-11)
{\color{blueline}\rule{5.5pt}{59pt}}
\end{textblock}
\begin{textblock}{183}(-30,-7)
\textblockcolour{}
\flushright
\fontsize{7}{7.5}\selectfont
\textbf{FACULTEIT WETENSCHAPPEN}\\
KU Leuven\\
Celestijnenlaan 200H - bus 2100\\
3001 LEUVEN (HEVERLEE), BELGI\"{E}\\
tel. +32 16 32 14 01\\
www.wet.kuleuven.be\\
\end{textblock}
\begin{textblock}{191}(155.5,-5.5)
\textblockcolour{}
\includegraphics*[height=16.5truemm]{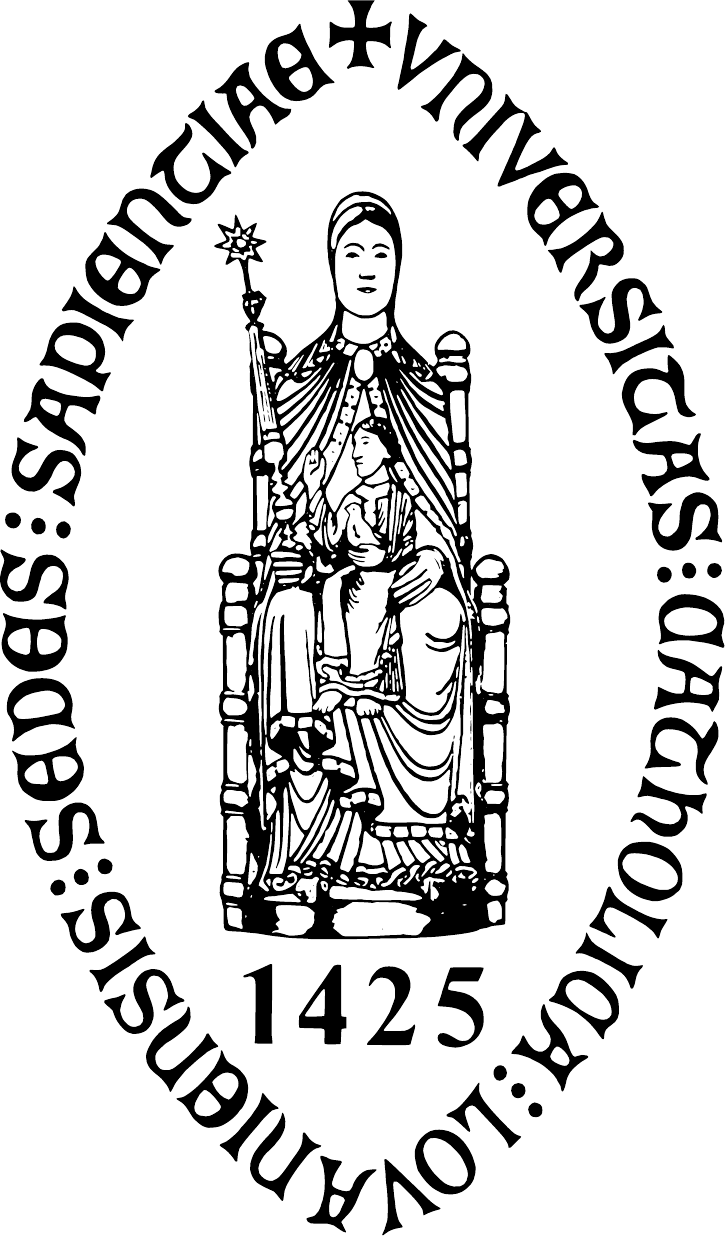}
\end{textblock}
\begin{textblock}{191}(-20,235)
{\color{bluetitle}\rule{544pt}{55pt}}
\end{textblock}
\end{document}